\documentclass{aastex}

\shorttitle{Optical Monitoring of 3C 273}
\shortauthors{Xiong et
al.}

\begin{document}
\title{Multi-color optical monitoring of the BL Lacertae object S5 0716+714 during the 2012 outburst}

\author{Shanwei Hong,}
\affil{Yunnan Observatories,
Chinese Academy of Sciences, 396 Yangfangwang, Guandu District, Kunming, 650216, P. R. China}
\affil{University of Chinese Academy of Sciences, Beijing 100049, China}
\affil{Department of Mathematics and Computer Science, Lijiang Teachers College, Lijiang 674199, China}

\author{Dingrong Xiong and Jinming Bai}
\affil{Yunnan Observatories,
Chinese Academy of Sciences, 396 Yangfangwang, Guandu District, Kunming, 650216, P. R. China}
\affil{Center for Astronomical Mega-Science, Chinese Academy of Sciences, 20A Datun Road, Chaoyang District, Beijing, 100012, P. R. China}
\affil{Key Laboratory for the Structure and Evolution of Celestial
Objects, Chinese Academy of Sciences, 396 Yangfangwang, Guandu District, Kunming, 650216, P. R. China}\email{xiongdingrong@ynao.ac.cn}

\begin{abstract}
We have monitored the BL Lacertae object S5 0716+714 in the optical bands during 2012 January and February with long time spans on intraday timescales ($>$5 hr) and high time resolutions. During this monitoring period, the object
shows violent flaring activity both in short and intraday timescales. The object has high value of duty cycle. The light curves detected as intraday variability (IDV) show variability of various shapes. The variability amplitude is from 12.81 per cent to 33.22 per cent, and the average value $19.92\pm5.87$ per cent. The overall magnitude variabilities are $\bigtriangleup B=1^{\rm m}.24$, $\bigtriangleup V=1^{\rm m}.42$, $\bigtriangleup R=1^{\rm m}.3$, $\bigtriangleup I=1^{\rm m}.23$ respectively. During the observations, the average change rate is $<CR>=0.035\pm0.009$ Mag/h during the ascent and $<CR>=0.035\pm0.014$ Mag/h during the descent. However, different cases are found on certain nights.  There are good inter-bands correlations but not significant time lags for intraday and short timescales. The results of the autocorrelation function show that the variability timescales range from 0.054 day to 0.134 day. Most of nights show bluer when brighter (BWB) chromatic trend; a weak redder with brighter (RWB) trend is found; a few nights show no correlations between magnitude and color index. The BWB trend appears in the short timescales. During the flare the spectral index exhibits a clockwise loop for inter-nights. A shock-in-jet model and the shock wave propagating along a helical path are likely to explain the variability and color index variability.

\end{abstract}


\section{INTRODUCTION}

BL Lacertae (BL Lac) objects are the most extreme subclass of active galactic nuclei (AGNs), hosted in massive elliptical galaxies, the emission of which is dominated by a relativistic jet closely aligned with
the line of sight. This implies the existence of a parent population of sources
with a misaligned jet that have been identified with low-power Fanaroff-Riley type I
radio galaxies (FR Is). A most distinctive characteristic of the class is the weakness or
absence of spectral lines that historically hindered the identification of their
nature and ever thereafter proved to be a hurdle in the determination of their
distance (Falomo et al. 2014; Angel \& Stockman 1980; Urry \& Padovani 1995). The spectrum of BL Lac objects, dominated by non-thermal emission over the whole electromagnetic range, together with bright compact radio cores, high
luminosity, rapid and large amplitude flux variability at all frequencies and
strong polarization make these sources become an optimal laboratory for high energy
astrophysics. The broadband spectral energy distributions (SEDs) of BL Lac objects have a double peaked structure. The low energy peak at the IR-optical-UV band is explained with the synchrotron emission of relativistic electrons, and the high energy peak at the GeV-TeV gamma-ray band due to the inverse Compton (IC) scattering (e.g., Dermer et al. 1995; Dermer et al. 2002; Bottcher 2007). The hadronic model is an alternative explanation for the high energy emissions from BL Lac objects (e.g., Dermer et al. 2012). According to the difference of synchrotron peak frequency, BL Lac objects can be divided into low-synchrotron-peaked (LSP, $\nu^s_{\rm{peak}}<10^{14}{\rm Hz}$), intermediate-synchrotron-peaked (ISP, $10^{14}{\rm Hz}<\nu^s_{\rm{peak}}<10^{15}{\rm Hz}$) and high-synchrotron-peaked (HSP, $10^{15}{\rm Hz}<\nu^s_{\rm{peak}}$) three categories (Abdo et al. 2010).

The variabilities of BL Lac objects at frequencies ranging from radio to TeV bands have been detected (e.g., Rani et al. 2013; Liao et al. 2014; Bartoli et al. 2016). The timescales of variabilities are from years down to minutes (e.g., Poon et al. 2009; Zhang et al. 2008a; Fan et al. 2005, 2009a, 2014). Brightness changes of a few tenths or hundredths of a magnitude during hours or less are often known as intraday variability (IDV) or microvariability (Wagner \& Witzel 1995). Short-term variability (STV) has timescales of days to weeks, even months, and long-term variability (LTV) ranges from months to years (Gupta et al. 2008a; Dai et al. 2015). IDV has been confirmed as intrinsic nature of the BL Lac objects and has become a subject of intense activity as its physical mechanisms are not understood well (e.g., Chandra et al. 2011; Fan et al. 2009b; Zhang et al. 2008b; Bai et al. 1998; Dai et al. 2015; Xiong et al. 2016). The shortest variability (microvariability) time-scales are important for understanding the
geometry of jets, the magnetic field and black hole mass, because it provides a possible minimum size of variation sources (e.g., Gupta et al. 2009; Rani et al. 2010; Fan et al. 2009b, 2014; Zhang et al. 2008b; Dai et al. 2015; Xiong et al. 2016). The spectral index (or color behavior) is an important but a simple factor to explore the variability mechanism (e.g., Gu \& Ai 2011; Zheng et al. 2008; Wu et al. 2007).

S5 0716+714 (R.A.=$07^{\rm h}21^{\rm m}53^{\rm s}.4$, decl.=$71^{\rm \circ}20^{\rm '}36^{\rm ''}$, J2000) is classified as an ISP BL Lac object ($\nu^s_{\rm{peak}}=10^{14.6}{\rm Hz}$; Abdo et al. 2010). It was discovered in the Bonn-NRAO Radio Survey of flat-spectrum radio sources with a 4.9 GHz flux greater than 1 Jy (Kuehr et al. 1981). Radio maps detected a compact core-jet structure and an extended emission resembling an FR II object (Antonucci et al. 1986; Gabuzda et al. 1998). The source has a featureless optical
continuum with the redshift $0.31\pm0.08$ derived by using the host galaxy as a standard candle (Nilsson
et al. 2008). Danforth et al. (2013) used intervening absorption systems to set the redshift range $0.2315<z<0.3407$. It is one of the brightest BL Lac objects which is highly variable from the radio to $\gamma$-ray bands with very high duty cycle (Wagner \& Witzel 1995). Because of its high power and lack of signs for ongoing accretion
or surrounding gas, the source is an ideal candidate to explore the multi-wavelength non thermal emission. The flux and spectral variations of S5 0716+714 have been extensively studied over the entire electromagnetic spectrum (Wagner \& Witzel 1995; Wagner et al. 1996; Heidt \& Wagner 1996; Ghisellini et al. 1997; Sagar et al. 1999; Qian et al. 2002; Raiteri et al. 2003; Wu et al. 2005, 2007; Nesci et al. 2005; Gu et al. 2006; Montagni et al. 2006; Ostorero et al. 2006; Foschini et al. 2006; Zhang et al. 2008b; Villata et al. 2000, 2008; Gupta et al. 2008b, 2009, 2012; Vittorini et al. 2009; Stalin et al. 2006, 2009; Anderhub et al. 2009; Carini et al. 2011; Zhang et al. 2012; Hu et al. 2014; Dai et al. 2013; Larionov et al. 2013; Liao et al. 2014; Rani et al. 2010, 2014; Chandra et al. 2011, 2015; Dai et al. 2015; Bhatta et al. 2015, 2016; Wierzcholska \& Siejkowski 2015, 2016; Man et al. 2016; Agarwal et al. 2016; Li et al. 2017). From the observations of Poon et al. (2009), the object showed four fast flares with amplitudes ranging from 0.3 to 0.75 mag. Typical timescales of microvariability ranged from 2 to 8 hr. Strong bluer-when-brighter (BWB) chromatism was found on internight timescales. However, a different spectral behavior was found on intranight timescales. A possible time lag of $\sim$11 minutes between
$B$ and $I$ bands was found on one night. The observations from Chandra et al. (2011) suggested that S5 0716+714 showed night-to-night and intra-night variabilities at various time scales. Wu et al. (2007) found that the source showed two strong flares, with variation amplitudes of about 0.8 and 0.6 mag in the $V$ band, respectively; strong BWB correlations were found for both internight and intranight variations; no apparent time lag was observed between the $V$ and $R$ band variations, and the observed BWB chromatism may be mainly attributed to the larger variation amplitude at shorter wavelength. Man et al. (2016) found that variations in the $R$ and $B$ bands were approximately 1.5 min lagging behind the $I$ band for the object. But from results of Carini et al. (2011), there are no significant lags between the $B$ and $I$-band flux variations. During nearly continuous multiband observations, Bhatta et al. (2016) presented that the source displayed pronounced variability and BWB spectral evolution. The results from Hu et al. (2014) showed a strong BWB trend on intranight time-scales and a mild bluer-when-brighter on internight time-scales for the source. Dai et al. (2013) and Dai et al. (2015) found that BWB chromatism was observed in long, intermediate, and short timescales. Stalin et al. (2006) and Agarwal et al. (2016) found no evidence of spectral changes with the source brightness on either internight or intranight time-scales for the BL Lac object S5 0716+714 even
when the target was in flaring state.

In order to further explore characteristics of IDV and STV timescales, and spectral properties on both intranight and internight timescales, we monitored the source in multi-color optical bands during the 2012 outburst. Due to the high temporal resolution and long time spans on intraday timescales, more accurate results can be obtained. This paper is organized as follows. The observations and data analysis are described in Section 2. Section 3 presents the results. Discussion and conclusions are reported in Section 4.

\section{OBSERVATIONS AND DATA ANALYSIS}

Our optical observations were carried out using a 60 cm BOOTES-4 auto-telescope which is located at the
Lijiang Observatory of Yunnan Observatories of Chinese Academy of Sciences, where the longitude is $100^\circ01^{'}51^{''}$E and the latitude $26^\circ42^{'}32^{''}$N, with an altitude of 3193 m. Its main objective is to observe the gamma-ray bursts (GRBs) and blazars. Further details about the telescope are given in Table 1. During our observations, the telescope was equipped with standard Johnson UBV and Cousins RI filters. The optical observations in the $B$, $V$, $R$ and $I$ bands were in a corresponding cyclic mode. Time resolutions for most of nights are less than six minutes, and time spans on a night more than five hours (Table 2). The time intervals between $V$ and $R$ bands range from 30 s to 131 s, and most of nights have time intervals less than 50 s. The typical exposure times in the $B$, $V$, $R$ and $I$ bands are 60 s, 40 s, 40 s, 40 s respectively. So our observations with high temporal resolution (few minutes) can be considered as quasi-simultaneous measurements. All images have been prereduced with the CCDRED package, and then observing data were processed using the photometric tool, DAOPHOT, in the
IRAF\footnote{\scriptsize{IRAF is distributed by the National Optical
Astronomy Observatories, which are operated by the Association of
Universities for Research in Astronomy, Inc., under cooperative
agreement with the National Science Foundation.}} software package. The flat-field images were taken at dusk and dawn when possible. The bias images were taken at the beginning and the end of the observation. In order to determine aperture radius, we used different aperture radii (1.5$\times$FWHM, 1.7$\times$FWHM, 2$\times$FWHM, 2.5$\times$FWHM, 3$\times$FWHM) to carry out aperture photometry for all observations on per night. We found less aperture radius with better mean S/N ratio on a night, i.e., 1.5$\times$FWHM had the best mean S/N ratio on a night. Moreover, we plotted the aperture radii (from 1$\times$FWHM to 4$\times$FWHM) versus magnitudes for different frames on a night. The results showed that the change rate of increasing brightness was rapid from 1$\times$FWHM to 1.7$\times$FWHM, and quickly slow down after 1.7$\times$FWHM. When the aperture radius was near 3$\times$FWHM, the brightness almost kept constant. So considering the best S/N ratio and constant brightness, we made a compromise, i.e., the aperture radius of 1.7$\times$FWHM was selected. In order to obtain pure skylight background, the target and other objects should be excluded from the sky annulus. The inner radius and width of sky annulus were 5$\times$FWHM and 2$\times$FWHM respectively. After correcting flat-field and bias, aperture photometry was performed, and then instrumental aperture magnitudes were obtained. The finding chart of S5 0716+714 was found in the webpage\footnote{https://www.lsw.uni-heidelberg.de/projects/extragalactic/charts/0716+714.html}. From the above finding chart, we chose the marked 3 and 5 stars as comparison stars because their magnitudes were similar to that of the blazar, and the differential magnitude between 3 and 5 stars was the smallest variations among the comparison stars. Following Zhang et al. (2004, 2008a), Fan et al. (2014), Bai et al. (1998), the source magnitude was given as the average of the values derived with respect to the two comparison stars ($\frac{m_3+m_5}{2}$, $m_3$ was the blazar magnitude obtained from standard star 3 and $m_5$ from standard star 5). The magnitudes of comparison stars in the field of S5 0716+714 were taken from Villata et al. (1998) and Ghisellini et al. (1997). The observing uncertainty on each night was the rms error of the differential magnitude between two comparison stars. The rms errors
of the photometry on a certain night were calculated from the
two comparison stars, star 3 and star 5, in the usual way:
\begin{equation}
\sigma=\sqrt{\sum \frac{(m_i-\overline{m})^2}{N-1}},~~~~i=1,2,3,...,N,
\end{equation}
where $m_i = (m_{\rm 3}-m_{\rm 5})_{i}$ is the differential
magnitude of stars $3$ and $5$, while
$\overline{m}=\overline{m_{3}-m_{5}}$ is the averaged
differential magnitude over the entire data set, and $N$ is the
number of the observations on a given night. The actual number of observations for S5 0716+714 is 13 nights obtaining 683 $B$-band, 871 $V$-band, 874 $R$-band and 878 $I$-band data points. The results of observations are given in Table $3-6$ for filters $B$, $V$, $R$ and $I$.

In order to quantify the IDV of the object, we have employed three statistical analysis techniques ($C$ test, $F$ test and the one-way analysis of variance (ANOVA); e.g., de Diego 2010; Goyal et al. 2012; Hu et al. 2014; Agarwal \& Gupta 2015; Dai et al. 2015; Xiong et al. 2016). The blazar is considered as variability (V) if the light curve satisfies the three criteria of $C$-test, $F$-test and ANOVA. The blazar is considered as probably variable (PV) if only one of the above three criteria is satisfied. The blazar is considered as non-variable (N) if none of the criteria are met. Romero et al. (1999) introduced the
variability parameter, $C$, as the average value between $C_1$ and
$C_2$:
\begin{equation}
C_1=\frac{\sigma(BL-StarA)}{\sigma(StarA-StarB)},
C_2=\frac{\sigma(BL-StarB)}{\sigma(StarA-StarB)},
\end{equation}
where (BL-StarA), (BL-StarB) and (StarA-StarB) are the differential
instrumental magnitudes of the blazar and comparison star A, the
blazar and comparison star B, and comparison stars A and B. $\sigma$
is the standard deviation of the differential instrumental
magnitudes. The adopted variability criterion requires $C\geq2.576$,
which corresponds to a 99 per cent confidence level. Despite the very common use of the $C$-statistics, de Diego
(2010) has pointed out that it has severe problems. The $F$-test is considered to be a proper statistics to  quantify the IDV (e.g., de Diego 2010; Joshi et al. 2011; Goyal
et al. 2012; Hu et al. 2014; Agarwal \& Gupta 2015; Xiong et al. 2016). $F$ value is
calculated as
\begin{equation}
F_1=\frac{Var(BL-StarA)}{Var(StarA-StarB)},
F_2=\frac{Var(BL-StarB)}{Var(StarA-StarB)},
\end{equation}
where Var(BL-StarA), Var(BL-StarB) and Var(StarA-StarB) are the
variances of differential instrumental magnitudes. The $F$ value
from the average of $F_1$ and $F_2$ is compared with the critical
$F$-value, $F^\alpha_{\nu_{bl},\nu_\ast}$, where $\nu_{bl}$ and
$\nu_{\ast}$ are the number of degrees of freedom for the blazar and
comparison star respectively ($\nu=N-1$), and $\alpha$ is the
significance level set as 0.01 (2.6$\sigma$). If the average $F$
value is larger than the critical value, the blazar is variable at a
confidence level of 99 per cent. de Diego (2010) reported that the ANOVA is a powerful and
robust estimator for microvariations. It does not rely on error
measurement but derives the expected variance from subsamples of the data. The one-way ANOVA test divides the data into many groups. Then it compares the variances between inter-groups and intra-groups. From Appendix A3 of de Diego (2010), the statistics
\begin{equation}
F=\frac{\sum^k_{j=1}\sum^{n_j}_{i=1}(\bar{y_j}-\bar{y})^2/(k-1)}{\sum^k_{j=1}\sum^{n_j}_{i=1}(y_{ij}-\bar{y_j})^2/(N-k)}=\frac{SS_{\rm G}/(k-1)}{SS_{\rm R}/(N-k)},
\end{equation}
where $y_{ij}$ represents the $\textit{i}$th observation (with $\textit{i}$= 1, 2, ..., $n_j$ ) on the $\textit{j}$th group (with $\textit{j}$= 1, 2, ..., k), $\bar{y}$ the mean of the whole data set, $\bar{y_j}$ the mean on the $\textit{j}$th group, $k$ the number of groups, $N$ the number of the whole data set. The $SS_{\rm G}$ and $SS_{\rm R}$ are between-groups sum of squares and within-groups sum of squares respectively. Considering the time of exposure, we bin the data in a group of five observations (see Xiong et al. 2016 and de Diego 2010 for detail). If the measurements in the last group are less than 5, then it is merged with the previous group. The critical value of ANOVA can be obtained by $F^\alpha_{\nu_1,\nu_2}$ in the $F$-statistics, where $\nu_1=k-1$, $\nu_2=N-k$ and $\alpha$ is the significance level set as 0.01. if $F$ value from equation (4) exceeds the
critical value $F^\alpha_{\nu_1,\nu_2}$, the blazar is variable at a
confidence level of 99 per cent. In order to further
quantify the reliability of variability, the value of $S_x$ can be calculated as (e.g., Hu et al. 2014)
\begin{equation}
S_x=m_i-\overline{m},~~~x=V,R,I,
\end{equation}
where $m_i$ and $\overline{m}$ are same with equation (1). The variability amplitude (Amp) can be calculated by (Heidt \&
Wagner 1996)
\begin{equation}
{\rm Amp}=100\times \sqrt{(A_{max}-A_{min})^2-2\sigma^2}~{\rm per
cent},
\end{equation}
where $A_{max}$ and $A_{min}$ are the maximum and minimum magnitude,
respectively, of  the light curve for the night being considered,
and $\sigma$ is rms errors. When estimating the variability amplitude, we only consider the nights detected as variability.

The duty cycle (DC) is calculated as (Romero et al. 1999; Stalin et
al. 2009; Hu et al. 2014)
\begin{equation}
{\rm DC}=100\frac{\sum^n_{i=1}N_i(1/\triangle
T_i)}{\sum^n_{i=1}(1/\triangle T_i)}{\rm per~cent},
\end{equation}
where $\triangle T_i=\triangle T_{i,obs}(1+z)^{-1}$, $z$ is the
redshift of the object and $\triangle T_{i,obs}$ is the duration of
the monitoring session of the \emph{ith} night. Note that since for
a given source the monitoring durations on different nights were not
always equal, the computation of DC has been weighted by the actual
monitoring duration $\triangle T_i$ on the \emph{ith} night. $N_i$
will be set to 1 if intraday variability is detected, otherwise
$N_i=0$ (Goyal et al. 2013).

\section{RESULTS} \label{bozomath}
\subsection{Variability}

The analysis results on intraday variability are shown in Table 7. From Table 7, it can be seen that IDV is found on seven nights with at least two bands detected as IDV on a night. The light curves detected as IDV are given in Fig. 1 which shows variability of various shapes. For the seven nights, we also check the color
variations on intraday timescales. However, the results from three statistical tests do not show significant color variations on intraday timescales. The rest of nights are considered as PV except for the $I$ band on February 6. Generally, On January 27, the brightness first increases and then decreases along the arc. On January 28, the brightness first increases and then decreases, and last increases. On January 29, the brightness first decreases and then increases along the symmetrical arc. On January 30, the brightness changes are similar with that on January 29 but not along the symmetrical arc. On February 1 and 8, the brightness continues to increase while on February 4, the brightness continues to decrease. The corresponding changes of magnitude and change rate are seen in Table 8. When calculating the change rate, we first determine increasing or decreasing time points and then use the slope from errors weighted linear regression analysis as change rate. The results from Table 8 show that the average change rate is $<CR>=0.035\pm0.009$ Mag/h during the ascent and $<CR>=0.035\pm0.014$ Mag/h during the descent. The average change rates between the ascent and descent are equal. However, different cases are found on certain nights. The corresponding times of variation range from 47 min to 274 min. On January 29, the increasing and decreasing change rates are close. The correlations between variability amplitude and the source average brightness are shown in Fig. 2. Although Fig. 2 shows trends/tendencies of larger amplitude with brighter magnitude for $I$ and $R$ bands, the analysis of non-parametric Spearman rank indicates that there are not significant correlations between variability amplitude and the source average brightness for different wavebands (significance level $P_{\rm I}=0.1$, $P_{\rm R}=0.1$, $P_{\rm V}=0.8$, $P_{\rm B}=0.7$). The variability amplitude is from 12.81 per cent to 33.22 per cent, and the average value $19.92\pm5.87$ per cent (Table 7 and Fig. 2). Making use of equation (6), we calculate DC of intraday variability. The value of DC is 44\% for S5 0716+714 (98\%, if PV cases are also included).

Short-term light curves and color index are given in Fig. 3. From Fig. 3, we can see that on the whole the BL Lac object continues to brighten with a remarkably first brightening and then darkening peak during the short term, and the average color index on each night remains approximately constant during this period. The overall magnitude variabilities are $\bigtriangleup B=1^{\rm m}.24$, $\bigtriangleup V=1^{\rm m}.42$, $\bigtriangleup
R=1^{\rm m}.3$, $\bigtriangleup I=1^{\rm m}.23$ respectively. The magnitude distributions in the $B$, $V$, $R$ and $I$ bands are $14^{\rm m}.89<B<13^{\rm m}.65$, $14^{\rm m}.45<V<13^{\rm m}.03$, $13^{\rm m}.94<R<12^{\rm m}.64$ and $13^{\rm m}.39<I<12^{\rm m}.16$ respectively. The average values of magnitude and color index are $<B>=14^{\rm m}.35\pm0^{\rm m}.31$, $<V>=13^{\rm m}.73\pm0^{\rm m}.36$, $<R>=13^{\rm m}.33\pm0^{\rm m}.34$, $<I>=12^{\rm m}.81\pm0^{\rm m}.33$ and $<V-R>=0^{\rm m}.39\pm0^{\rm m}.04$ respectively.

\subsection{Cross-correlation Analysis and Variability Timescales}

Following Giveon et al. (1999), Wu et al. (2007), Liu et al. (2008), Liao et al. (2014) and Xiong et al. (2016), we use the $z$-transformed discrete correlation function (ZDCF; Alexander 1997) to perform the inter-bands correlation analysis and search for the possible inter-bands time delay. The ZDCF code of Alexander et al. (1997) can automatically
set how many bins are given, and be used to calculate the inter-bands correlation and the ACF. In order to achieve statistical significance, the minimal number of points per bin is 10. The results of ZDCF for all data are given in Fig. 4. As an illustration, the results of ZDCF on four nights are given in Fig. 4. The results of ZDCF show that there are good inter-bands correlations but not significant time lags.

The zero-crossing time of the autocorrelation function (ACF) is a
well-defined quantity and used as a characteristic variability
timescale (e.g., Alexander 1997; Giveon et al. 1999; Netzer et al.
1996; Liu et al. 2008). The zero-crossing time is the shortest time it takes the ACF to fall to zero (Alexander 1997). If there is an underlying signal in the
light curve, with a typical timescale, then the width of the ACF
peak near zero time lag will be proportional to this timescale
(Giveon et al. 1999; Liu et al. 2008). The width of the ACF may be related to a
characteristic size scale of the corresponding emission region (Chatterjee
et al. 2012). Another function used in variability studies to estimate the variability
timescales is the first-order structure function (SF; e.g., Trevese et al. 1994). There is a simple relation between the ACF and the SF (Giveon et al. 1999). We therefore perform only
an ACF analysis on our lightcurves. The ACF was estimated by ZDCF. We only analyze
the nights detected as intraday variability. The results of ACF analysis are given in Fig. 5. The results of ACF analysis on February 04 and 08 are not shown in Fig. 5 because all of the ACF values of the two nights are more than zero. Following Giveon et al. (1999), we then use a least-squares procedure to fit a fifth-order polynomial to the ACF, with the constraint that ACF($\tau=0$)=1. From the fitting results, it is seen that the variability timescales range from 0.054 day to 0.134 day.

\subsection{Correlation between Magnitude and Color Index}

For the color index, we used the correction factors from Schlafly \& Finkbeiner (2011) to correct the Galactic extinction. In order to minimize the bias introduced by the dependence of the color index on the magnitudes, brightness was calculated by averaging the magnitudes of the two bands used to calculate the index (Dai et al. 2015). We concentrate on $V-R$ index and $(V+R)/2$ magnitude because $V-R$ index is frequently studied. Fig. 6 shows the correlations between $V-R$ index and $(V+R)/2$ magnitude on intraday and short timescales. The results of analysis of Spearman rank are given in Table 9. The Table 9 shows that 8 nights have the significant correlations between $V-R$ index and $(V+R)/2$ magnitude, and 1 night has significant anti-correlation between $V-R$ index and $(V+R)/2$ magnitude (significance level $P<0.05$ confidence level). Moveover, 4 nights have not significant correlations between $V-R$ index and $(V+R)/2$ magnitude ($P>0.05$). For short timescales, the significant correlation between $V-R$ index and $(V+R)/2$ magnitude is found (Table 9). Therefore, during the outburst period, most of nights show bluer when brighter (BWB) chromatic trend; a weak redder with brighter (RWB) trend is found; a few nights show no correlations between magnitude and color index. The BWB trend appears in the short timescales. The spectral index versus the flux in the flare event is given in Fig. 7. It is seen that during the flare the spectral index exhibits a clockwise loop for inter-nights.

\section{DISCUSSION AND CONCLUSIONS} \label{bozomath}
\subsection{Variability}

We have observed the BL Lac object S5 0716+714 with long time spans on intraday timescale ($>$5 hr) and high time resolutions. During this monitoring period, the object shows violent flaring activity both in short term and intraday timescales. The value of DC is 44\% for S5 0716+714 (98\%, if PV cases are also included). If we only consider the $F$ test and ANOVA, the value of DC is 74\%. So, similar to the previous results (e.g., Agarwal et al. 2016; Hu et al. 2014; Chandra et al. 2011; Stalin et al. 2009), the object has high value of DC. The light curves detected as IDV show variability of various shapes. The variability amplitude is from 12.81 per cent to 33.22 per cent, and the average value $19.92\pm5.87$ per cent. The overall magnitude variabilities are $\bigtriangleup B=1^{\rm m}.24$, $\bigtriangleup V=1^{\rm m}.42$, $\bigtriangleup R=1^{\rm m}.3$, $\bigtriangleup I=1^{\rm m}.23$ respectively. During the observations, the average change rate is $<CR>=0.035\pm0.009$ Mag/h during the ascent and $<CR>=0.035\pm0.014$ Mag/h during the descent, i.e., the average change rates between the ascent and descent are equal. However, different cases are found on certain nights. On January 29, the brightness first decreases and then increases along the symmetrical arc. Also the increasing and decreasing change rates are close on the night. From previous observations (Agarwal et al. 2016; Man et al. 2016; Dai et al. 2015; Hu et al. 2014; Wu et al. 2007, 2012; Chandra et al. 2011; Stalin et al. 2009; Poon et al. 2009; Sasada et al. 2008; Zhang et al. 2008b; ), many of the common cases of variability in the object are as follows: (i) the brightness continues to increase/decrease; (ii) the brightness first decreases and then increases, or vice versa, but not along the symmetrical arc; (iii) the brightness first increases and then decreases, and last increases; (iv) the brightness first increases and then decreases along the arc.  These cases of variability have also been found in Fig. 1. Compared with these cases of variability, the type of variability on January 29 is rarely found on intraday timescales. From Fig. 3, we can obtain that the brightness is likely to increase from January 28 to January 29, and then decrease from January 29 to January 30. In addition, the color index on January 29 has BWB trend. Therefore, the variability on the night is likely related with relativistic jet activities (also see the following discussions).

For blazar, these components (jet, accretion disk and host galaxy) could contribute the emission of optical band. The host galaxy is more than 4 mag fainter than the brightness of S5 0716+714 (Nilsson et al. 2008). During the outburst period, the accretion disk radiation is always overwhelmed by the Doppler boosting flux from the relativistic jet. Then the variability of S5 0716+714 in the outburst state is likely to have an association with jet activities. The shock-in-jet model is often used to explain the IDV/short variability. The shocks propagate down relativistic jets, sweeping emitting
regions. If the emitting regions have intrinsic variations (magnetic field, particle velocity/distribution, a large number of new particles injected), then we could see the IDV/short variability (Marscher \& Gear 1985; Xiong et al. 2017). In addition to intrinsic variations, geometrical variations also could bring in flux variations. Rani et al. (2015) presented a high-frequency very long baseline interferometry (VLBI) kinematical study of the BL Lac object S5 0716+714. They found repetitive optical/$\gamma$-ray flares and the curved trajectories of the associated components, and suggested that the shock front propagates along a bent trajectory or helical path. In order to explain the multi-frequency behavior of an optical--$\gamma$-ray outburst in 2011, Larionov et al. (2013) also suggest a shock wave propagating along a helical path in the blazar¡¯s jet. The helical jet structure may cause the Doppler factor change on a very short variability timescale (Gopal-Krishna \& Wiita 1992). The variability may also be explained by turbulence behind an outgoing shock along the jet or the magnetic reconnections (Agarwal et al. 2016; Chandra et al. 2015). The symmetric/asymmetric light curves can be interpreted as arising from light-travel time effect (Chiaberge \& Ghisellini 1999; Chatterjee et al. 2012). Chiaberge \& Ghisellini (1999) presented that the symmetric shapes of the light curves strongly constrain the injection and cooling timescales. When the cooling time of the electrons is much shorter than the light crossing time, the light curves are symmetric. When the cooling time of the electrons is much larger than the light crossing time, the light curves are asymmetric.

The results of ACF show that the variability timescales range from 0.054 day to 0.134 day. If we consider the variability timescales as the light crossing time of the emitting blob, the range of the emission region in the jet is from $R\leq2.14\times10^{15}$ cm to $R\leq5.3\times10^{15}$ cm ($R\leq c \delta \Delta t_{min}/(1+z)$, where $\delta=20$ from Nesci et al. (2005) and the redshift $z=0.31$). In addition, there are good inter-bands correlations but not significant time lags for intraday and short timescales, and not significant correlations between variability amplitude and the source average brightness at individual bands. However, we still need more data to further confirm the correlations between variability amplitude and the source average brightness at individual bands due to the small sample size at individual bands in this work.

\subsection{Spectral Properties}
Generally, the BWB (bluer when brighter) chromatic trend is dominant for most of blazars while redder when brighter (RWB) trend is also found, especially for FSRQ (e.g., Gu et al. 2006; Guo \& Gu 2016). The BWB behavior is most likely to be explained by the shock-in-jet model. According to the shock-in-jet model, as the shock propagating down the jet strikes a region with a large electron population, radiations at different visible colors are produced at different distances behind the shocks. High-energy photons from the synchrotron mechanism typically emerge sooner and closer to the shock front than the lower frequency radiation, thus causing color variations (Agarwal \& Gupta 2015; Xiong et al. 2017). The BWB trend could be explained by two different jet components, i.e., flare component has a higher synchrotron peak frequency than the underlying component (Ikejiri et al. 2011). Assuming that the optical/UV variability is triggered by
fluctuations, Guo \& Gu (2016) presented that the RWB trend can likely be explained if the fluctuations occur first in the outer disk region, and the inner disk region has not yet fully responded when the fluctuations are being propagated inward. In contrast, the common BWB trend implies that the fluctuations likely more often happen first in the inner disk region. Gu et al. (2006) proposed that the different relative contributions of the thermal versus non-thermal radiation to the optical emission may be responsible for the different trends of the colour index with brightness in FSRQs and BL Lac objects. Moveover, the achromatic trend could be due to a Doppler factor variation on a spectrum
slightly deviating from a power law (Villata et al. 2004). For our results, the average color index on each night remains approximately constant during this period. Most of nights show bluer when brighter (BWB) chromatic trend; a weak redder with brighter (RWB) trend is found; a few nights show no correlations between magnitude and color index. The BWB trend appears in the short timescales. As discussed above, the variability and color index of S5 0716+714 in the outburst state are likely to have an association with jet activities. The shock-in-jet model can explain the BWB chromatic trend. The BWB or flatter when brighter could be due to the injection of fresh electrons, with an energy distribution harder than that of the previous cooled ones (e.g., Kirk et al. 1998; Mastichiadis \& Kirk 2002). However, if the injection of fresh electrons have an energy distribution softer than that of the previous cooled ones, the weak RWB may be seen. When a shock wave propagates along a helical path, the achromatic trend could be found. Therefore, a shock-in-jet model and the shock wave propagating along a helical path are likely to explain the variability and color index variability.

Finally, during the flare the spectral index exhibits a clockwise loop for inter-nights. This type of variability pattern represents a sort of hysteresis cycle in the scatter plot between the energy index and the flux. It may arise whenever the spectral slope is controlled by cooling processes (Fiorucci et al. 2004). The clockwise direction is due to changes in the injection of accelerated particles, propagating from high to low energies (Kirk et al. 1998).

\begin{acknowledgements}
We sincerely thank the referee for valuable comments and
suggestions. We acknowledge the support of the staff of the
Lijiang 2.4m and BOOTES-4 telescopes. Funding for the two telescopes has been provided by the Chinese Academy of Sciences and the Peoples Government of Yunnan Province. This work is financially supported by the National Nature Science Foundation of China (11433004, 11133006, 11673057, 11361140347), the Key Research Program of the Chinese Academy of Sciences (grant No. KJZD-EW-M06), the Strategic Priority Research Program ``The emergence of Cosmological Structures'' of the Chinese Academy of Sciences (grant No. XDB09000000), the Chinese Western Young Scholars Program
and the ¡°Light of West China¡± Program provided by CAS (Y7XB018001, Y5XB091001), the science and technology project for youth of Yunnan of China (2104FD059). This research has made use of the
NASA/IPAC Extragalactic Database (NED), that is operated by Jet
Propulsion Laboratory, California Institute of Technology, under
contract with the National Aeronautics and Space Administration.
\end{acknowledgements}

\begin{figure}
\begin{center}
\includegraphics[width=18cm,height=20cm]{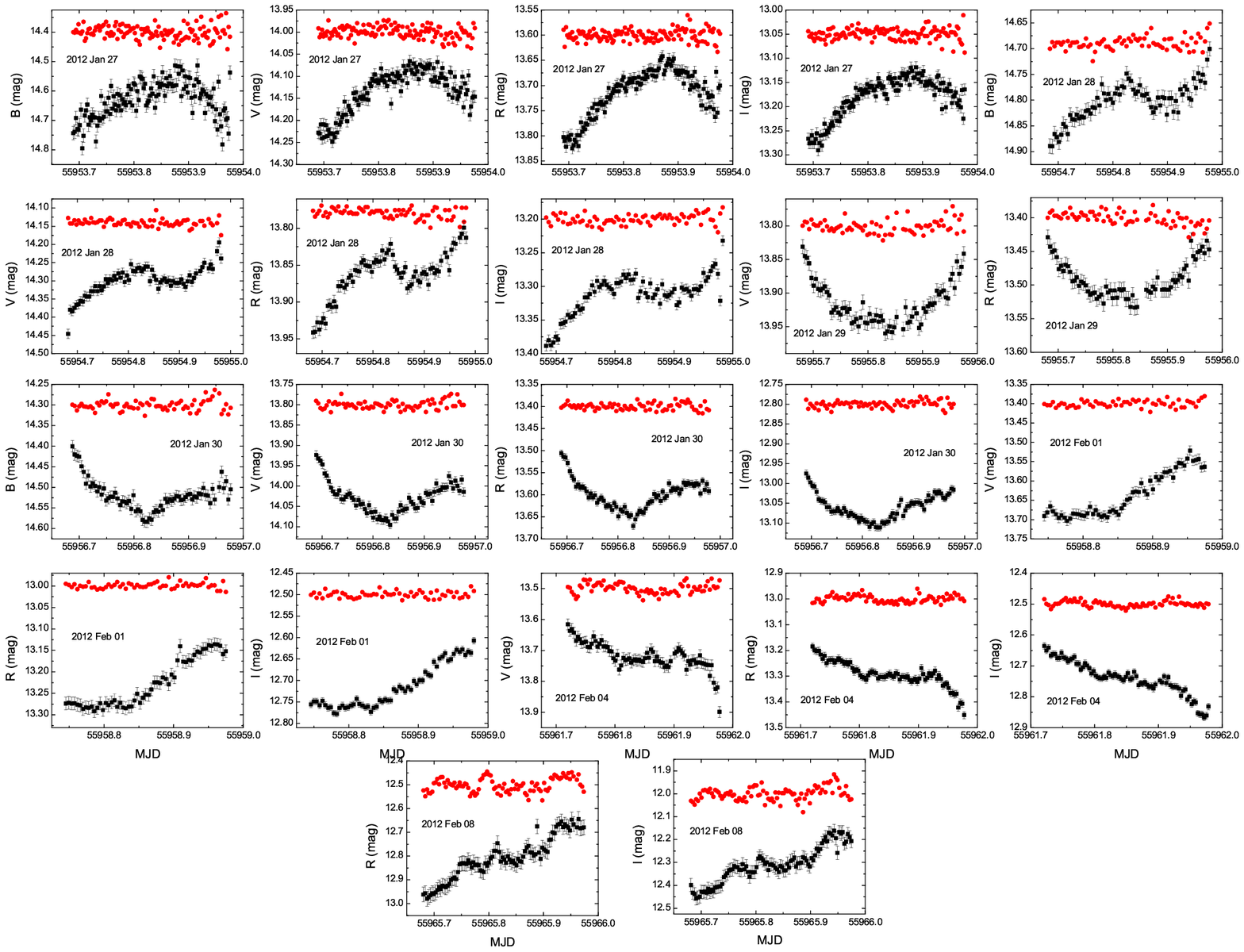}
\caption{Light curves of intraday variability for S5 0716+714. The black
squares are the light curves for S5 0716+714. The red
circles are the variations of $S_I$, $S_R$, $S_V$ and $S_B$. The
light curves of $S_I$, $S_R$, $S_V$ and $S_B$ are offset to avoid their
eclipsing with light curves of S5 0716+714. \label{fig3}}
\end{center}
\end{figure}

\begin{figure}
\begin{center}
\includegraphics[width=14cm,height=14cm]{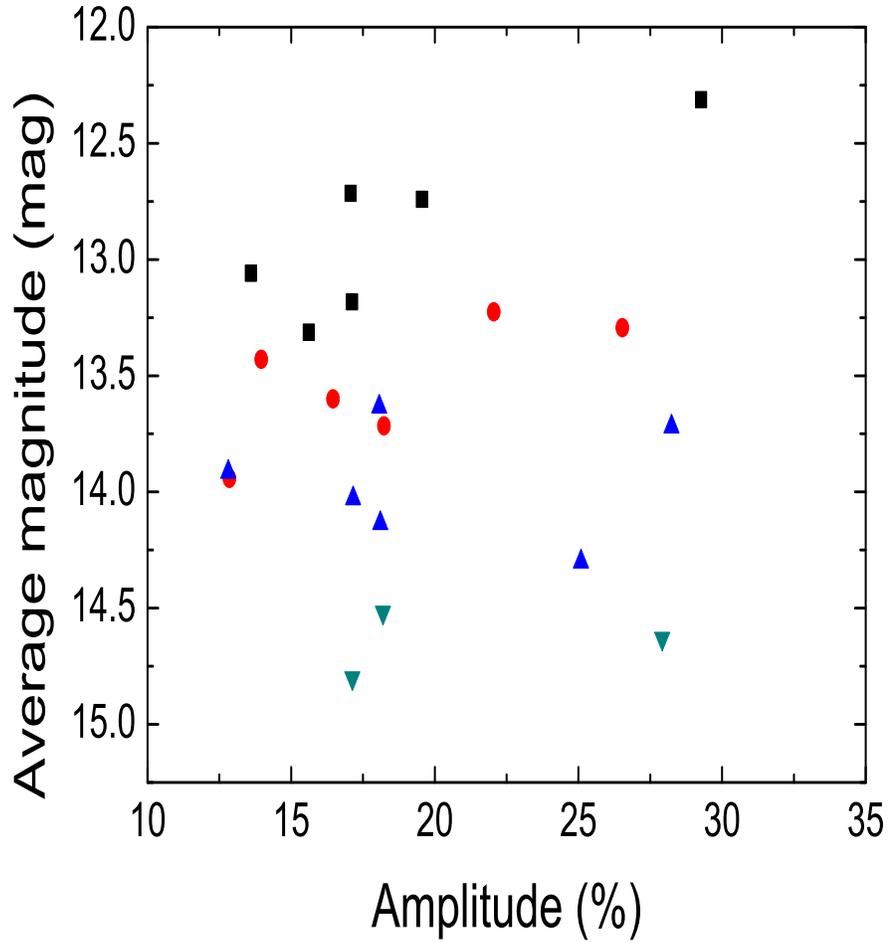}
\caption{The variability amplitude versus the average brightness. Different symbols stand for different optical bands. The black rectangles stand for $I$ band, red circles for $R$ band, positive triangles for $V$ band and anti-triangles for $B$ band.}
\end{center}
\end{figure}

\begin{figure}
\begin{center}
\includegraphics[angle=0,width=0.48\textwidth]{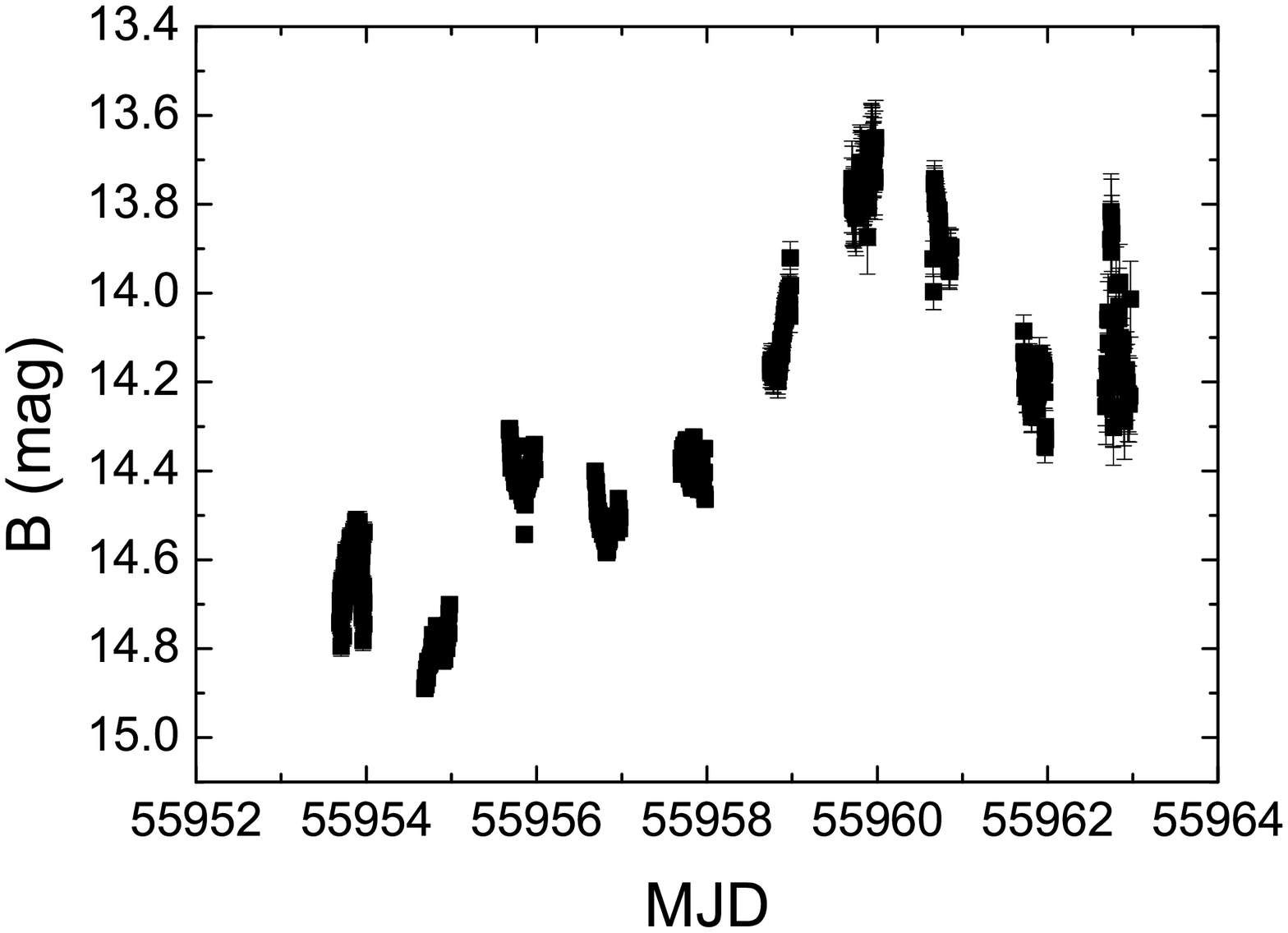}
\includegraphics[angle=0,width=0.48\textwidth]{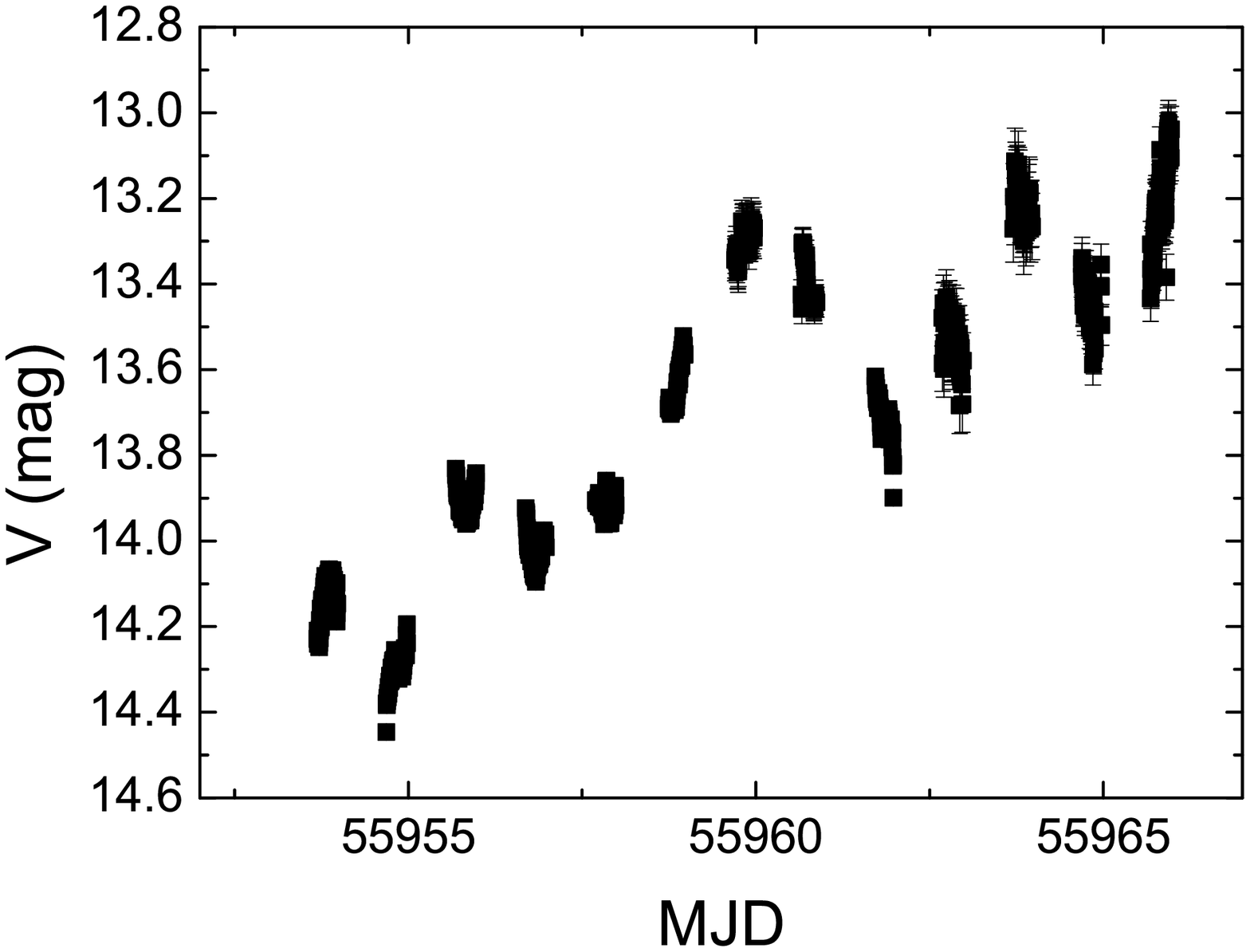}
\includegraphics[angle=0,width=0.48\textwidth]{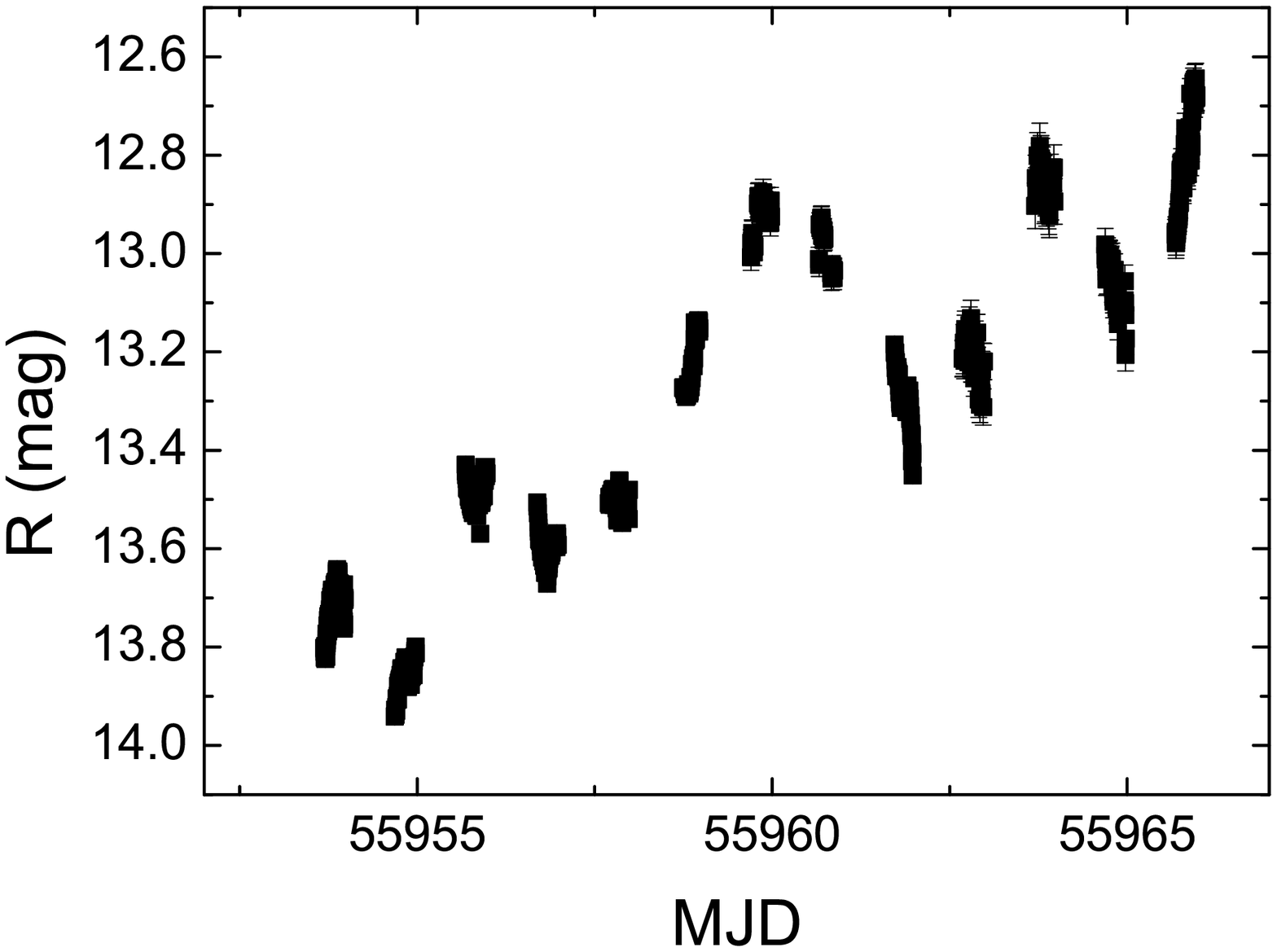}
\includegraphics[angle=0,width=0.48\textwidth]{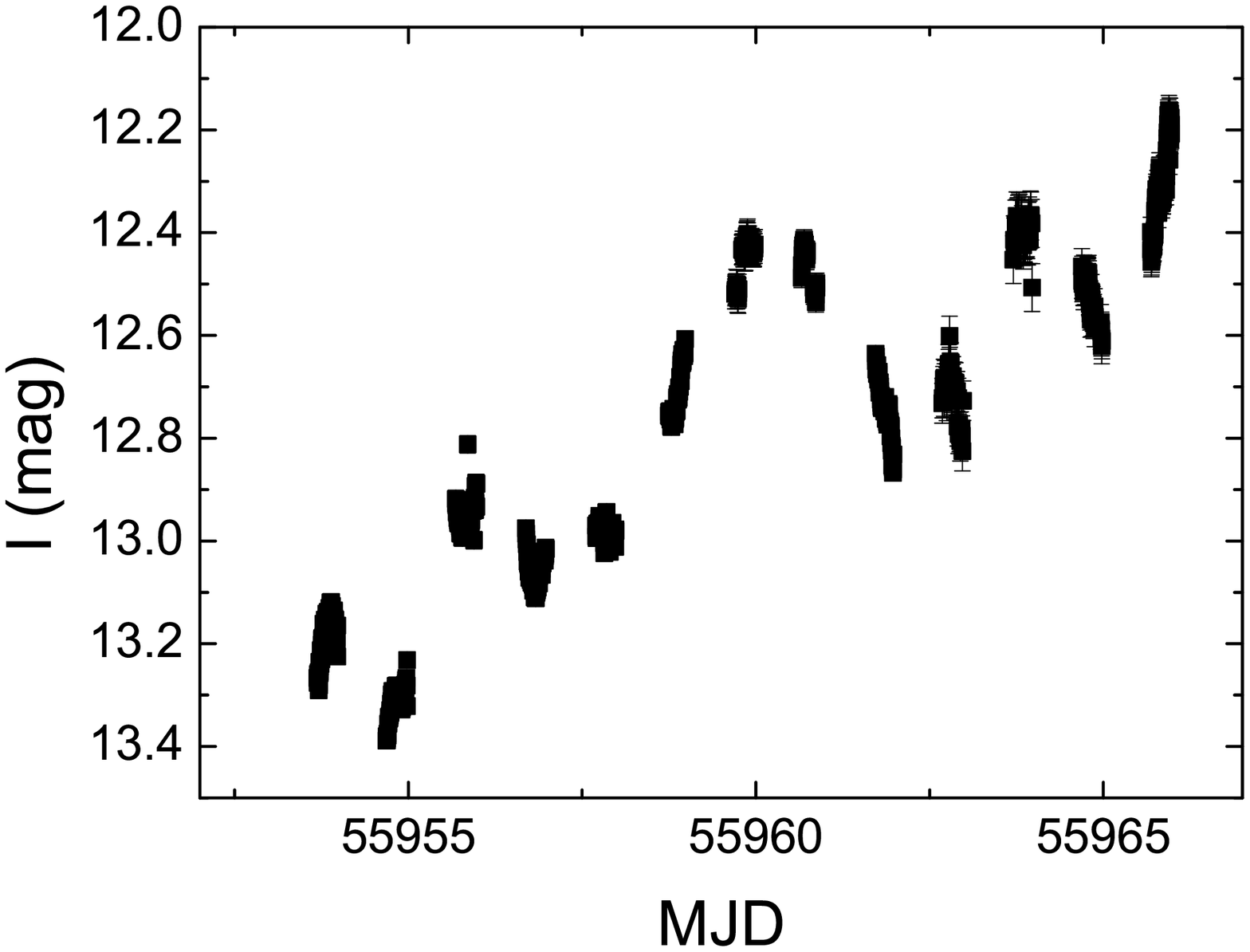}
\includegraphics[angle=0,width=0.48\textwidth]{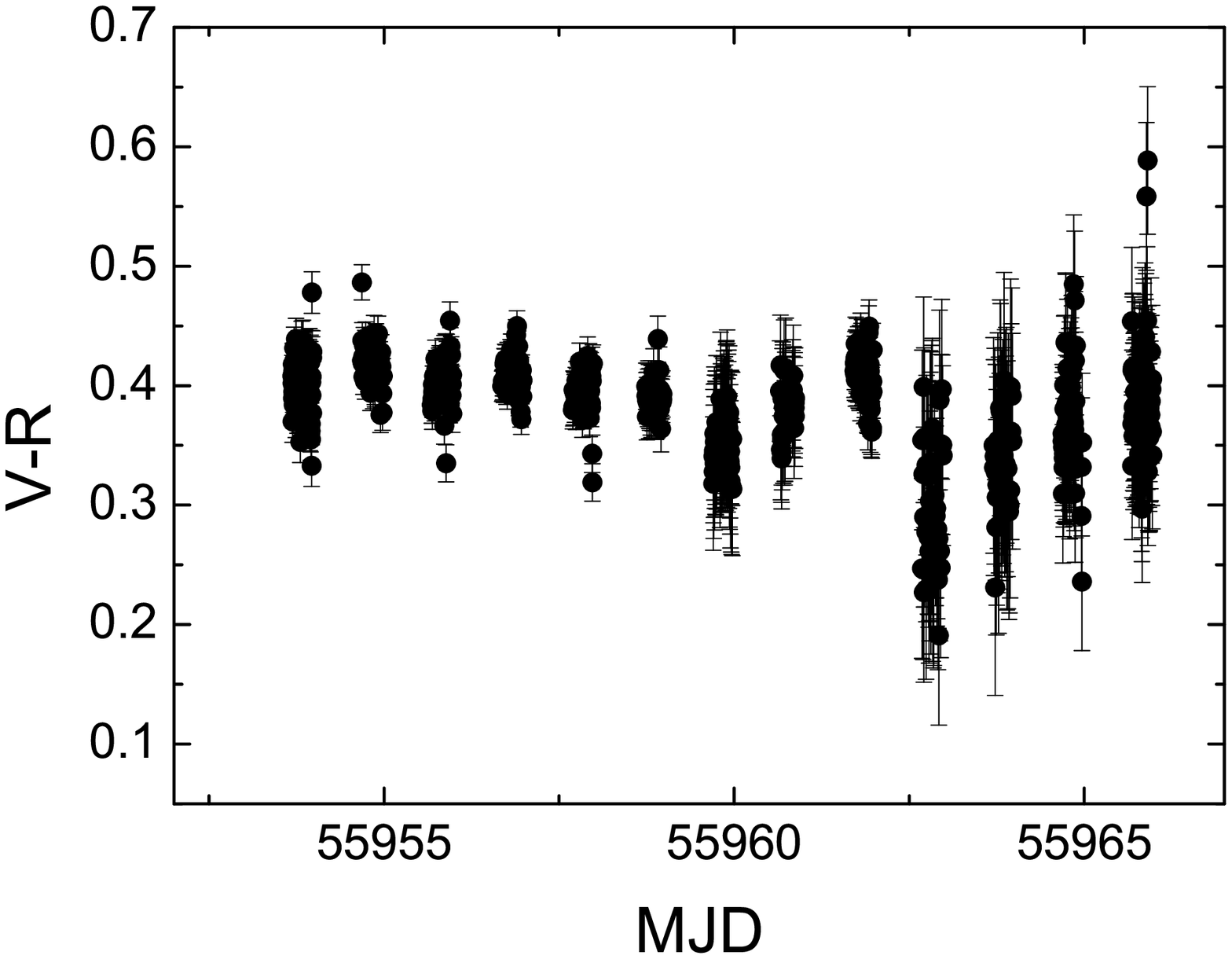}
\caption{Short-term light curves of S5 0716+714 in $B$, $V$, $R$ and $I$ bands and color index $V-R$.\label{fig3}}
\end{center}
\end{figure}

\begin{figure}
\begin{center}
\includegraphics[angle=0,width=0.32\textwidth]{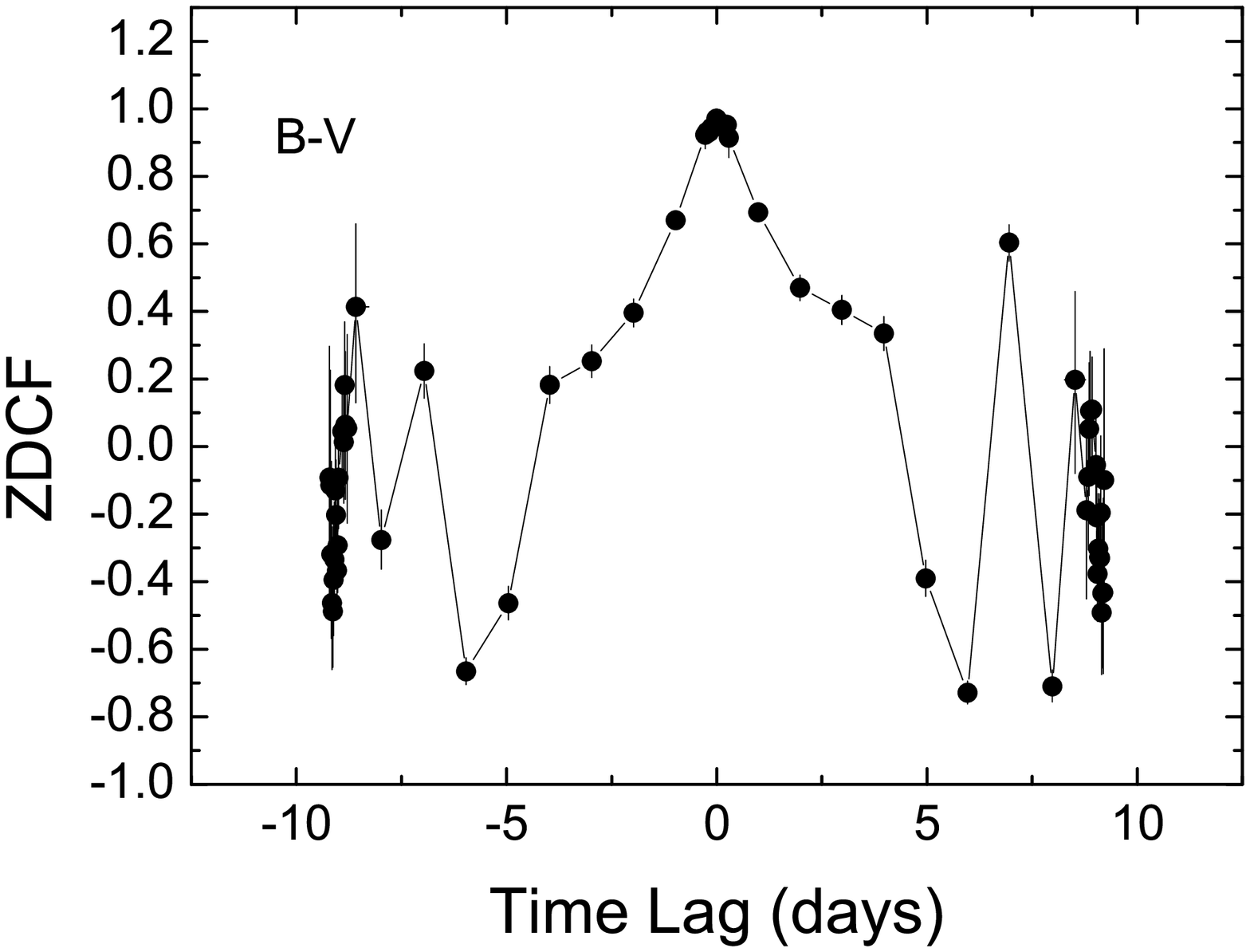}
\includegraphics[angle=0,width=0.32\textwidth]{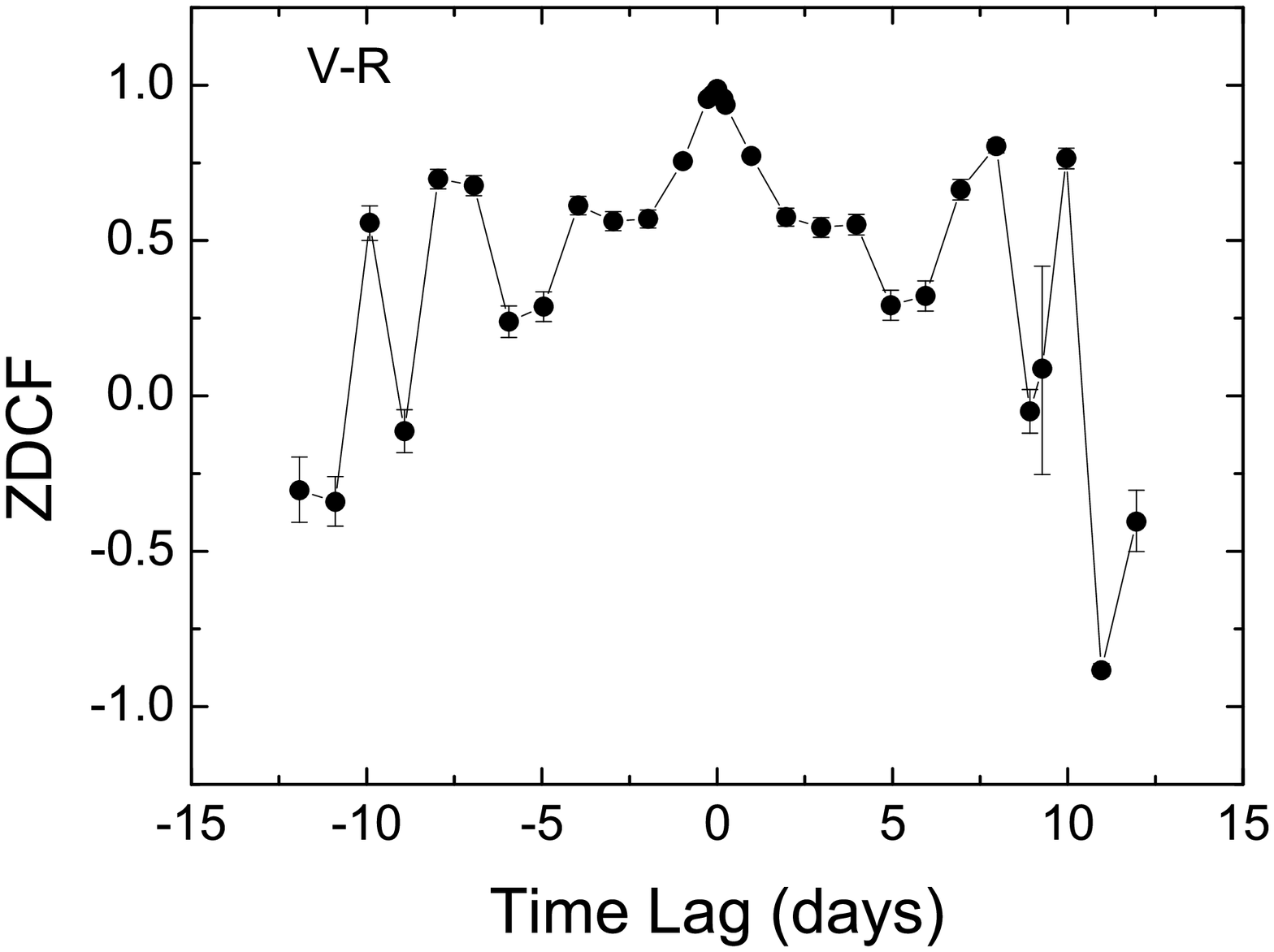}
\includegraphics[angle=0,width=0.32\textwidth]{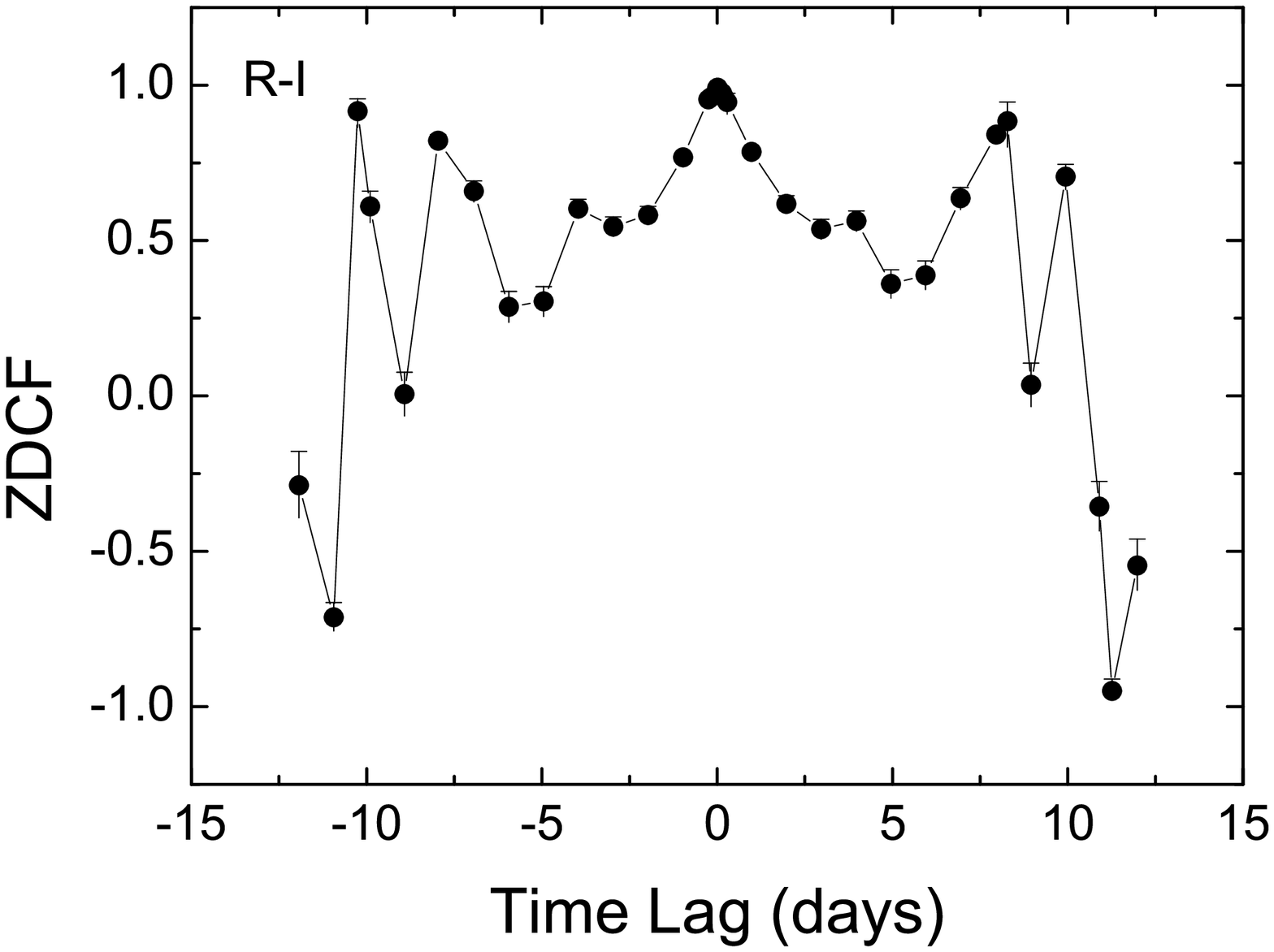}
\includegraphics[angle=0,width=0.32\textwidth]{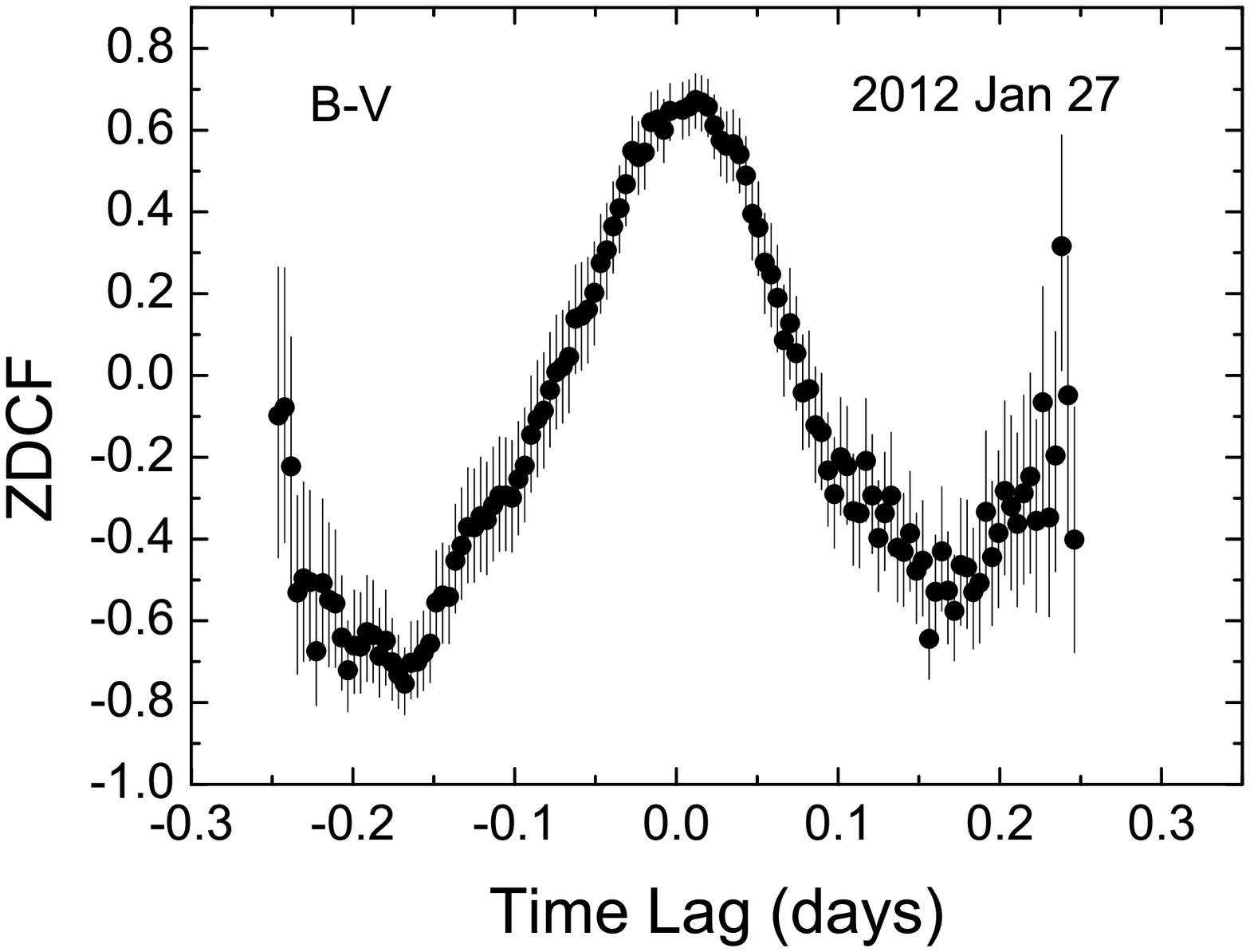}
\includegraphics[angle=0,width=0.32\textwidth]{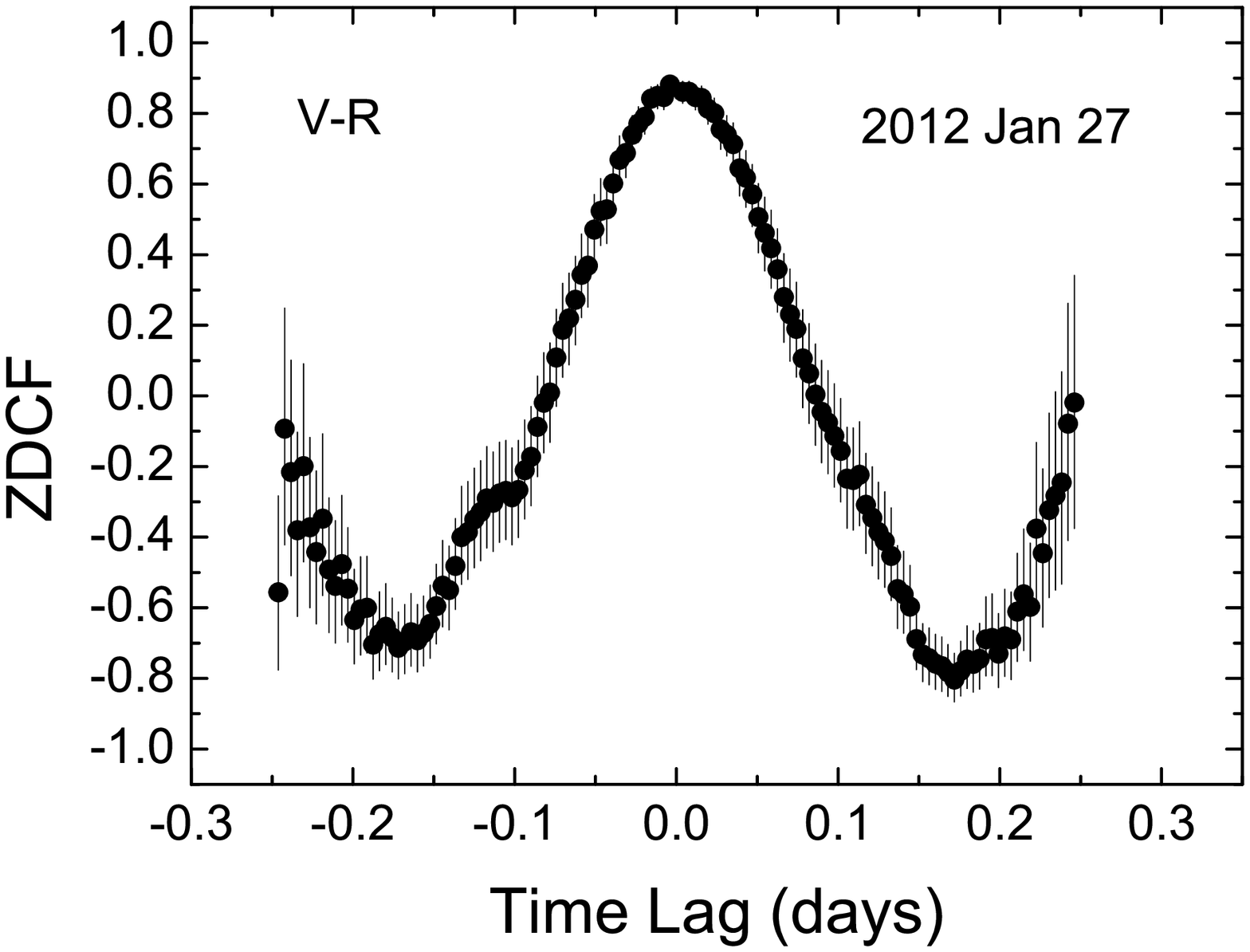}
\includegraphics[angle=0,width=0.32\textwidth]{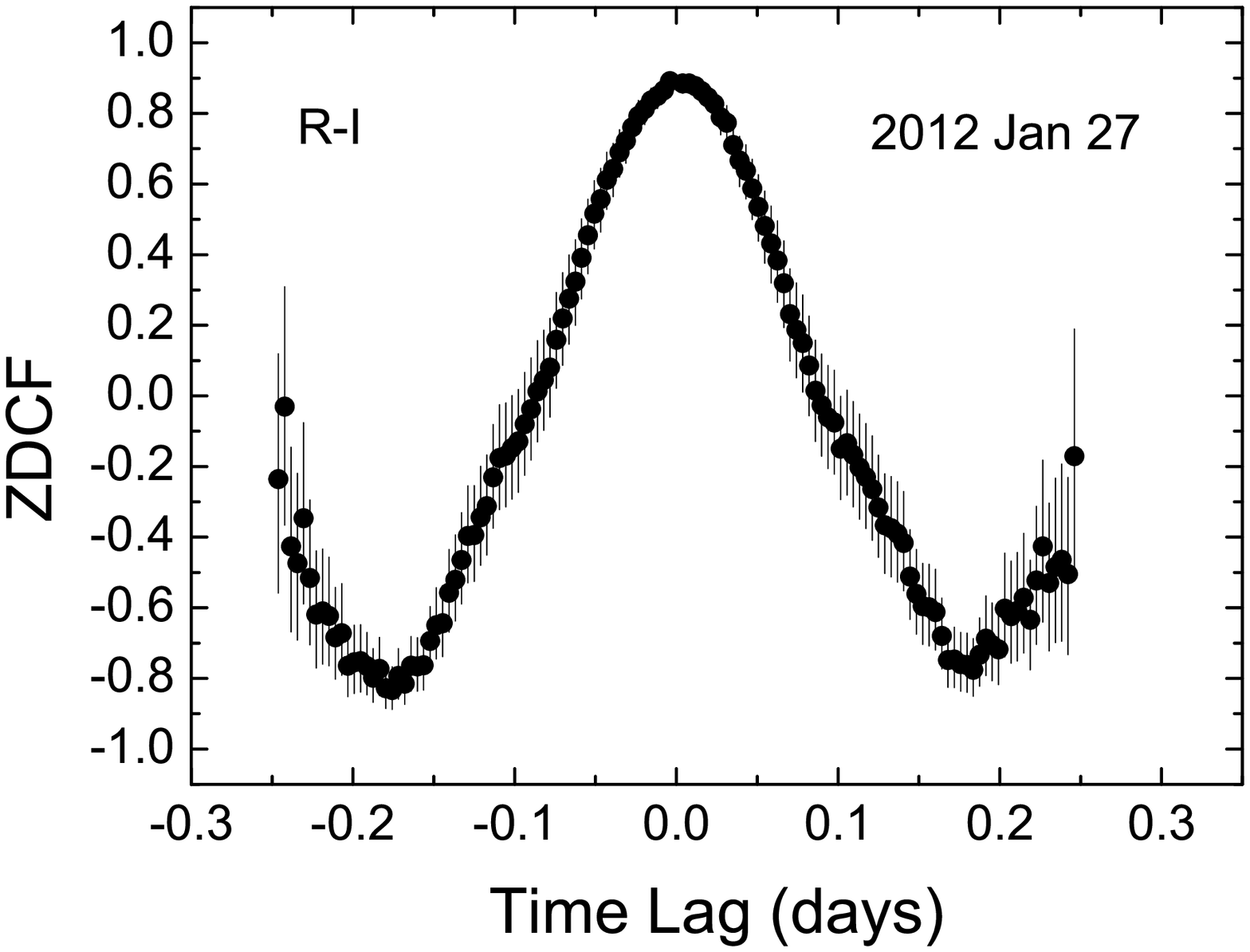}
\includegraphics[angle=0,width=0.32\textwidth]{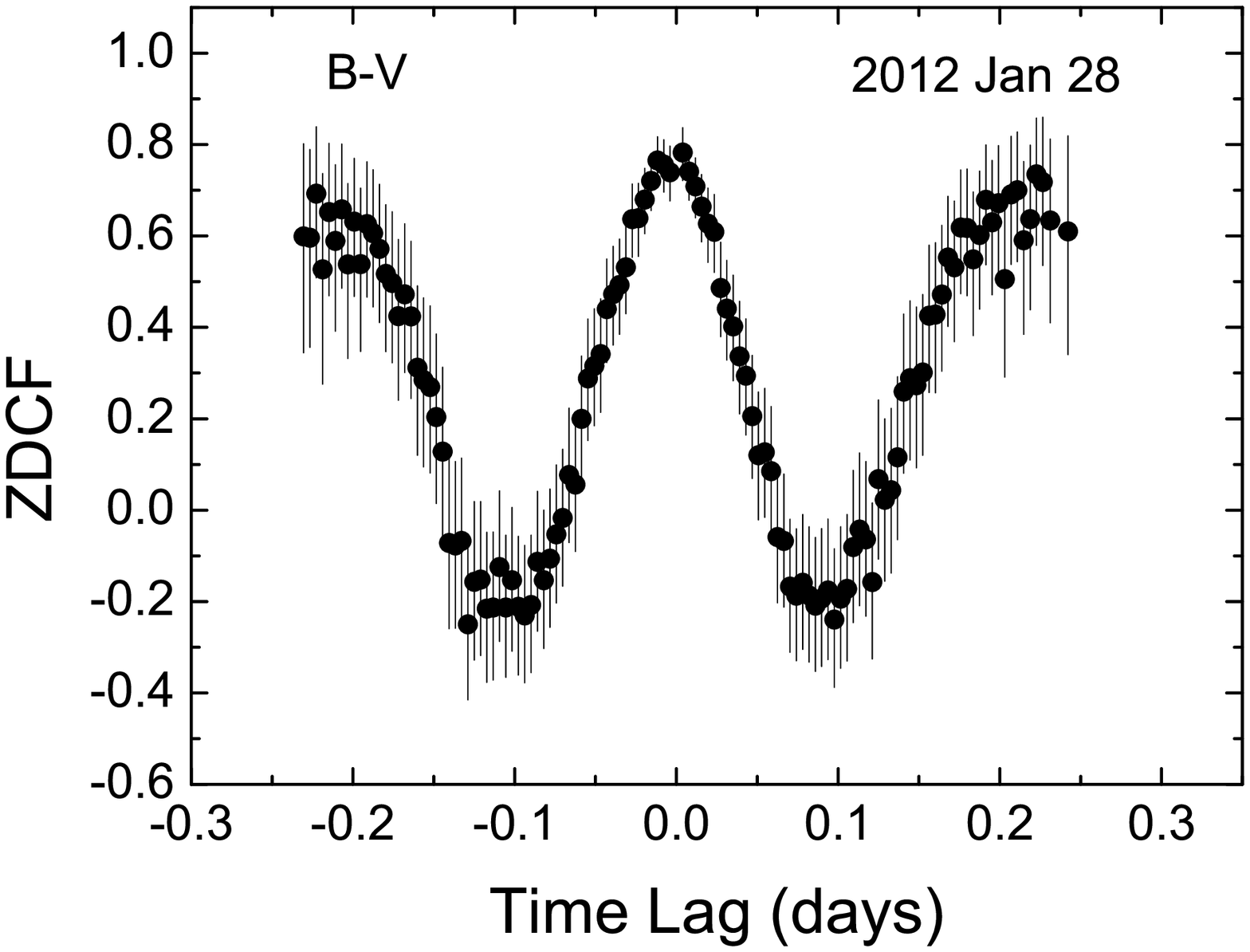}
\includegraphics[angle=0,width=0.32\textwidth]{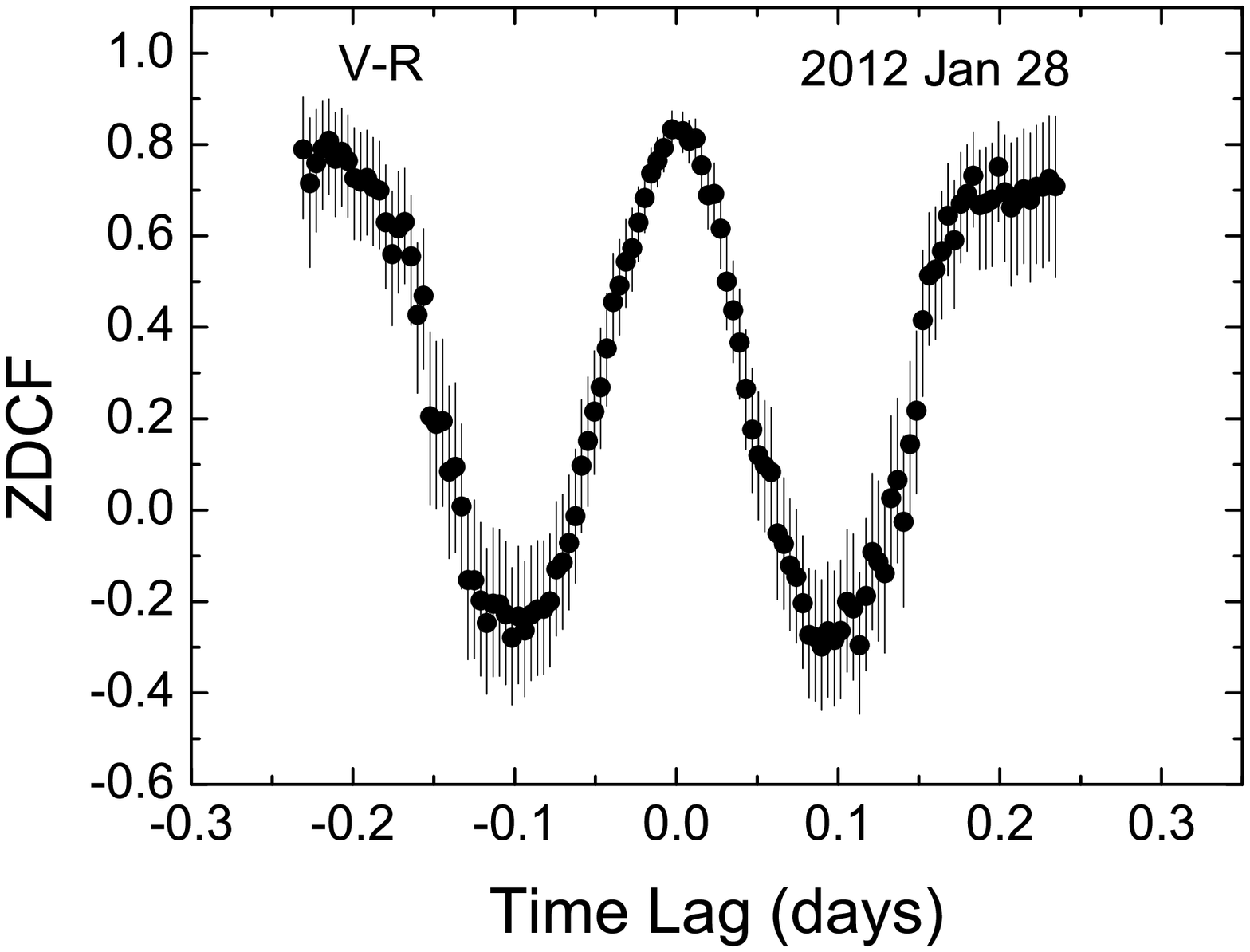}
\includegraphics[angle=0,width=0.32\textwidth]{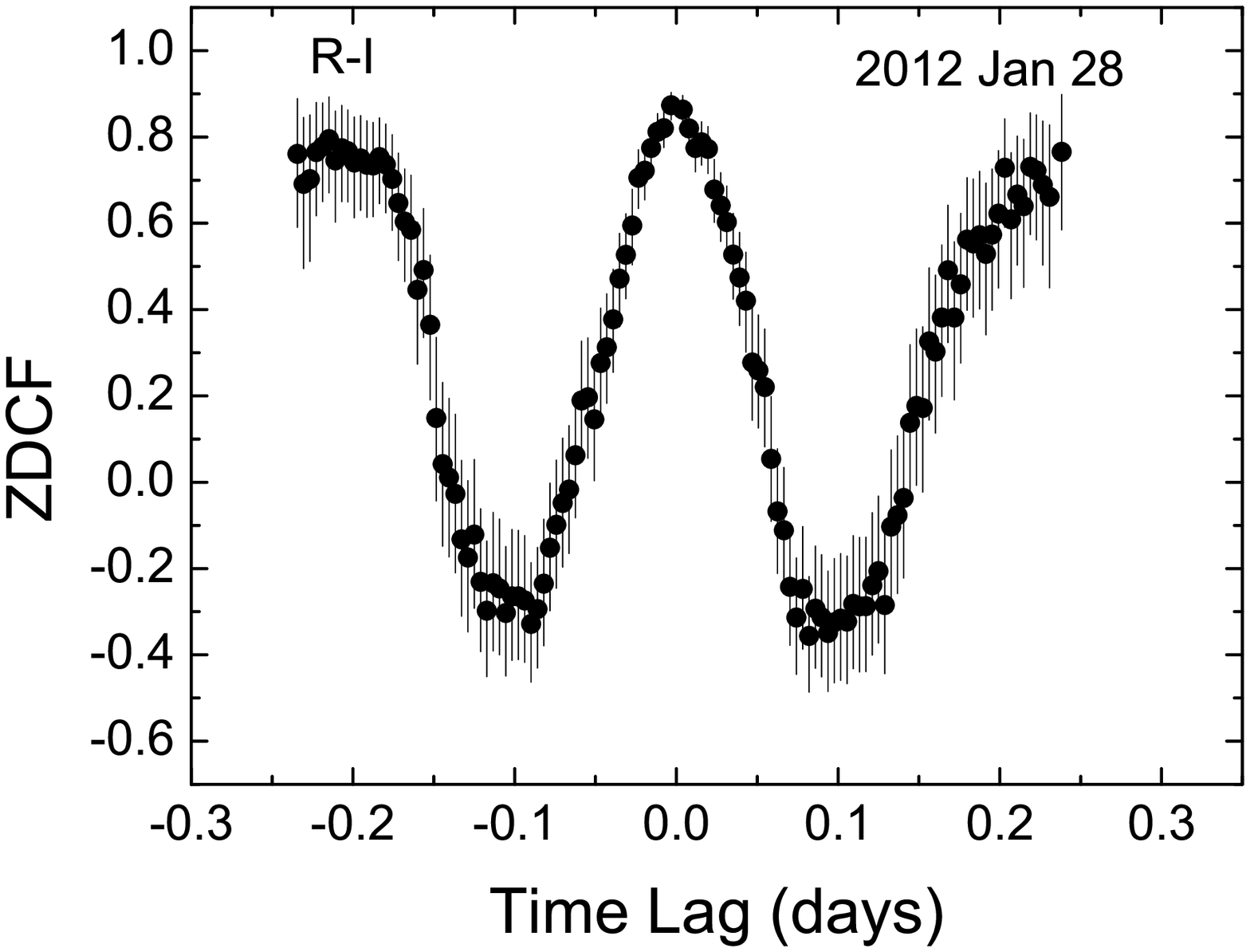}
\includegraphics[angle=0,width=0.32\textwidth]{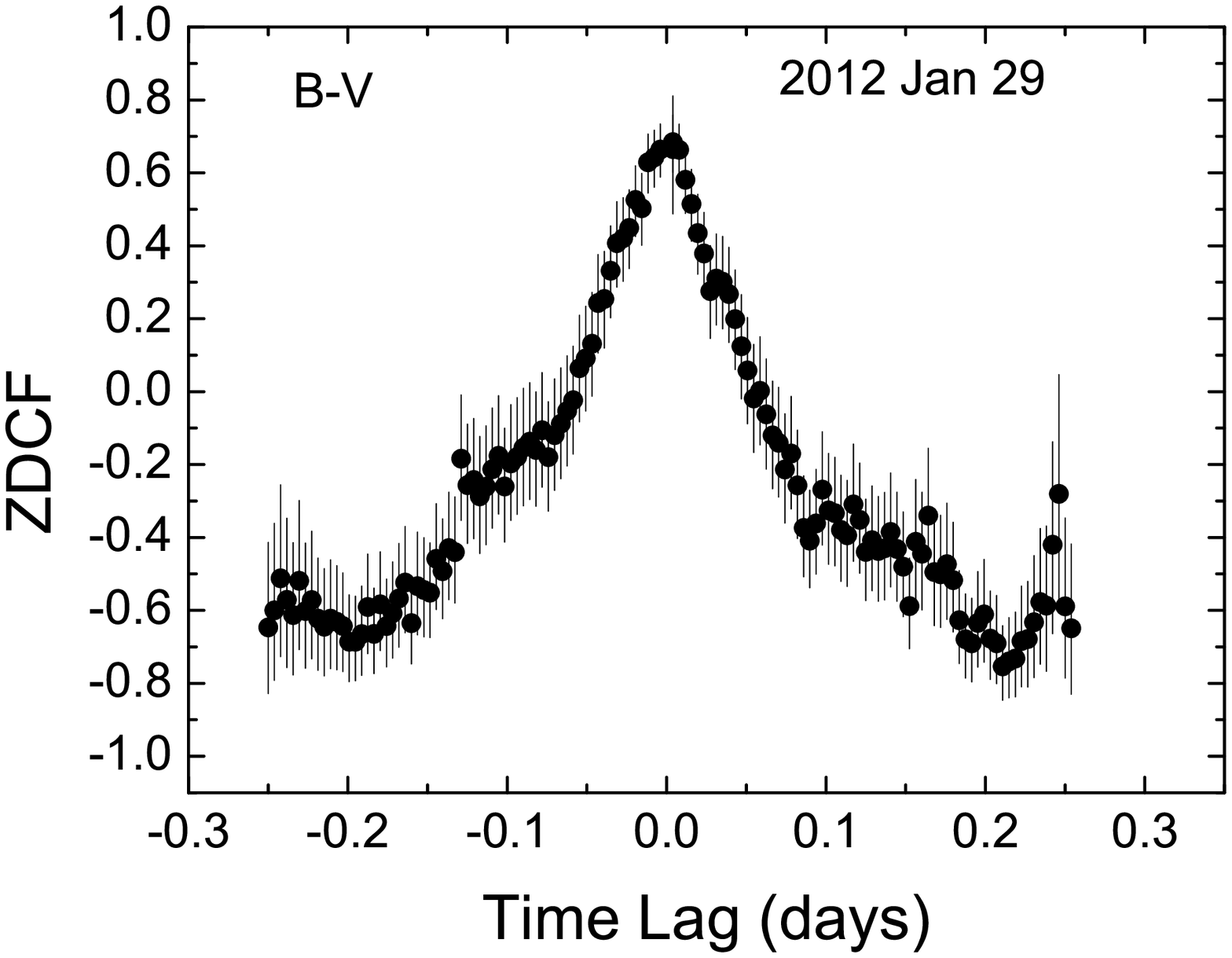}
\includegraphics[angle=0,width=0.32\textwidth]{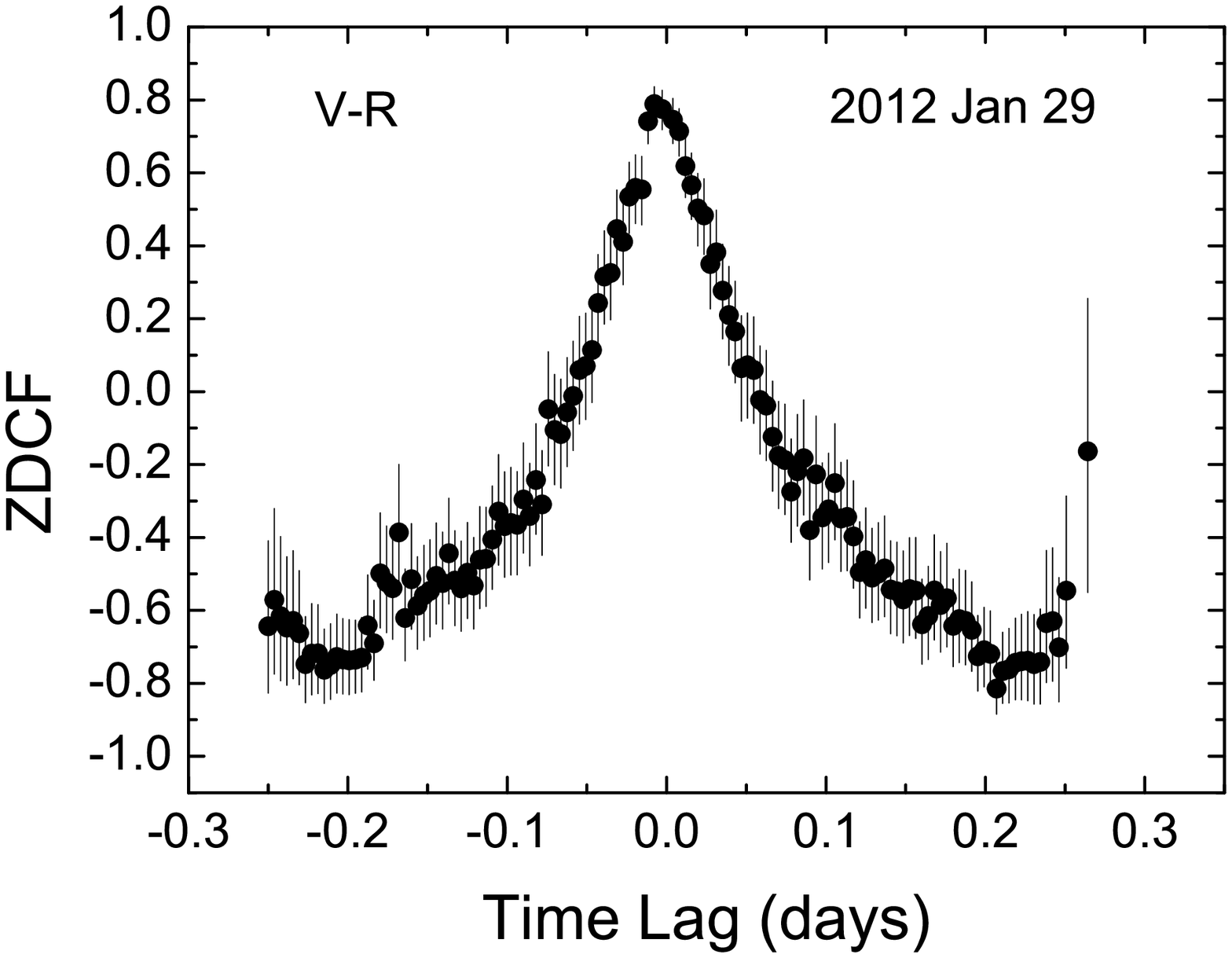}
\includegraphics[angle=0,width=0.32\textwidth]{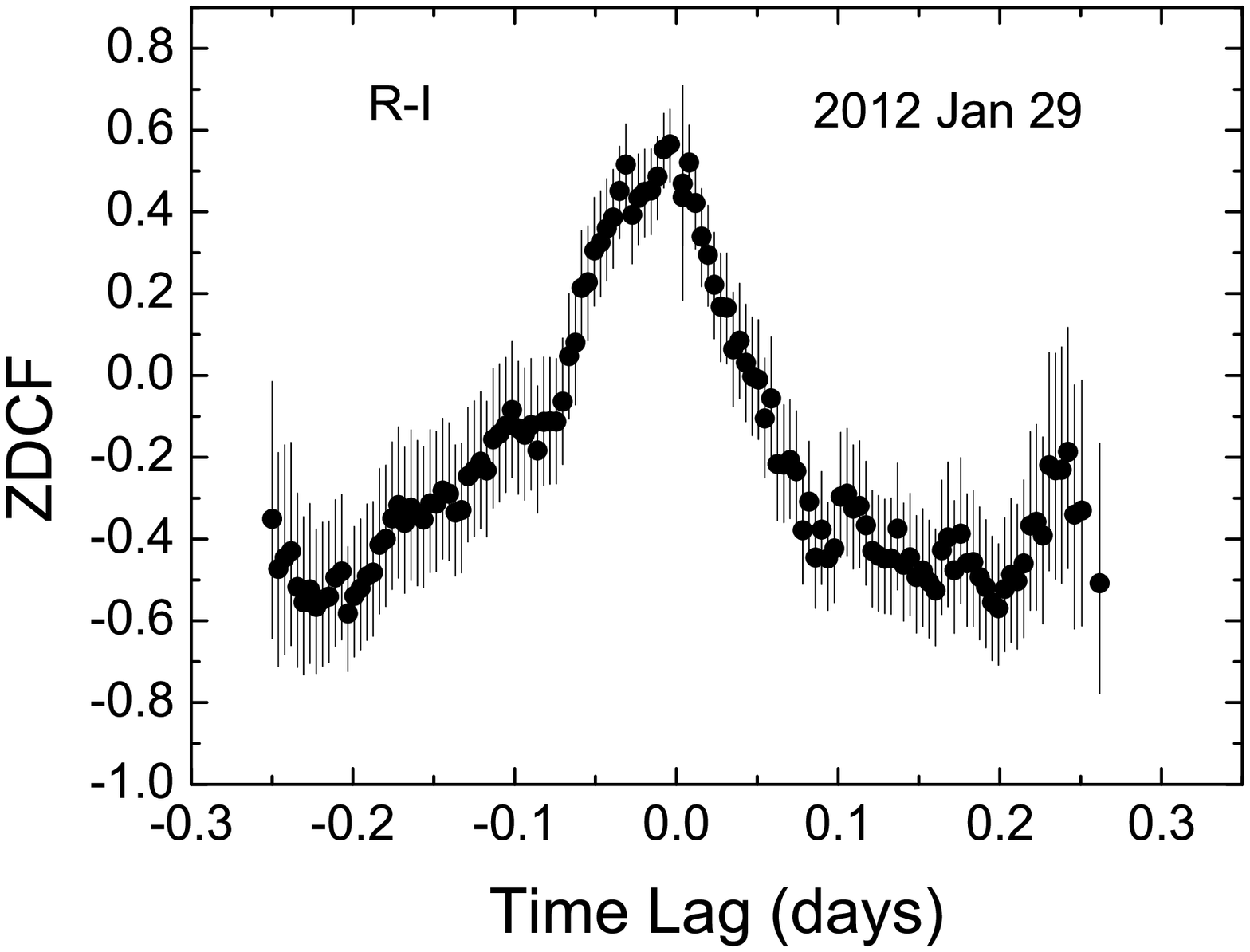}
\includegraphics[angle=0,width=0.32\textwidth]{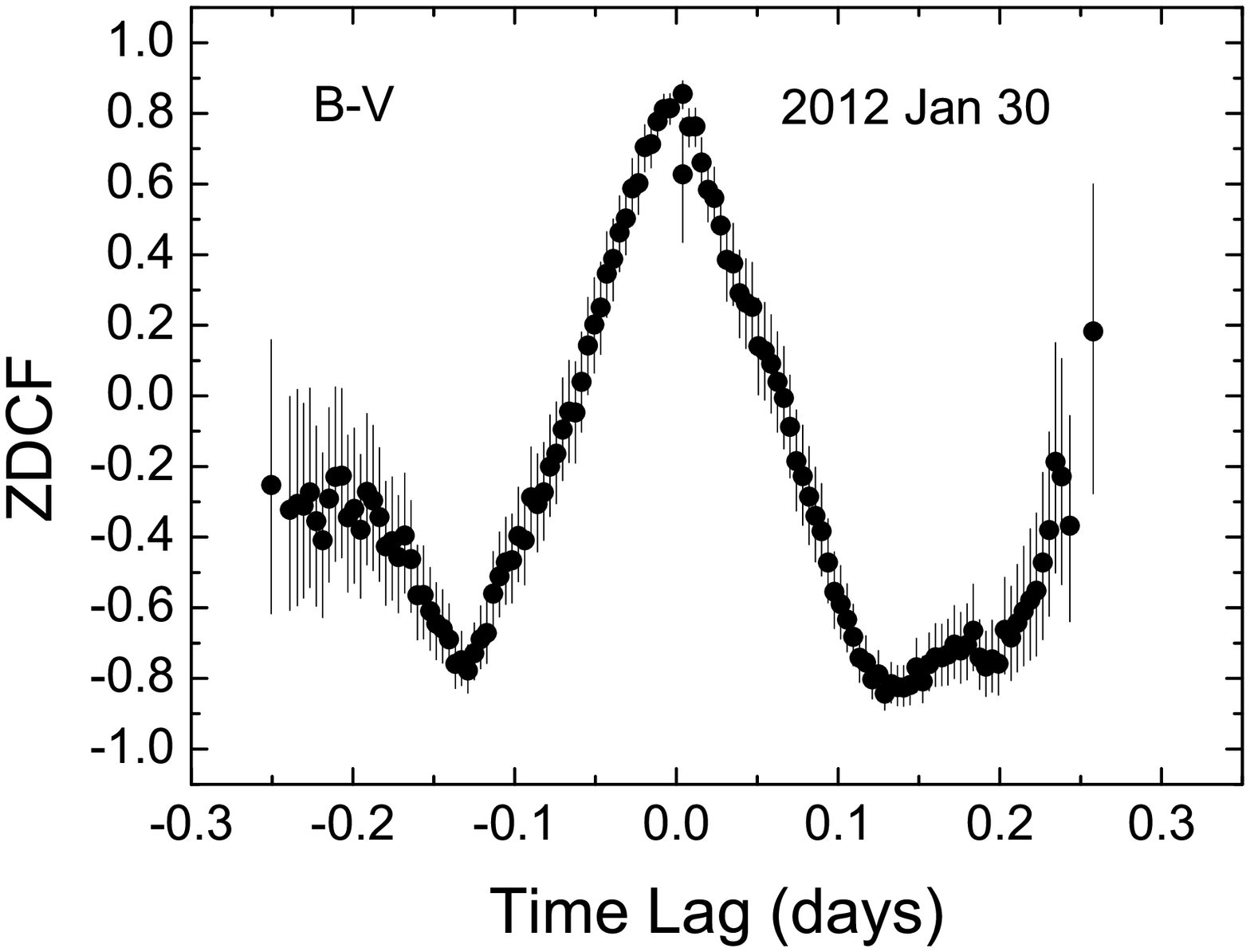}
\includegraphics[angle=0,width=0.32\textwidth]{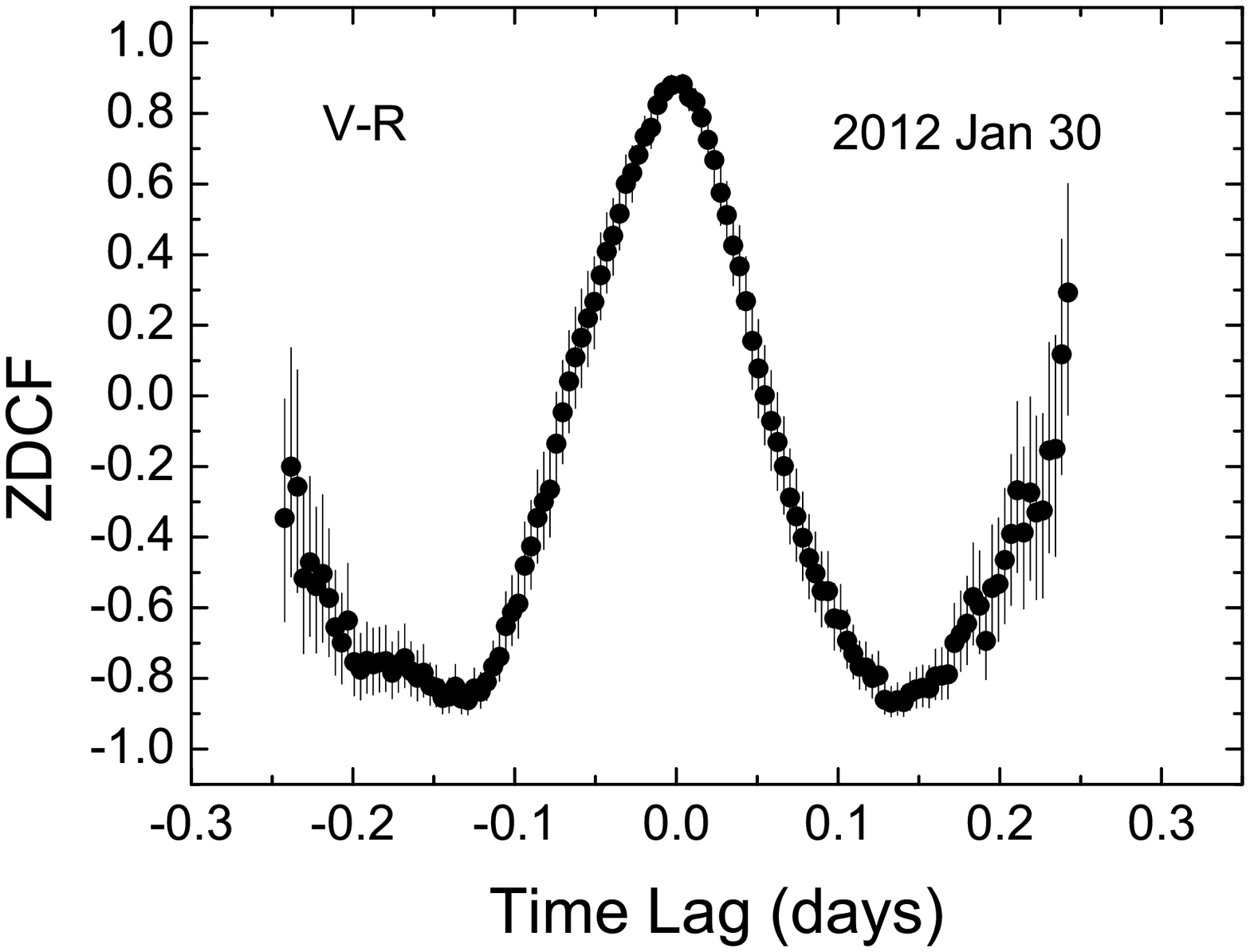}
\includegraphics[angle=0,width=0.32\textwidth]{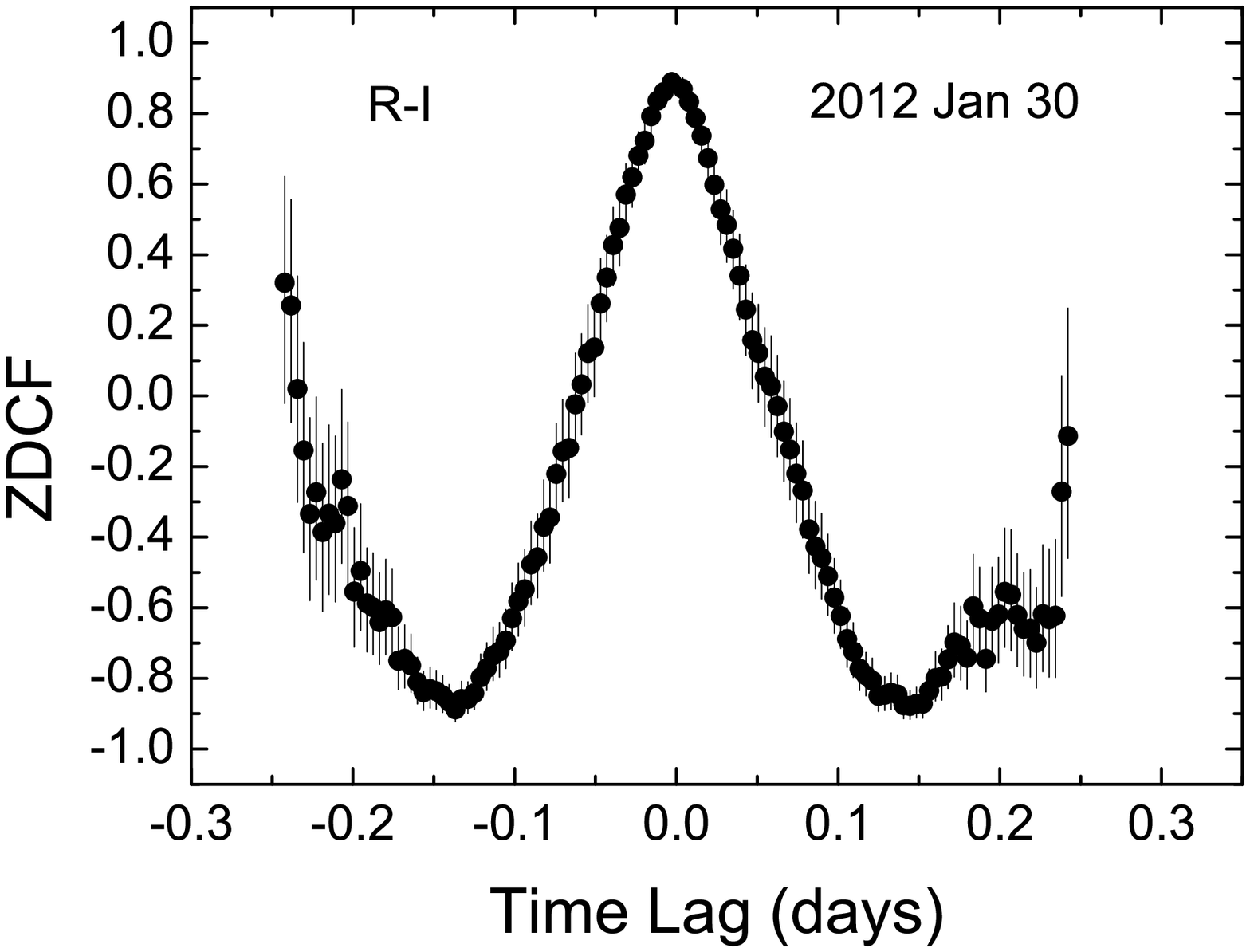}
\caption{The ZDCF plots. The first panel is the results of ZDCF for all data. The rest of panels are the results of ZDCF for intraday timescales. \label{fig3}}
\end{center}
\end{figure}

\begin{figure}
\begin{center}
\includegraphics[angle=0,width=0.25\textwidth]{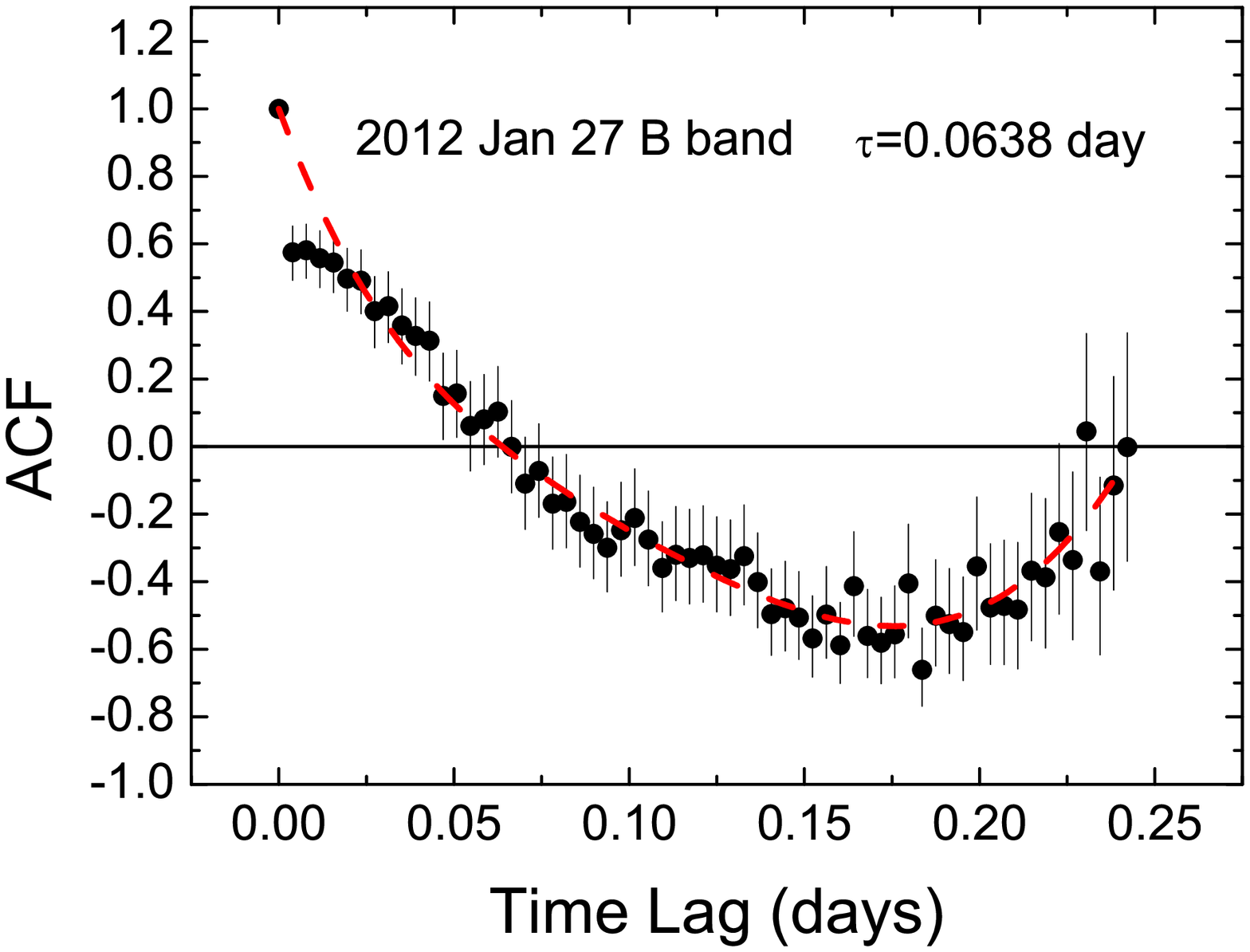}
\includegraphics[angle=0,width=0.25\textwidth]{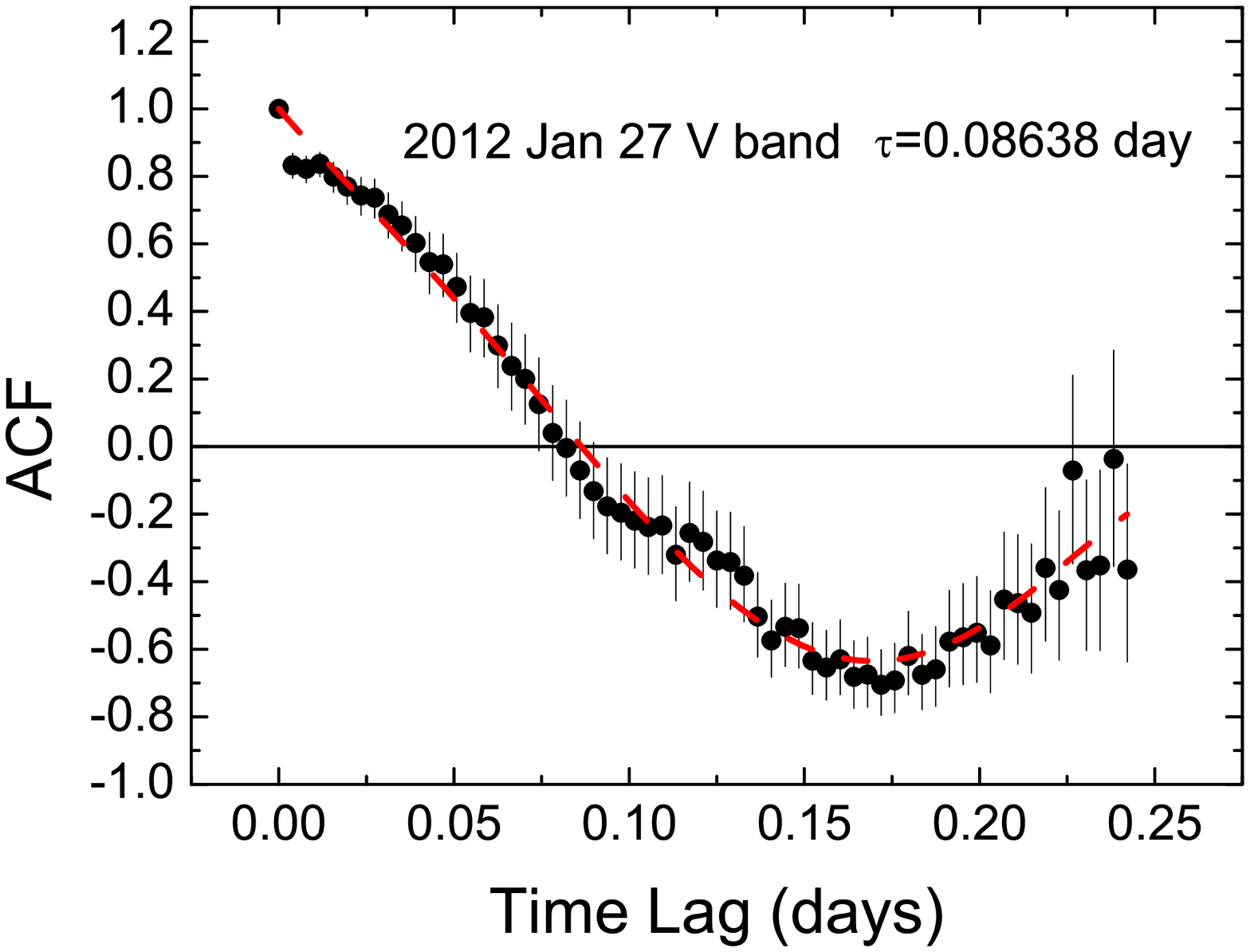}
\includegraphics[angle=0,width=0.25\textwidth]{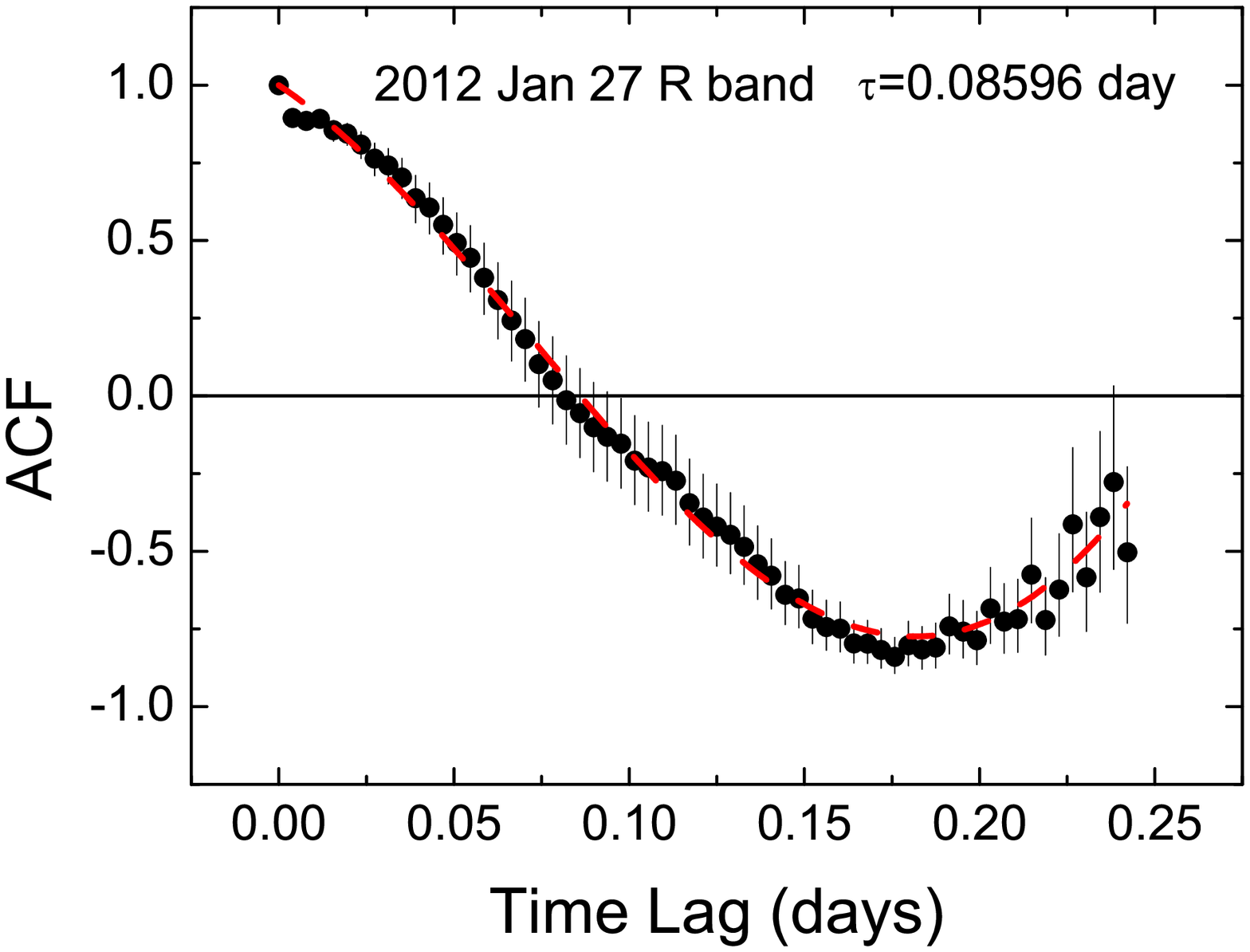}
\includegraphics[angle=0,width=0.25\textwidth]{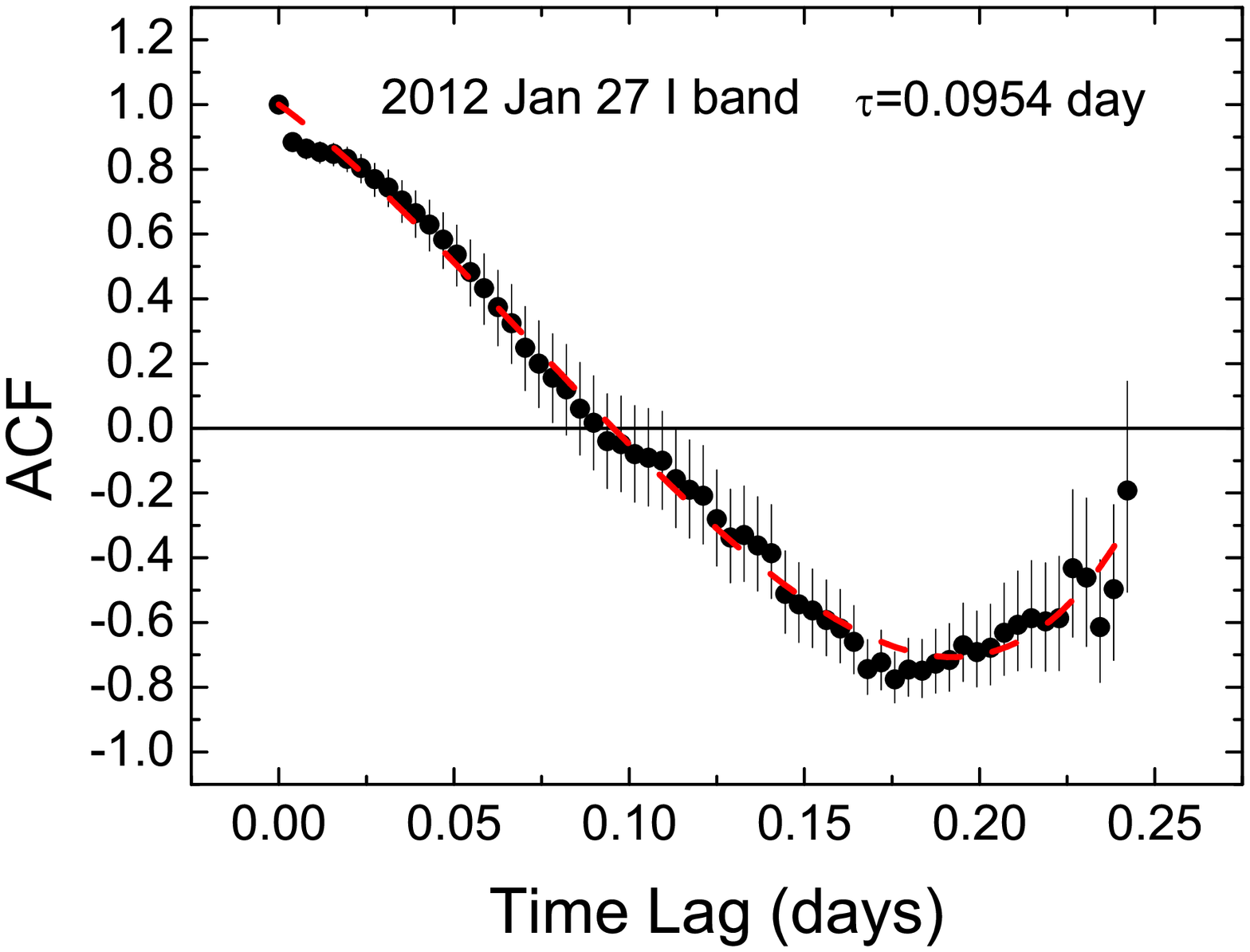}
\includegraphics[angle=0,width=0.25\textwidth]{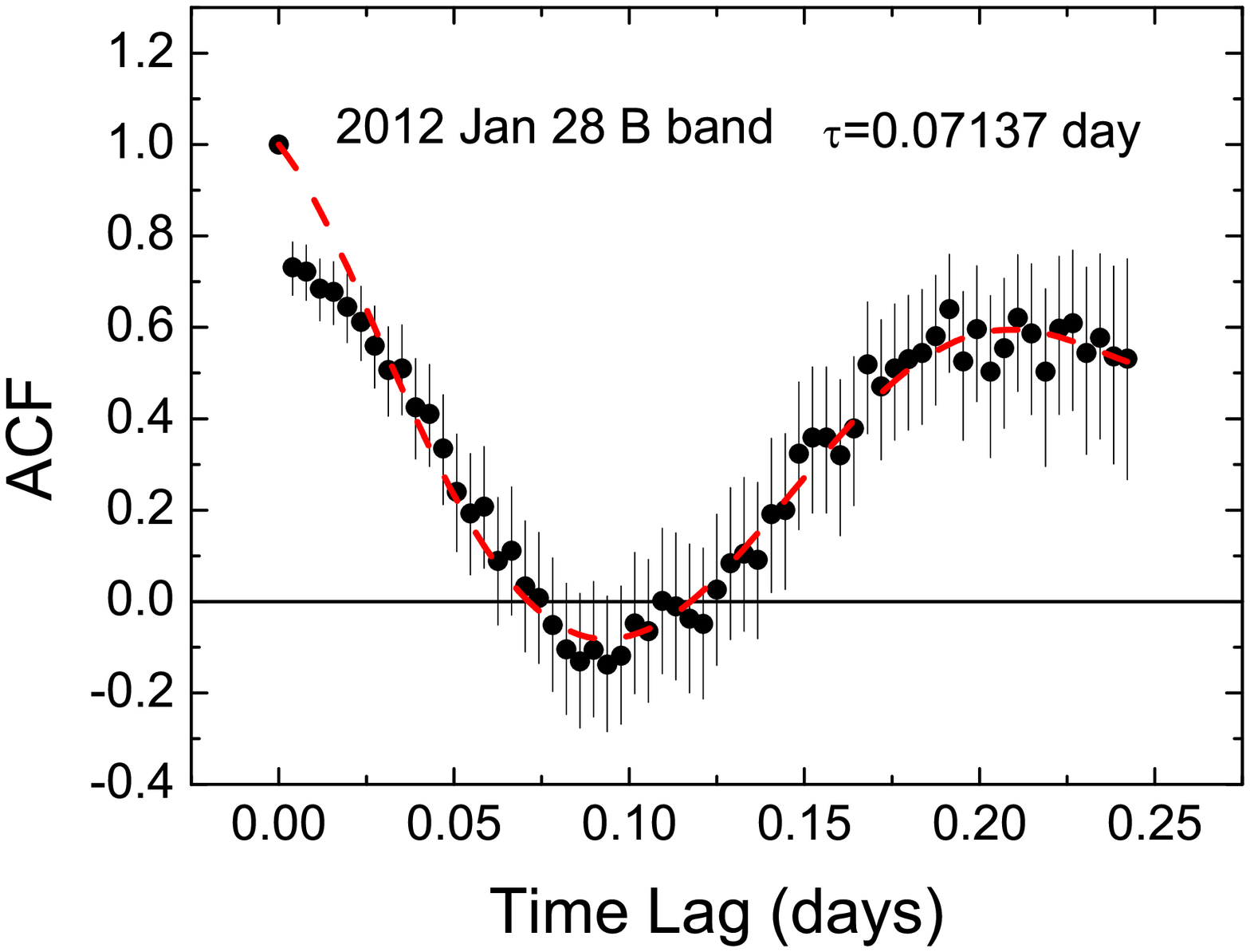}
\includegraphics[angle=0,width=0.25\textwidth]{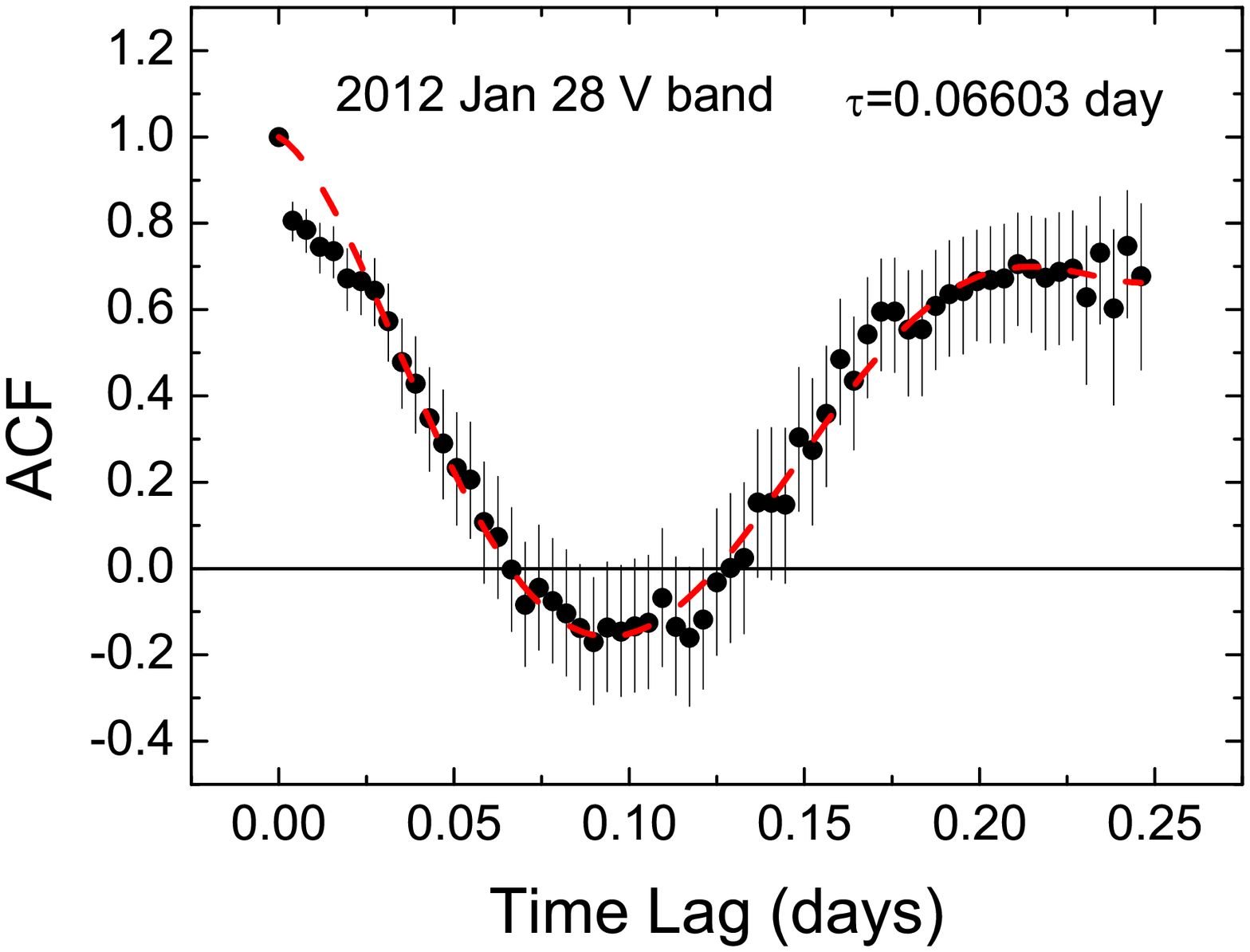}
\includegraphics[angle=0,width=0.25\textwidth]{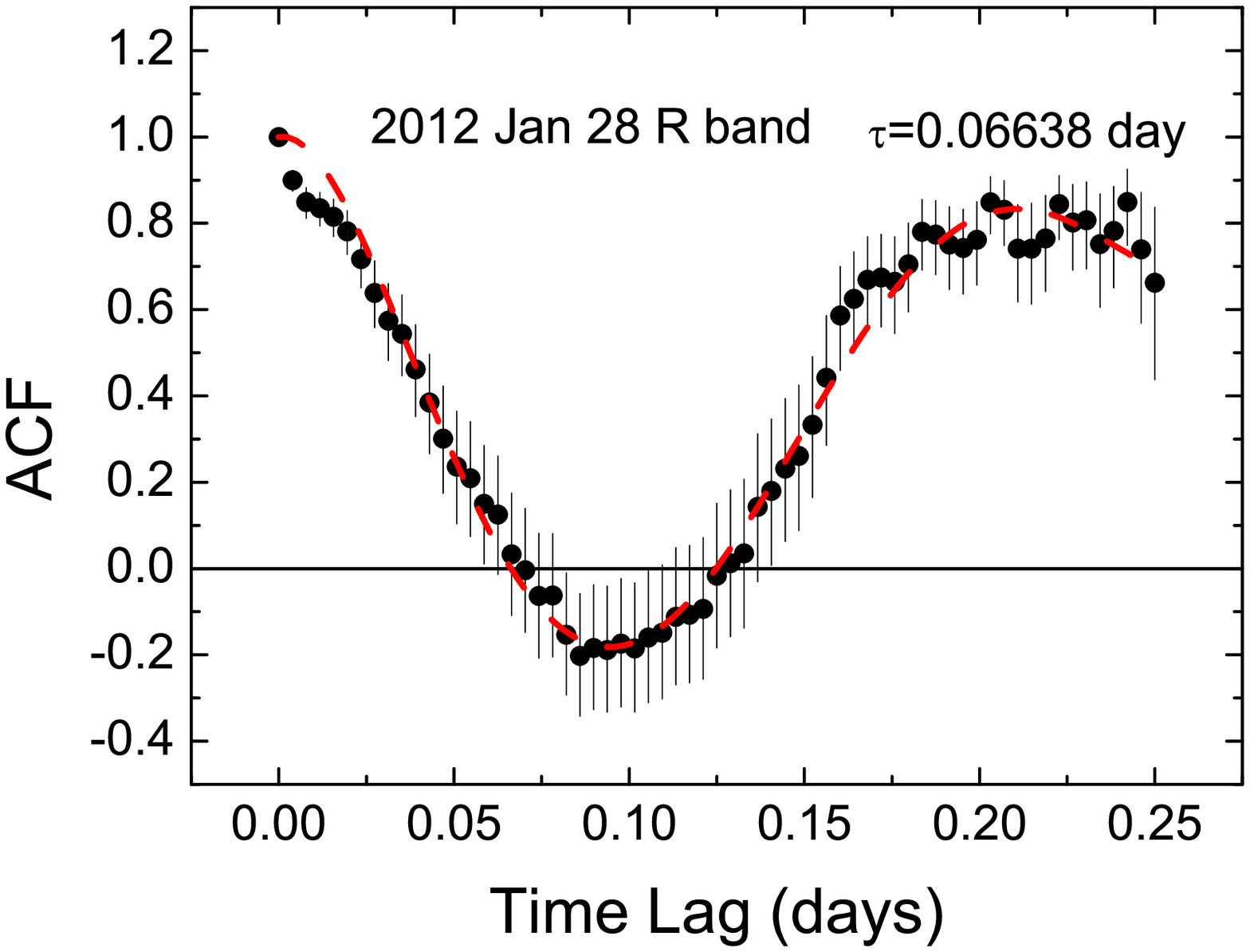}
\includegraphics[angle=0,width=0.25\textwidth]{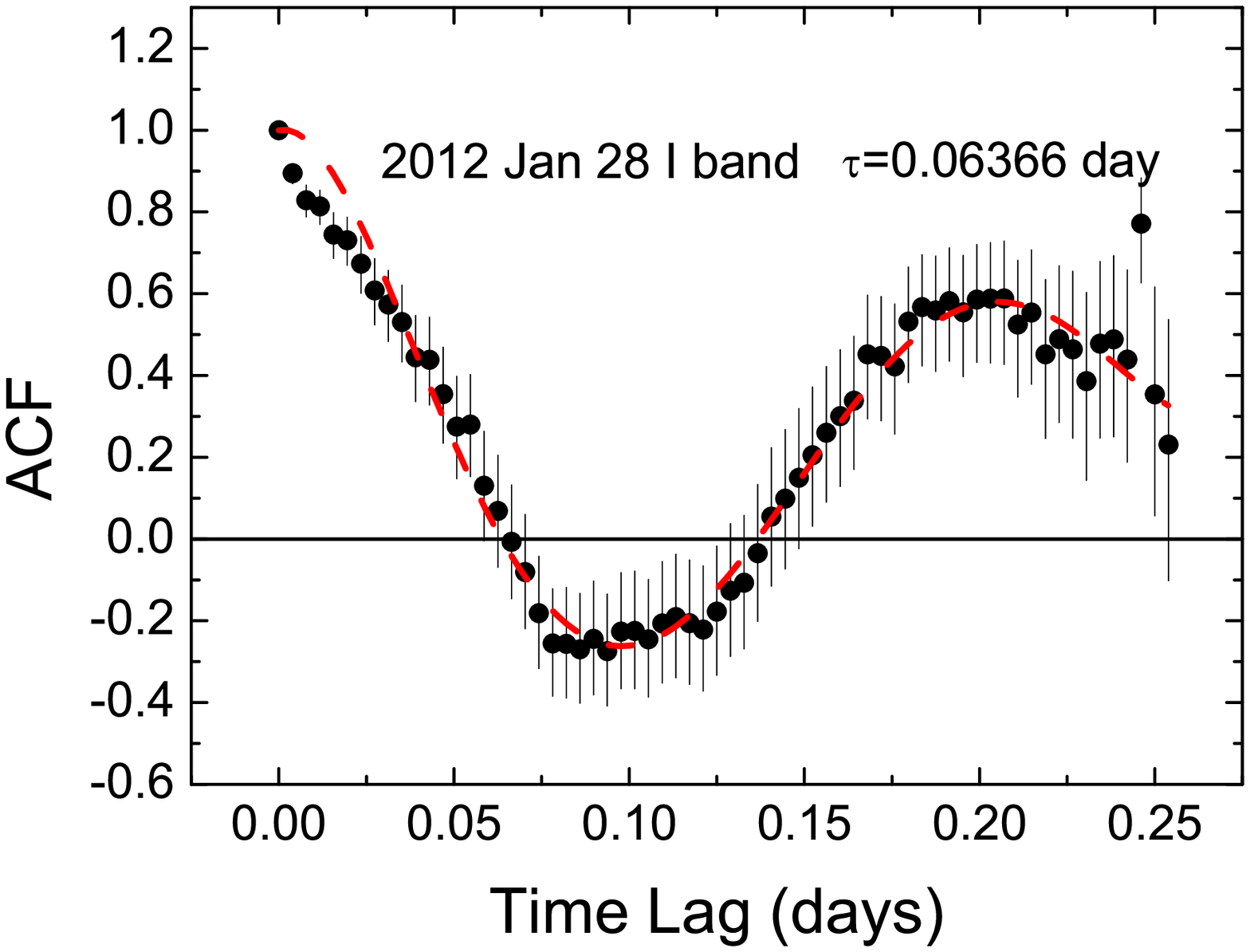}
\includegraphics[angle=0,width=0.25\textwidth]{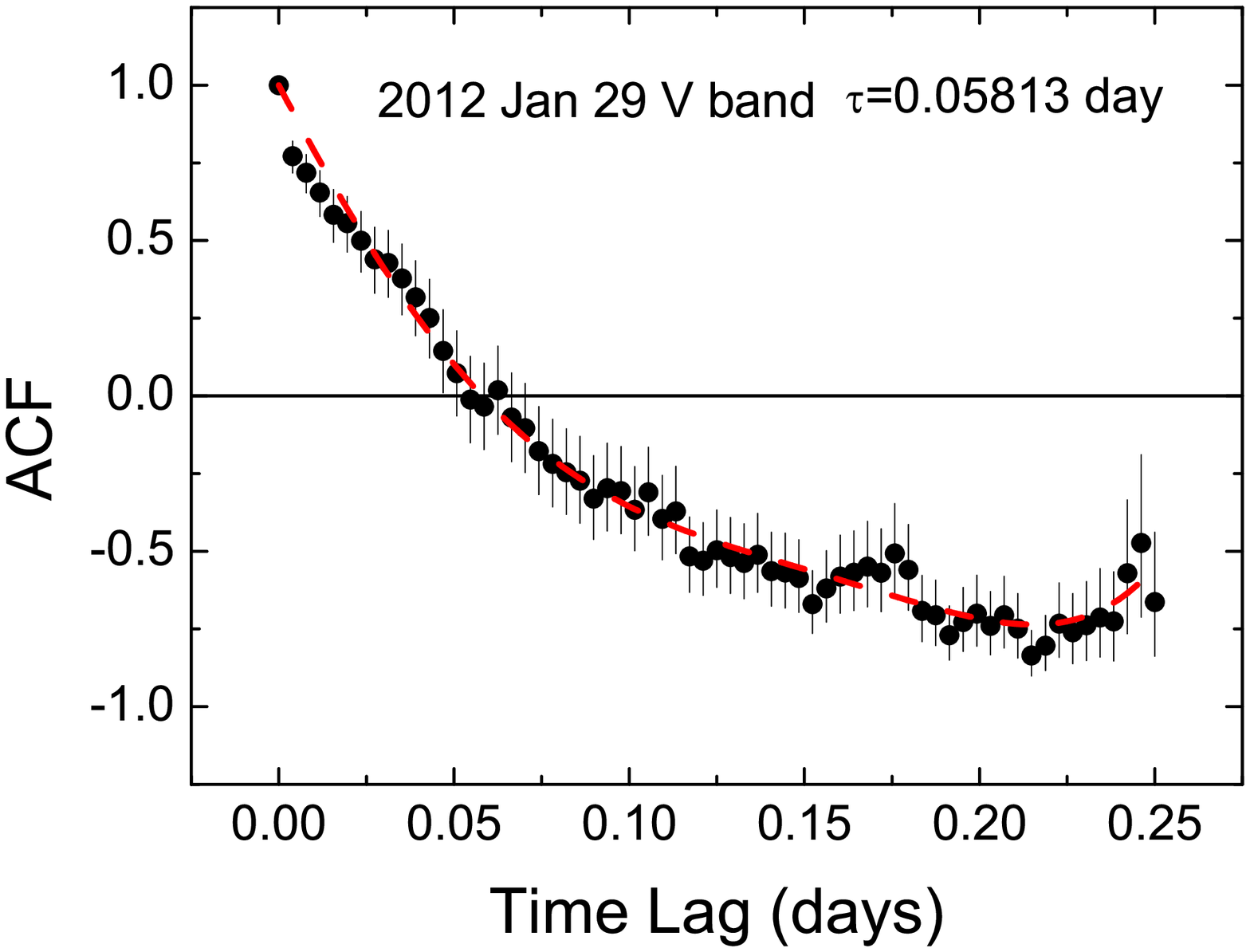}
\includegraphics[angle=0,width=0.25\textwidth]{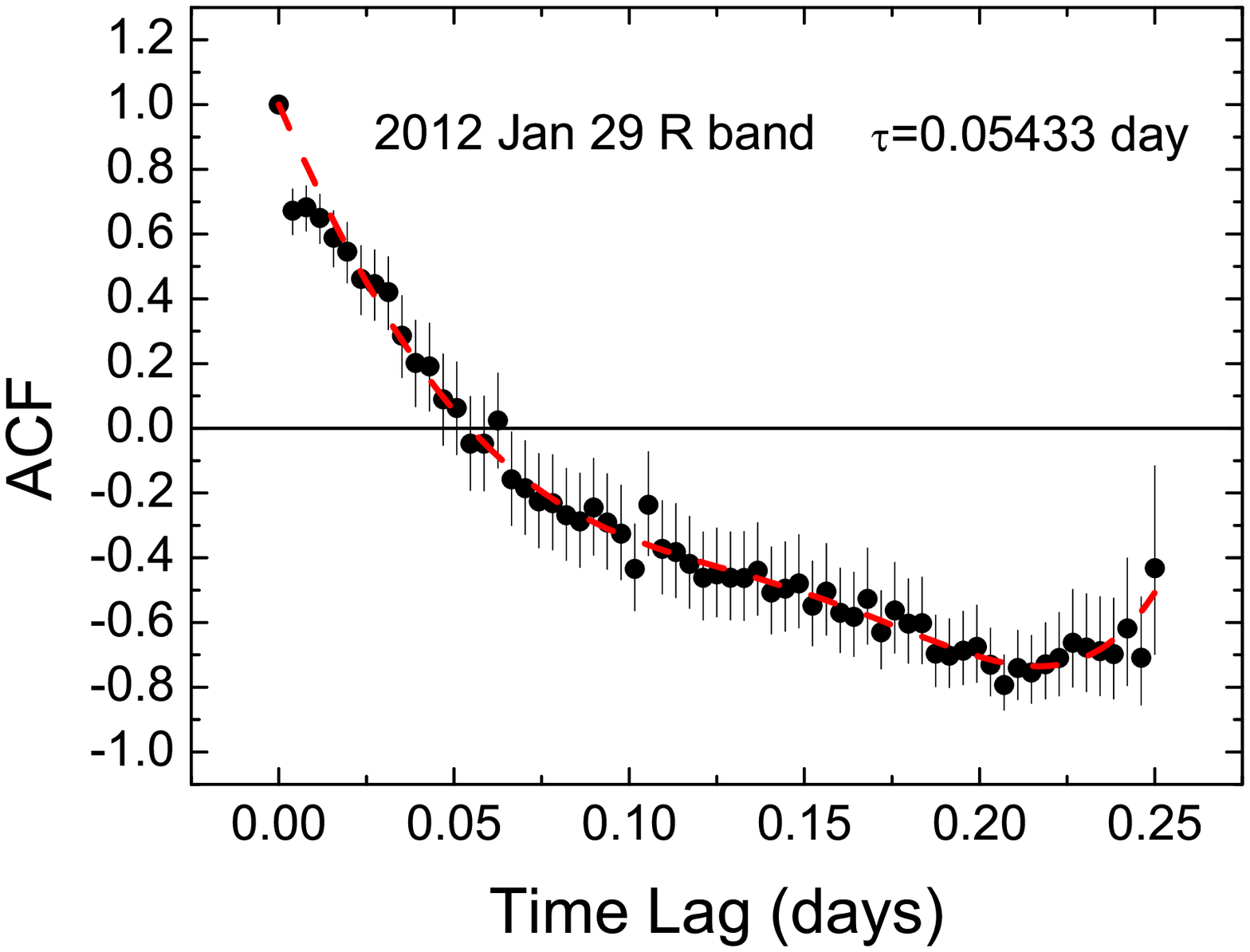}
\includegraphics[angle=0,width=0.25\textwidth]{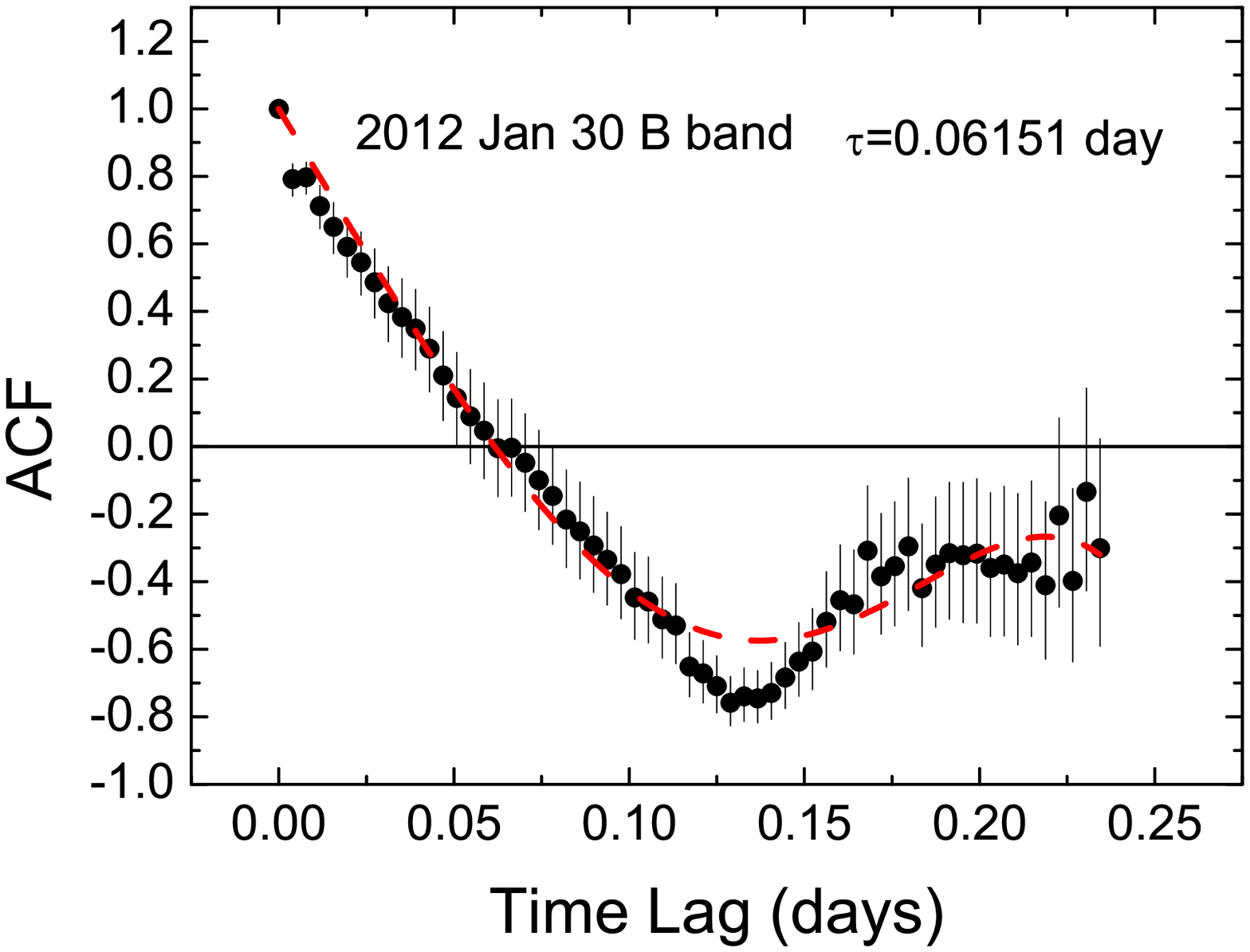}
\includegraphics[angle=0,width=0.25\textwidth]{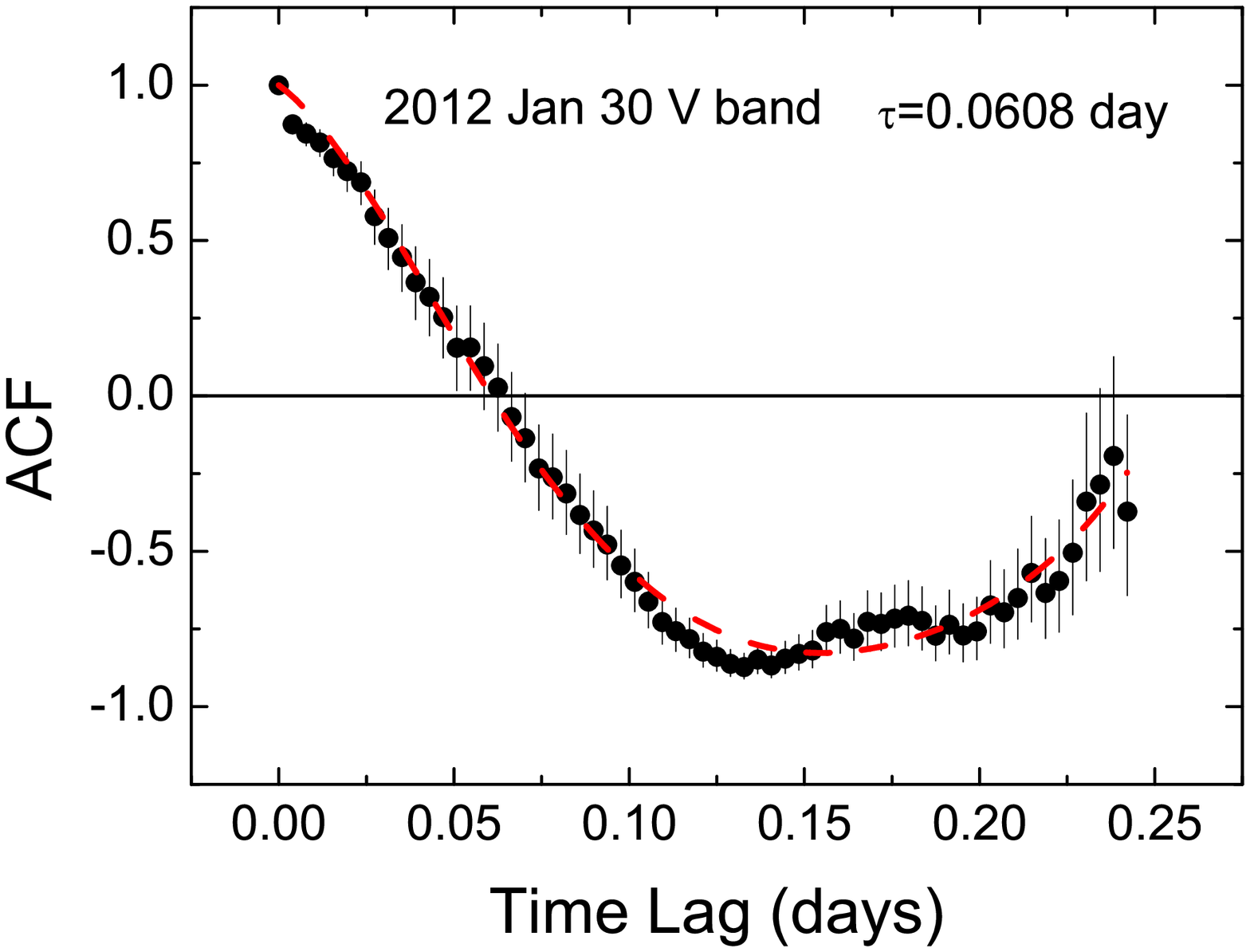}
\includegraphics[angle=0,width=0.25\textwidth]{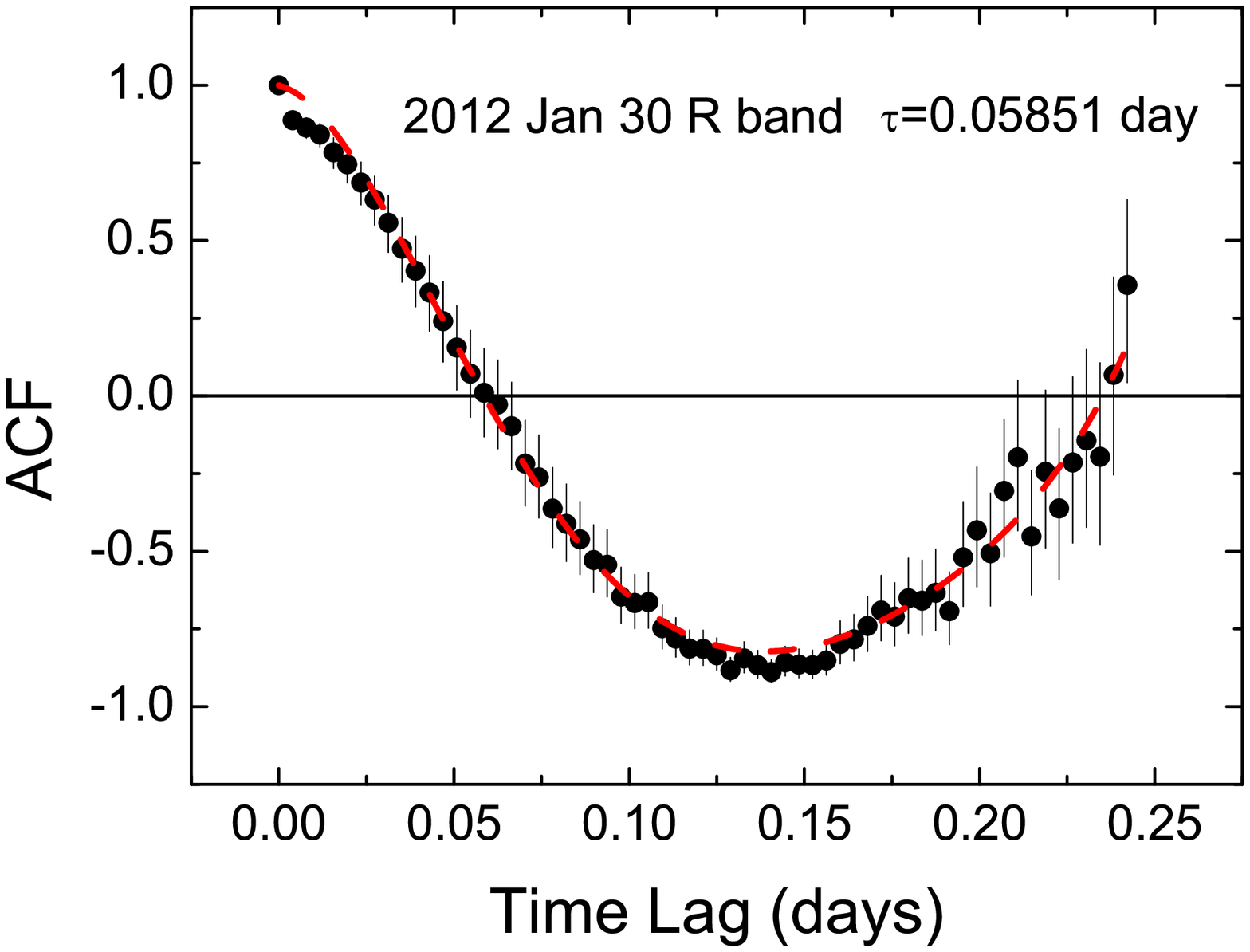}
\includegraphics[angle=0,width=0.25\textwidth]{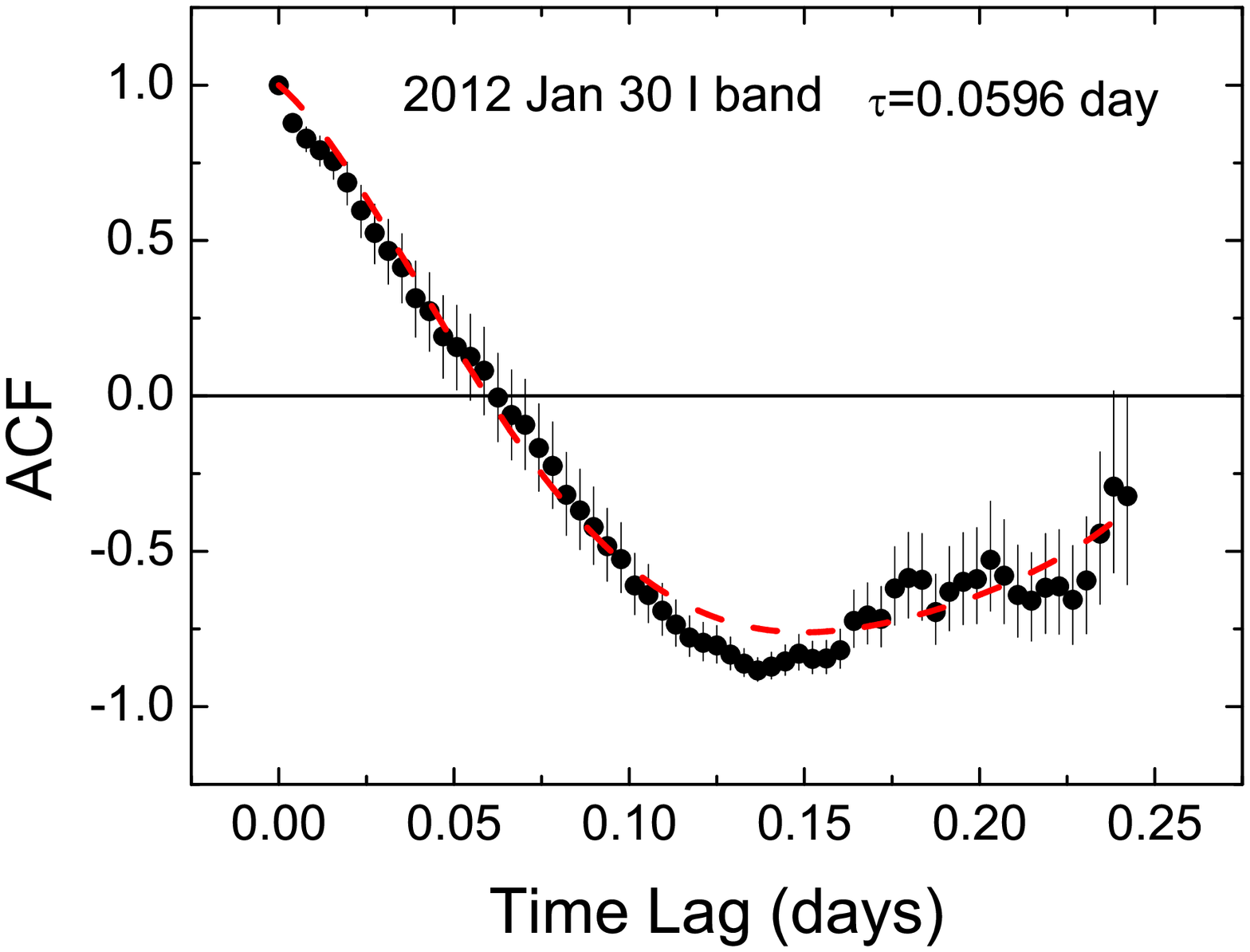}
\includegraphics[angle=0,width=0.25\textwidth]{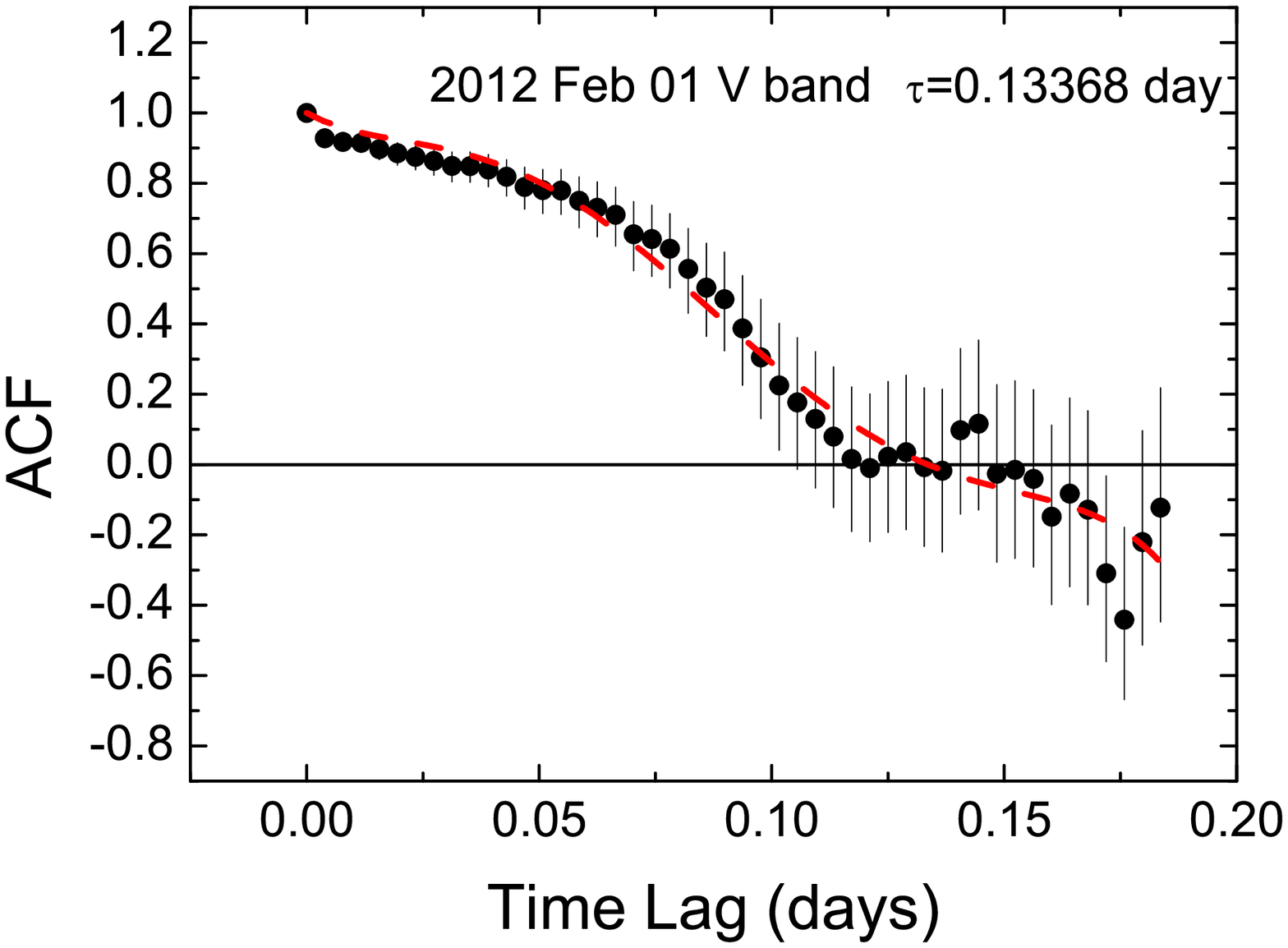}
\includegraphics[angle=0,width=0.25\textwidth]{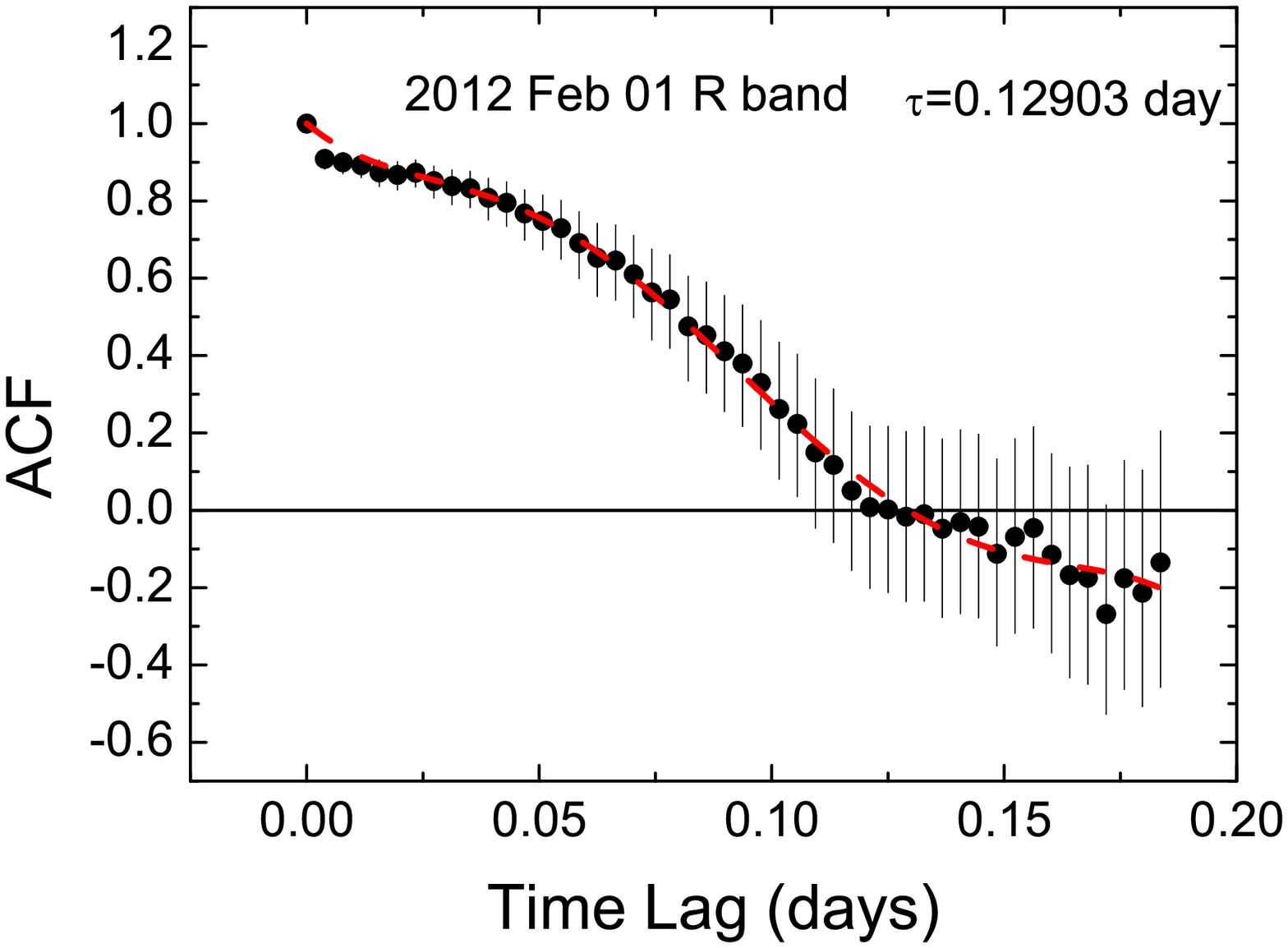}
\includegraphics[angle=0,width=0.25\textwidth]{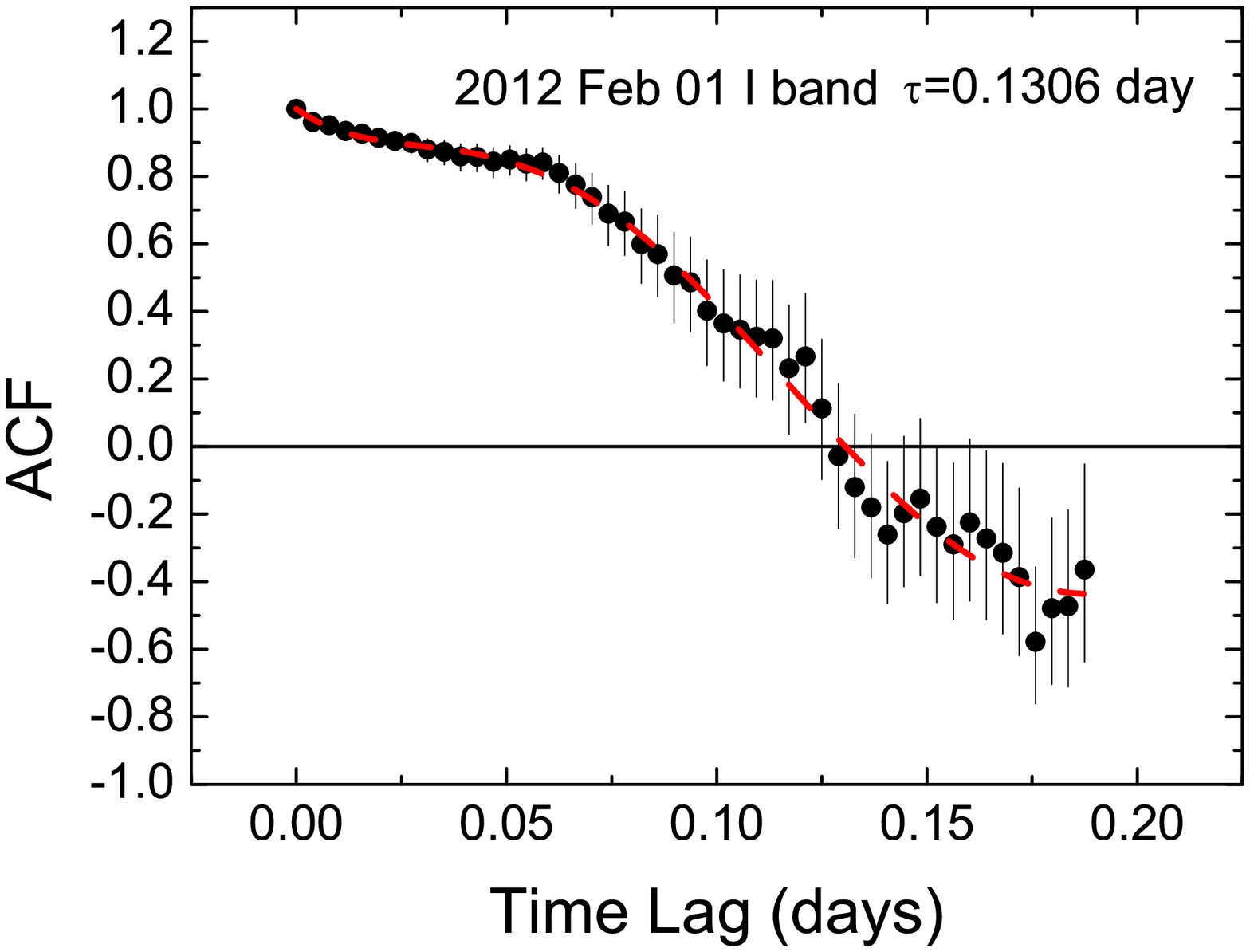}
\caption{The results of ACF analysis. The red dashed line is a fifth-order polynomial
least-squares fit.}
\end{center}
\end{figure}

\begin{figure}
\begin{center}
\includegraphics[angle=0,width=0.32\textwidth]{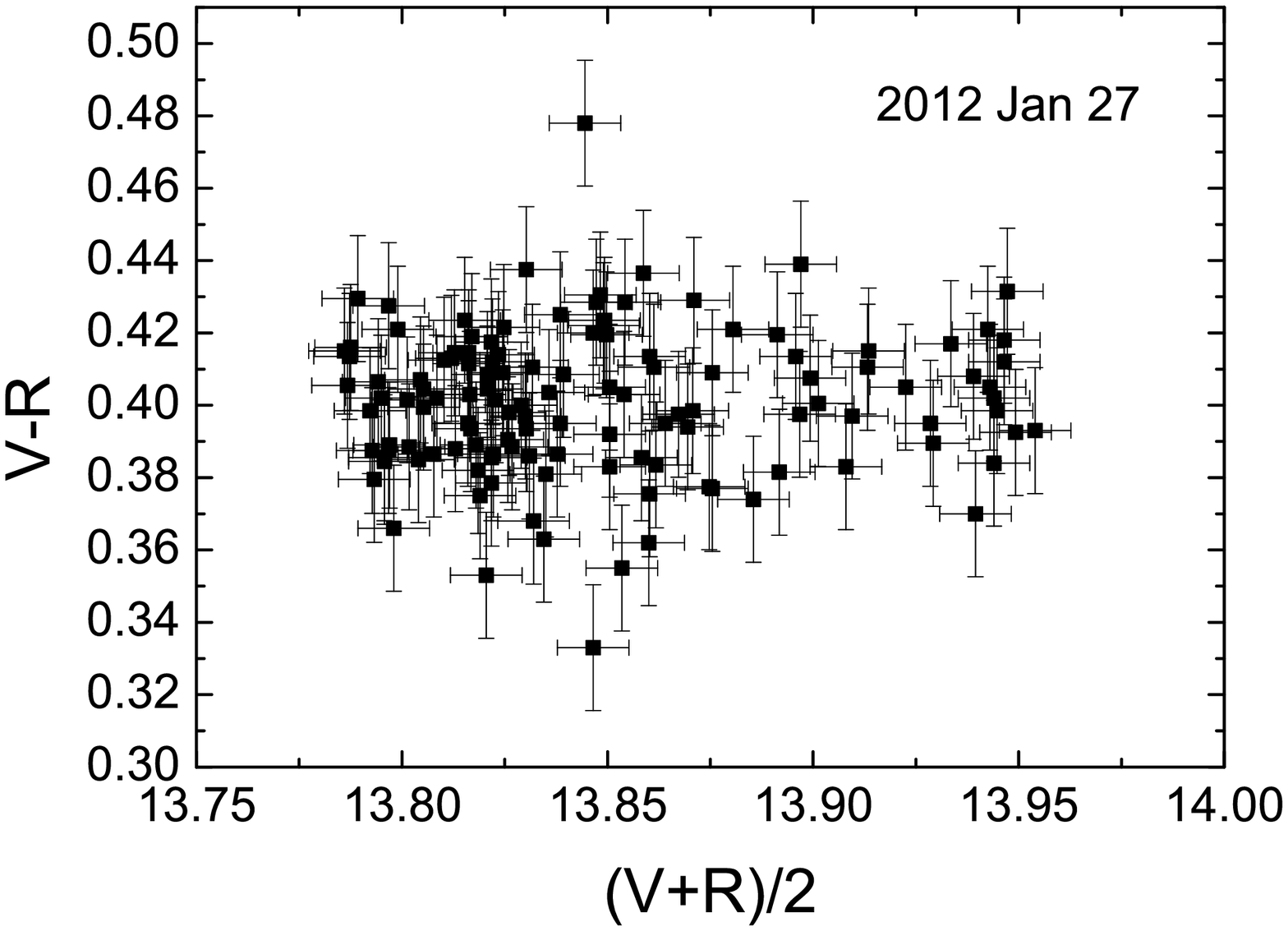}
\includegraphics[angle=0,width=0.32\textwidth]{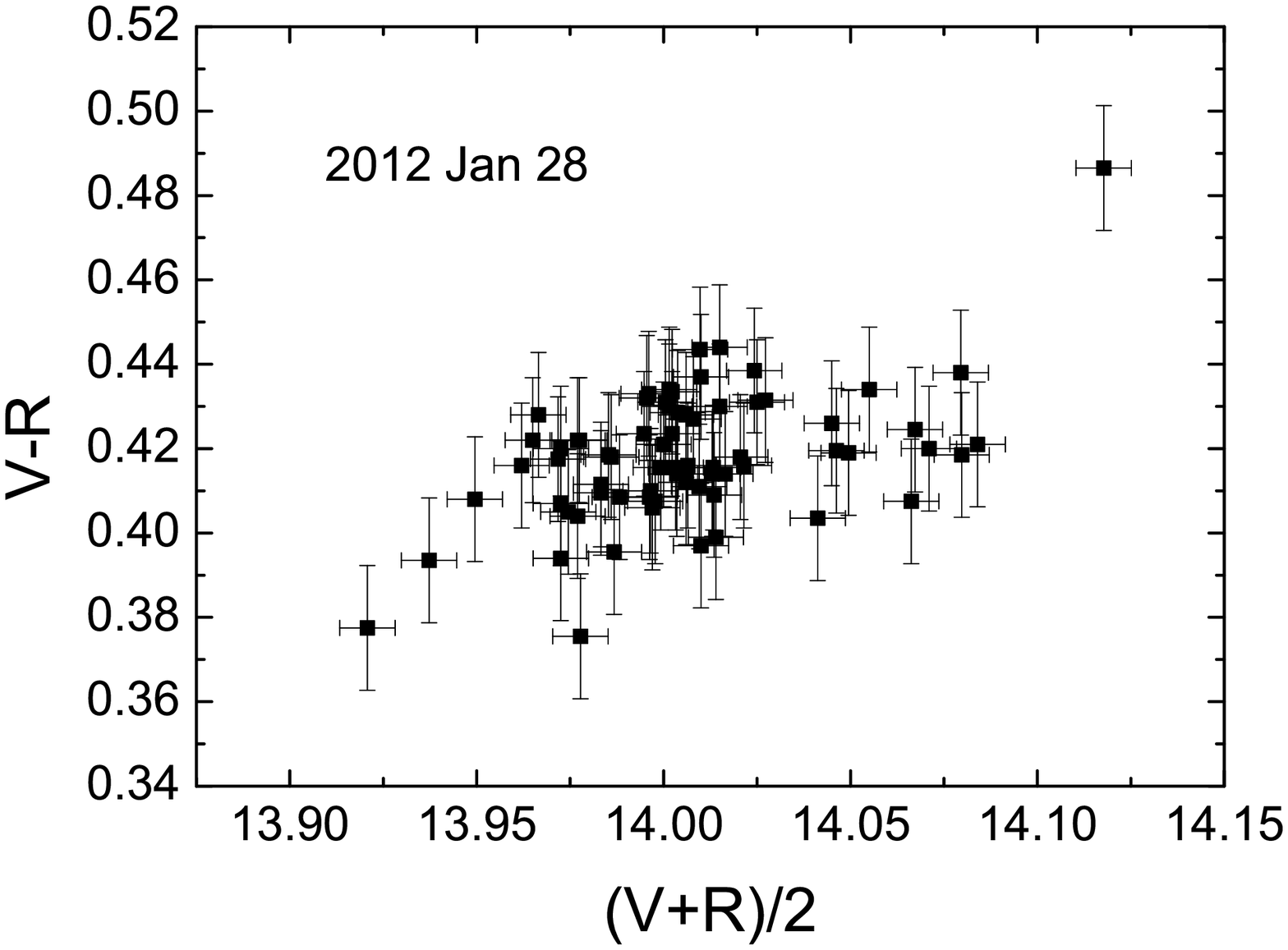}
\includegraphics[angle=0,width=0.32\textwidth]{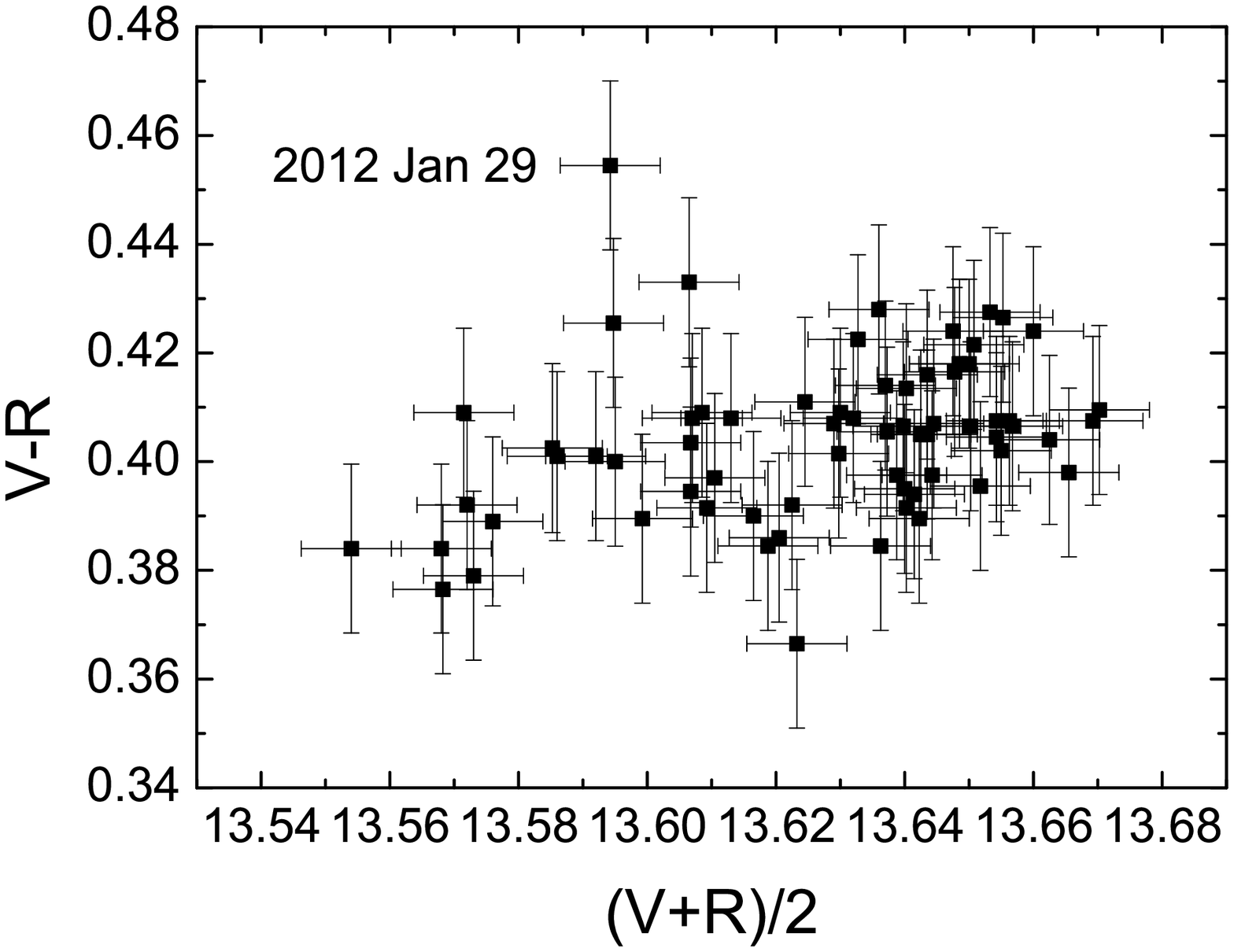}
\includegraphics[angle=0,width=0.32\textwidth]{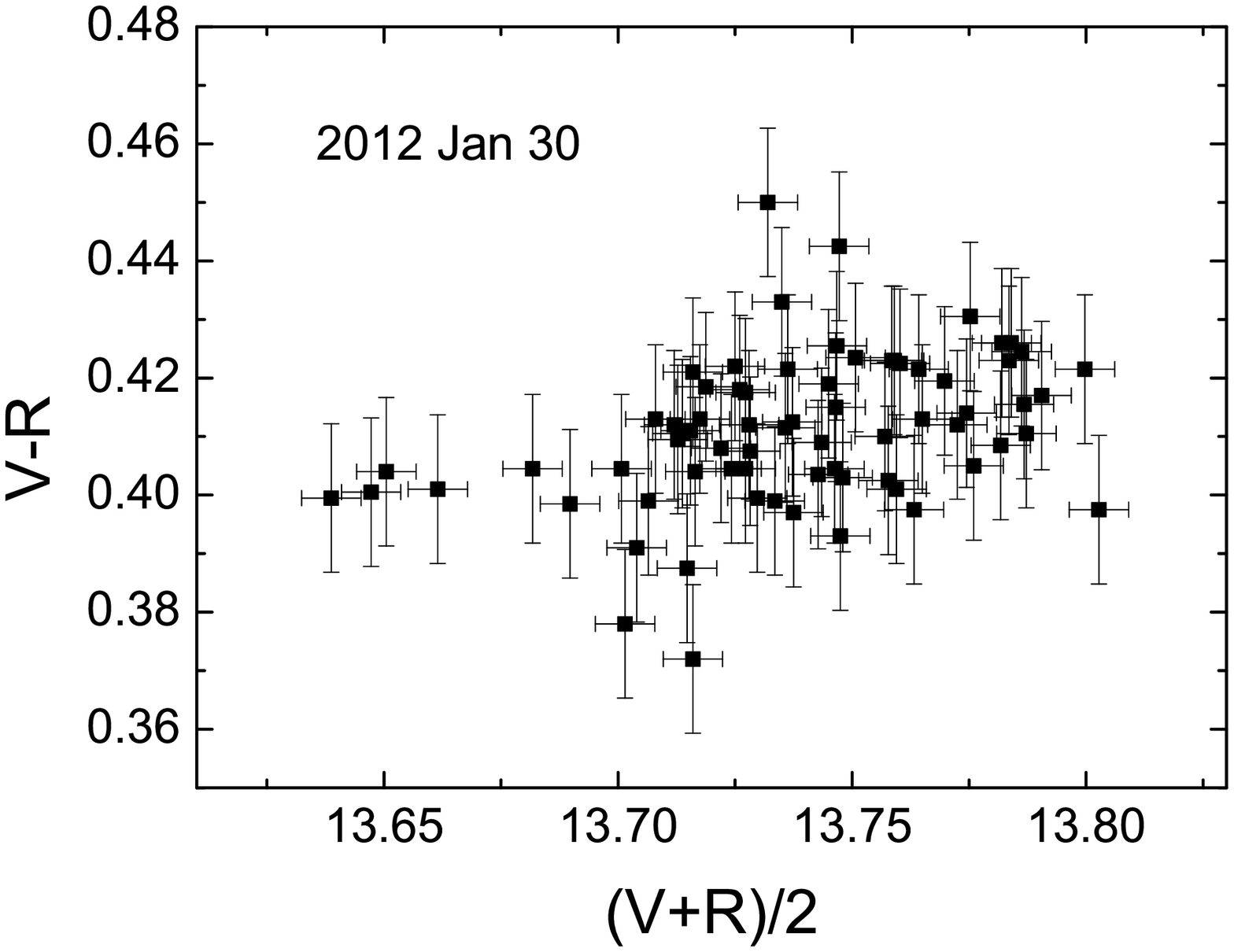}
\includegraphics[angle=0,width=0.32\textwidth]{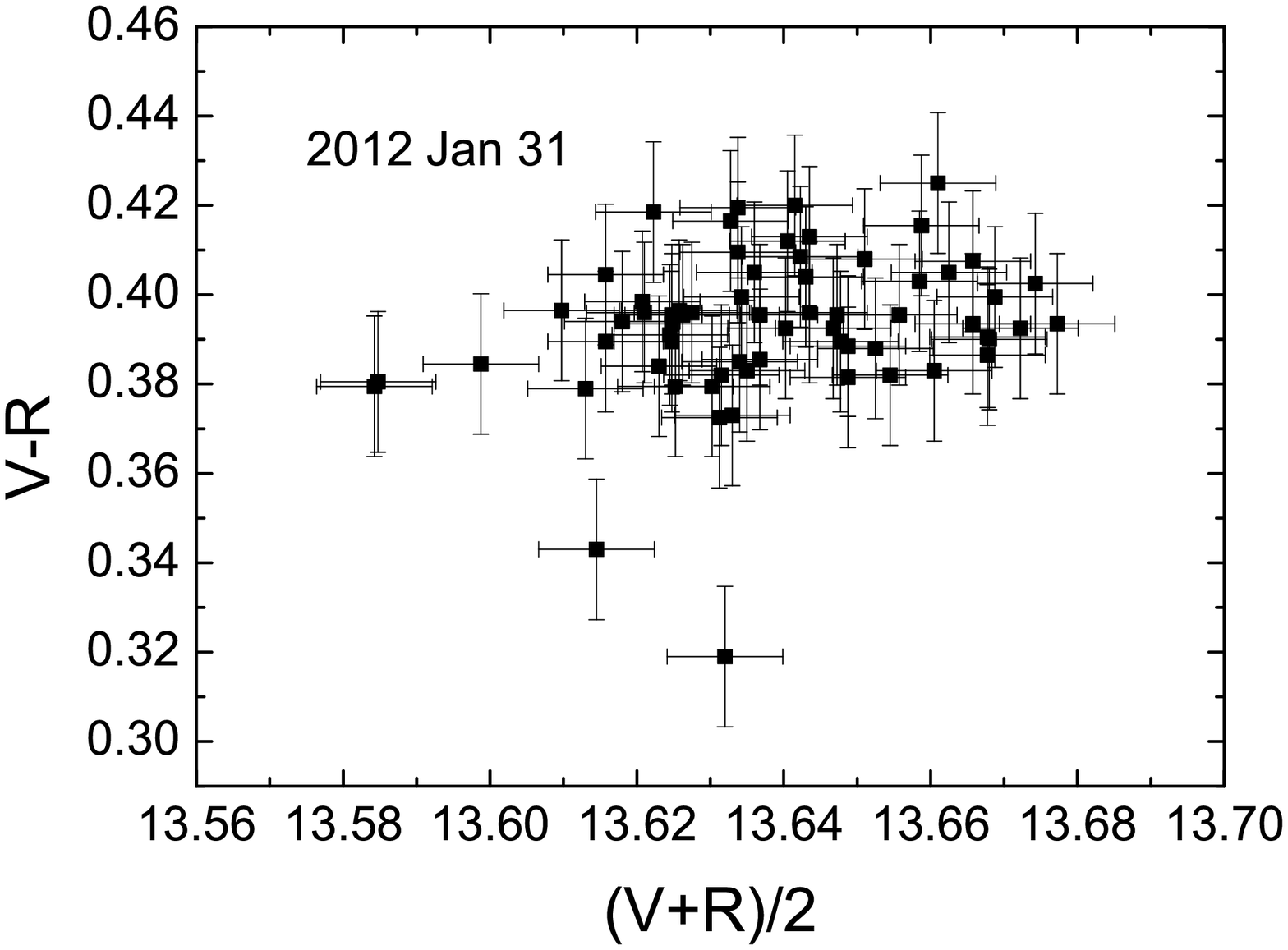}
\includegraphics[angle=0,width=0.32\textwidth]{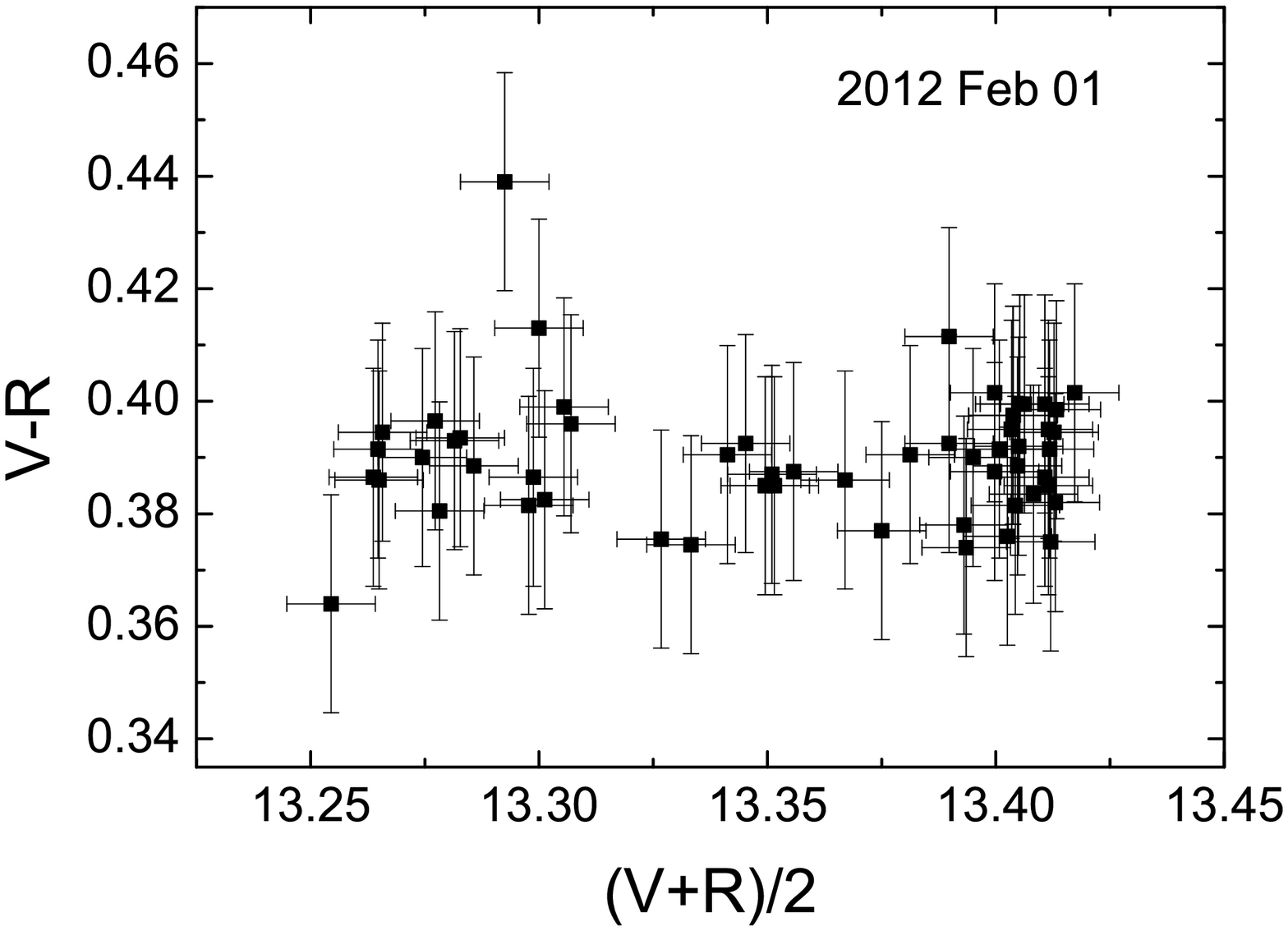}
\includegraphics[angle=0,width=0.32\textwidth]{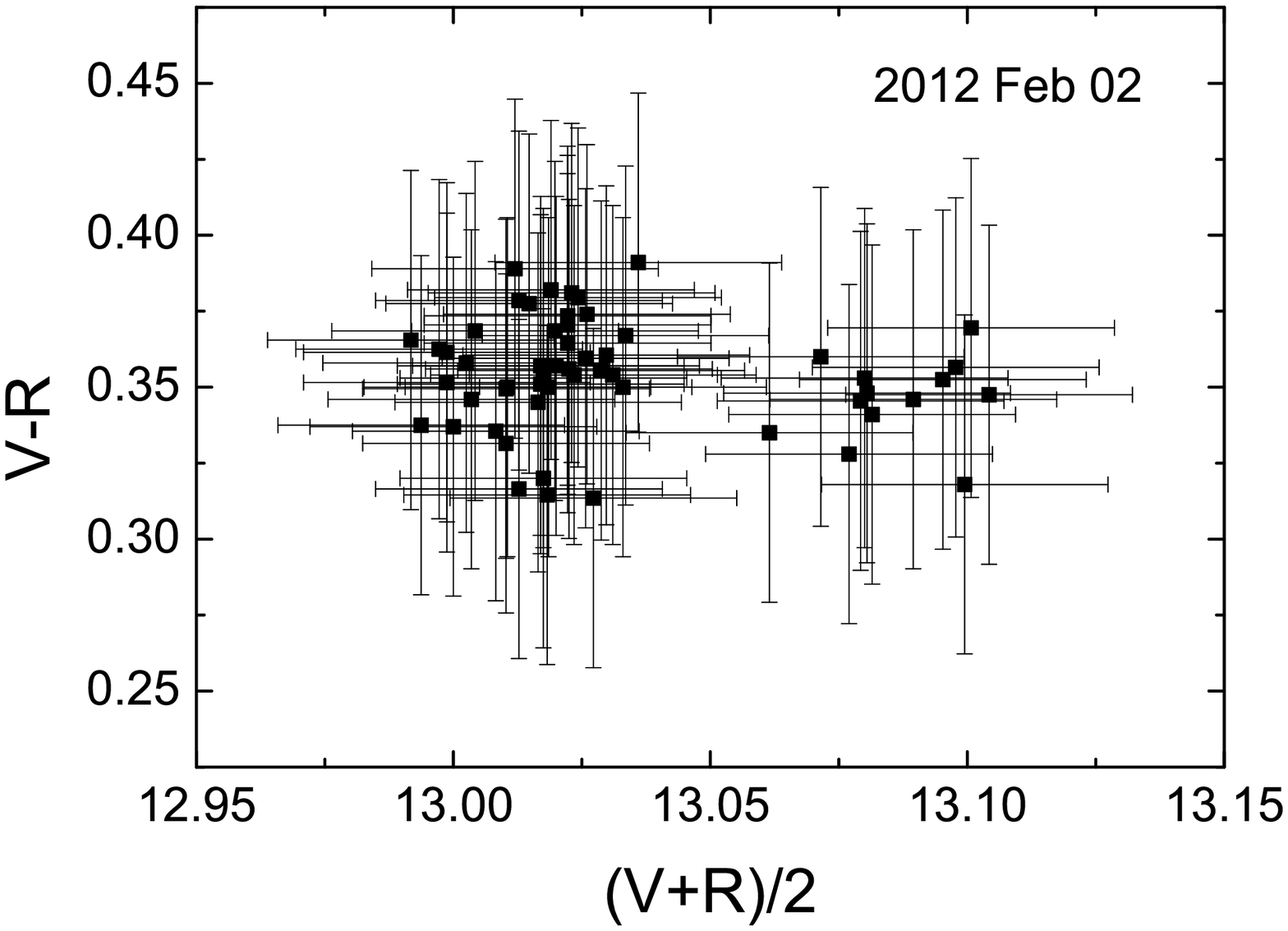}
\includegraphics[angle=0,width=0.32\textwidth]{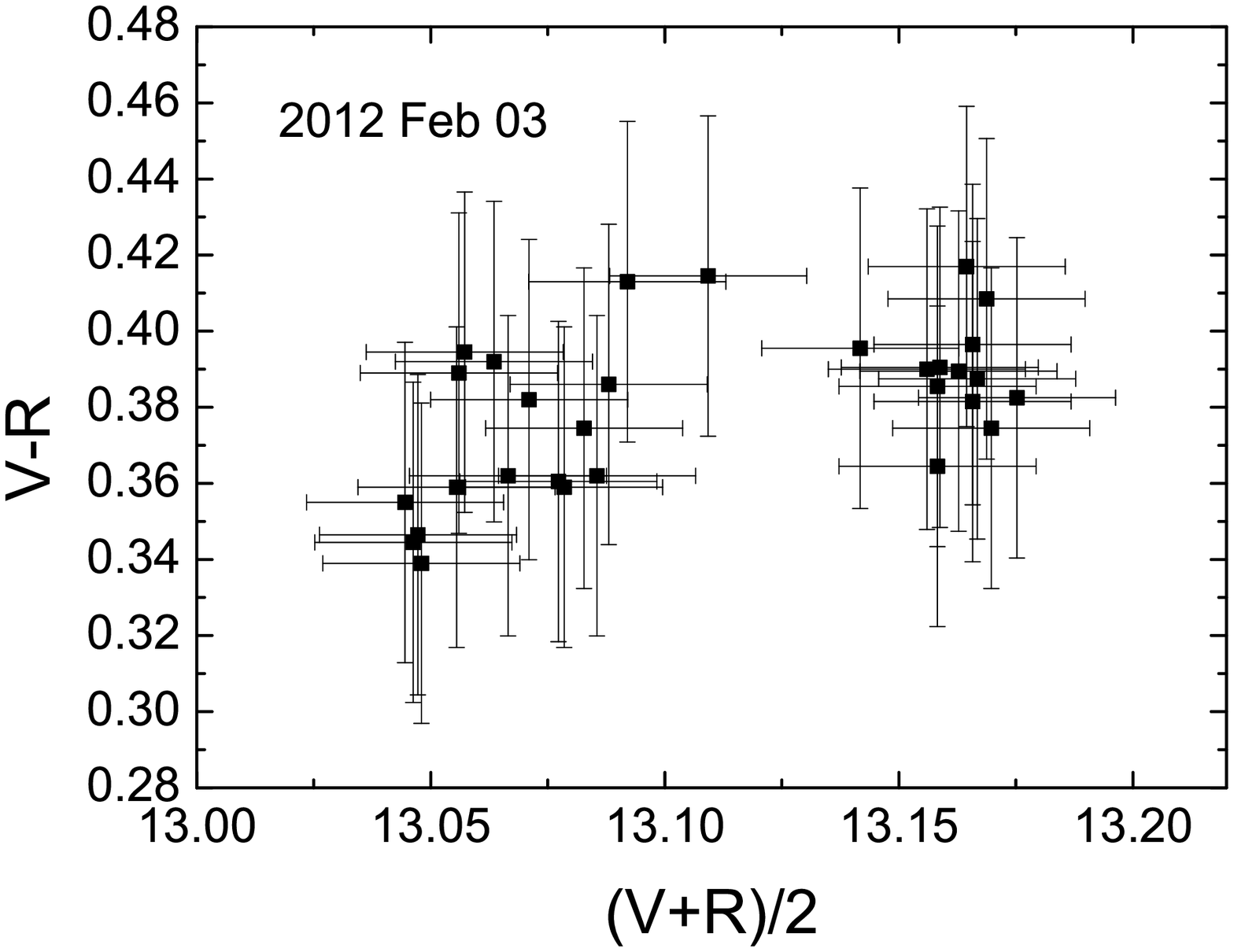}
\includegraphics[angle=0,width=0.32\textwidth]{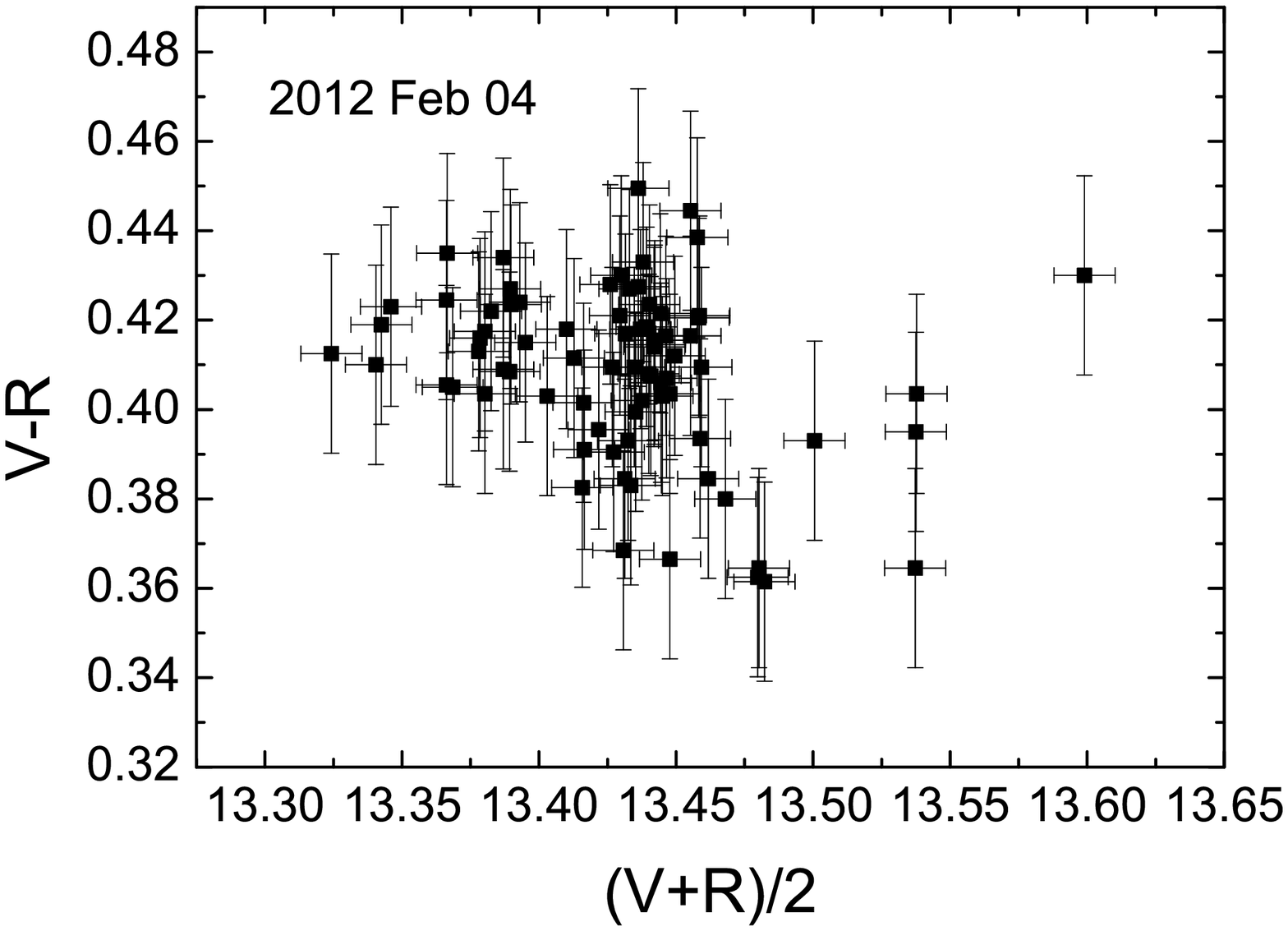}
\includegraphics[angle=0,width=0.32\textwidth]{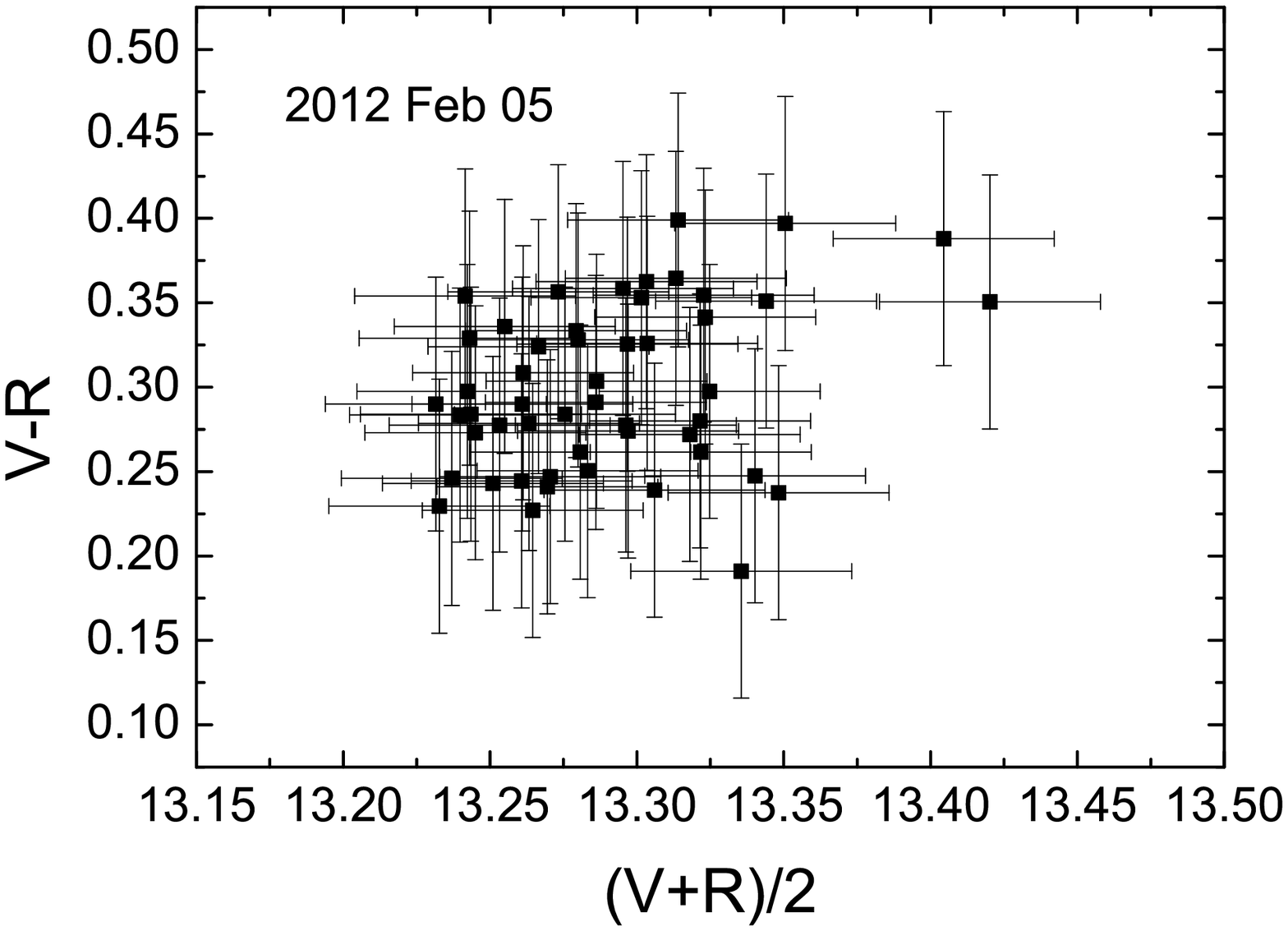}
\includegraphics[angle=0,width=0.32\textwidth]{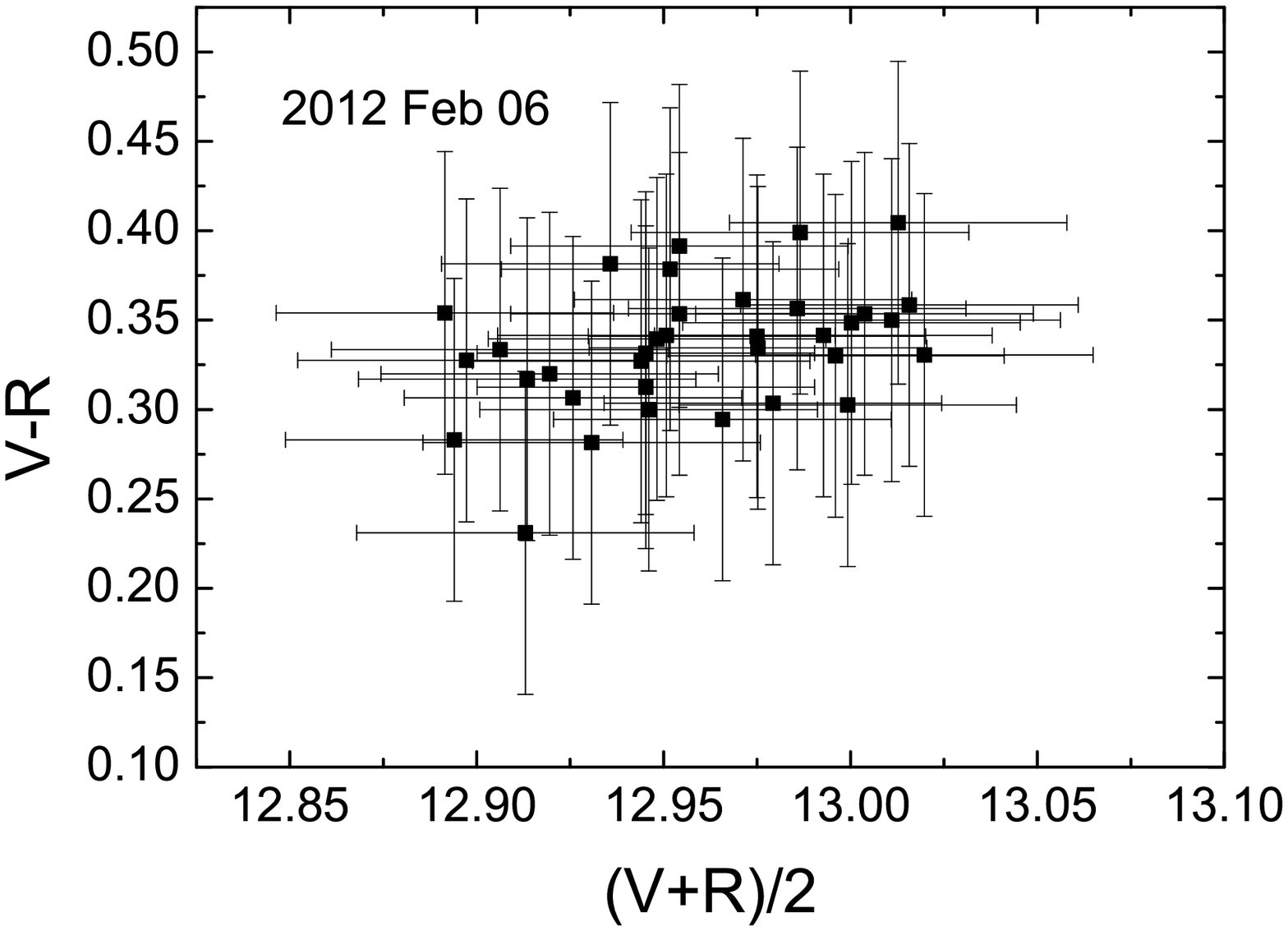}
\includegraphics[angle=0,width=0.32\textwidth]{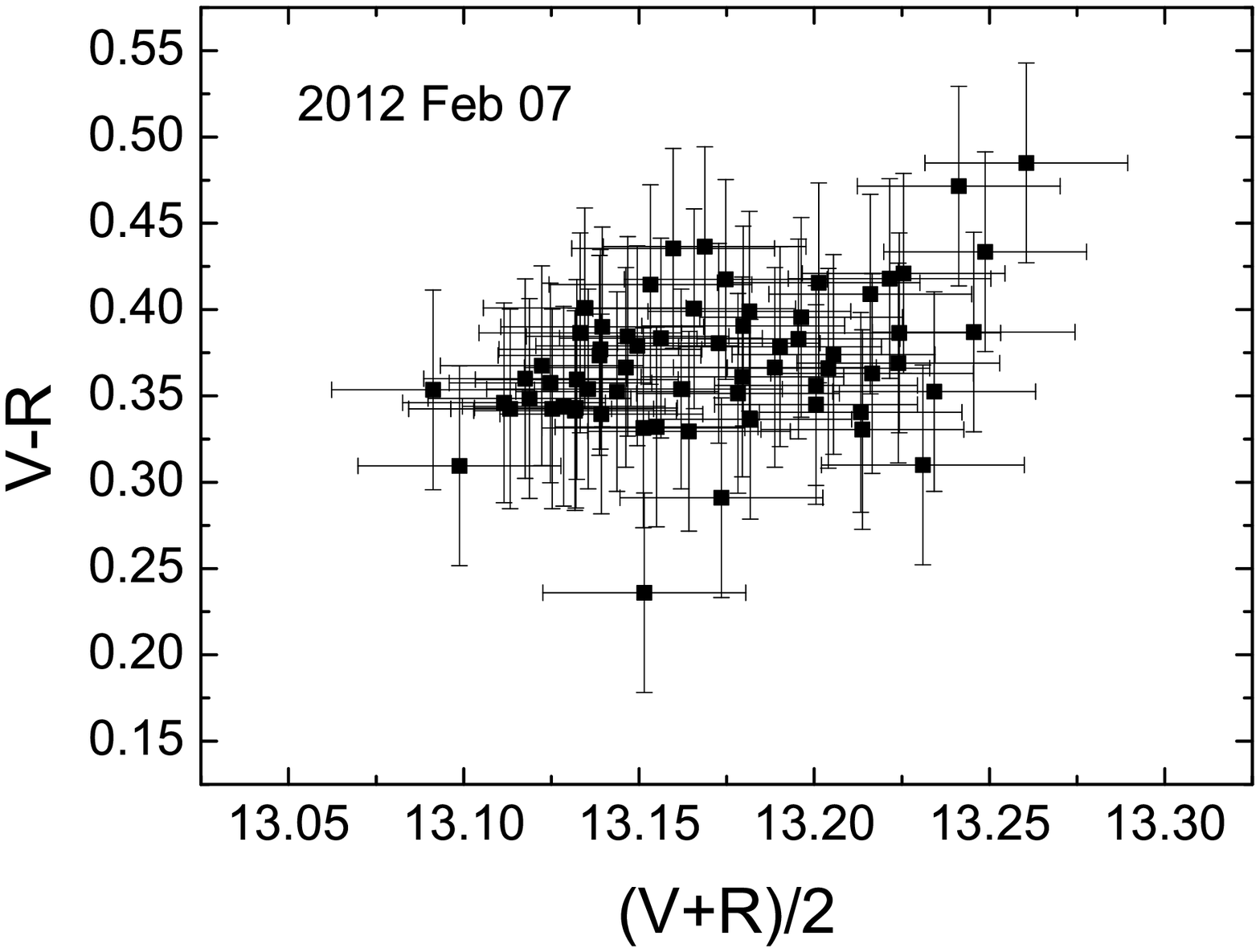}
\includegraphics[angle=0,width=0.32\textwidth]{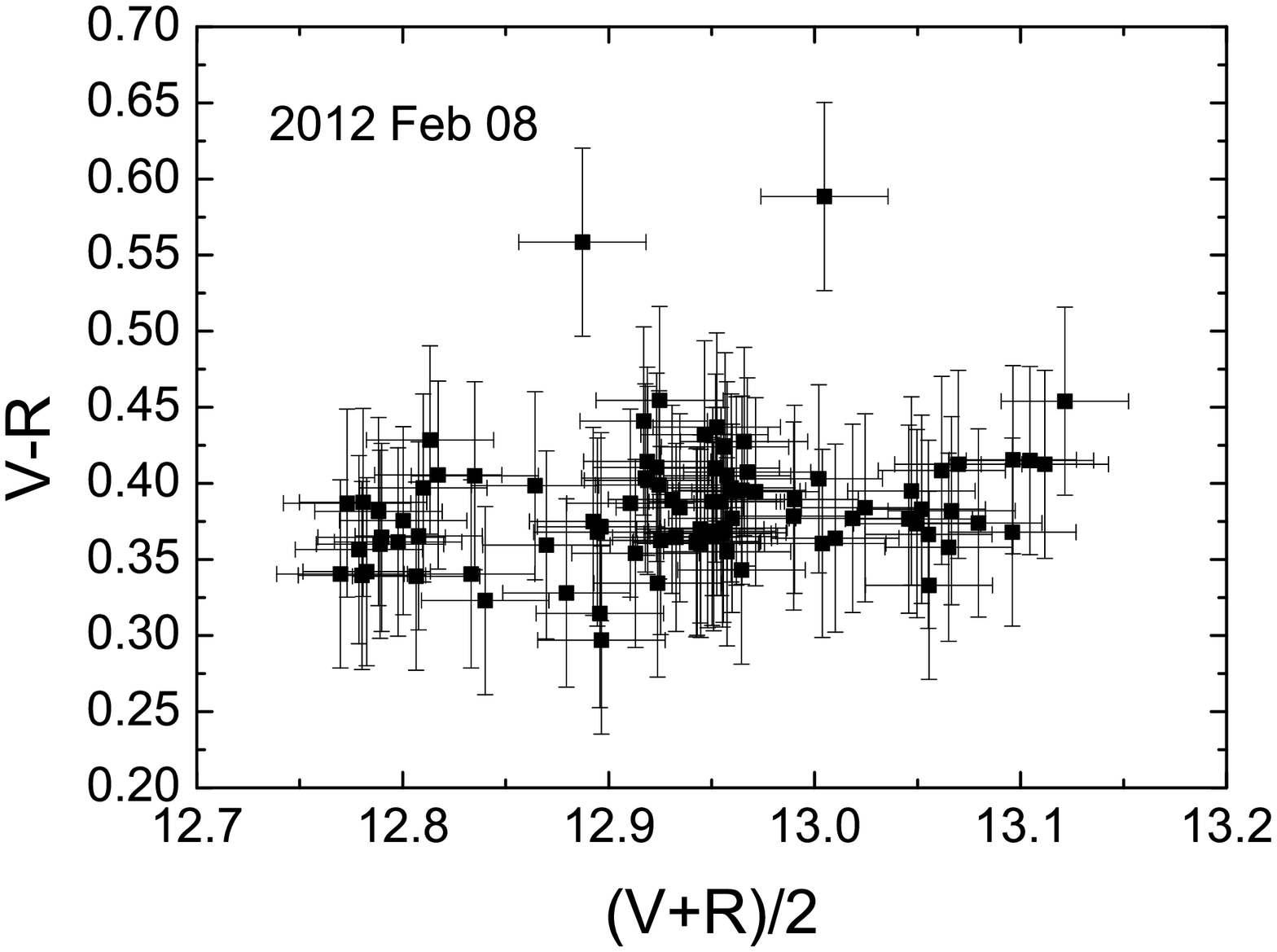}
\includegraphics[angle=0,width=0.45\textwidth]{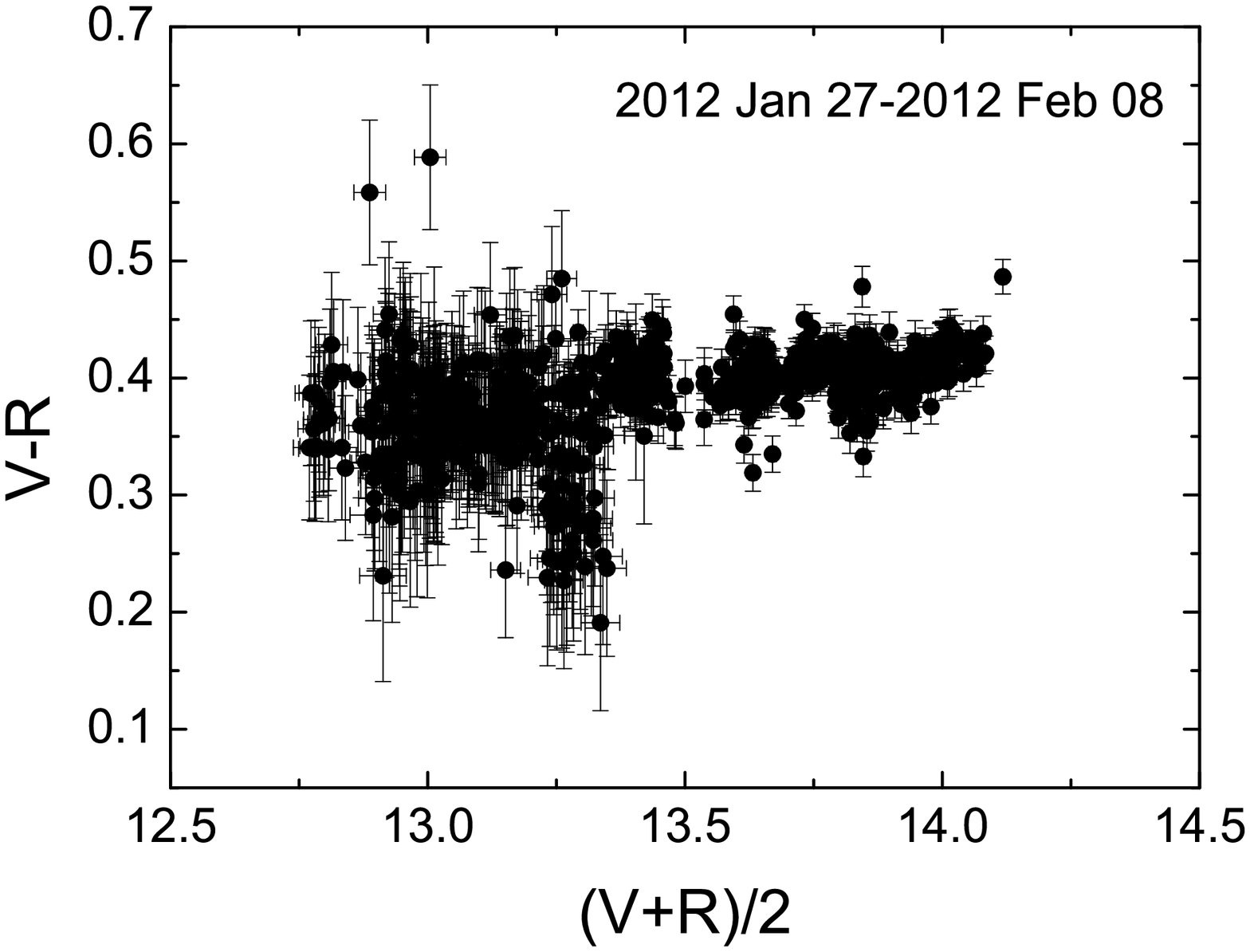}
\caption{The correlations between $V-R$ index and $(V+R)/2$ magnitude on intraday and short timescales.}
\end{center}
\end{figure}

\begin{figure}
\begin{center}
\includegraphics[width=14cm,height=14cm]{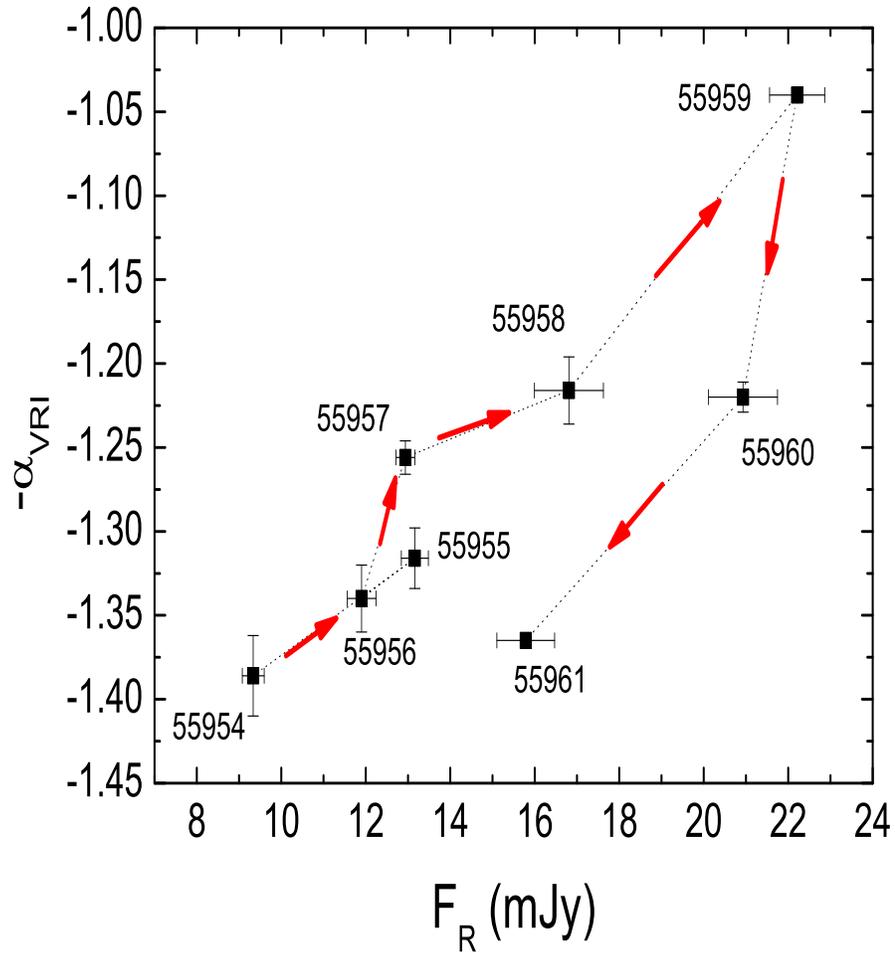}
\caption{The spectral index versus the flux in the flare event. The flare extended from Jan. 28 to Feb. 04.}
\end{center}
\end{figure}

\clearpage

\begin{deluxetable}{cr}
\tablecaption{Details of telescopes and instruments.\label{tbl-3}} \tablewidth{0pt}
\tablehead{\colhead{Telescope} & \colhead{60-cm BOOTES-4}}
\startdata
CCD model:	&	SBIG 1001 E	\\
Chip size:	&	$1024\times1024$ pixels	\\
Pixel size:	&	$24.6\times24.6$ $\mu m$	\\
Scale:	&	$1.07^{''}$ pixel$^{-1}$	\\
Field:	&	$18^{'}\times18^{'}$	\\
Gain:	&	1.75 $e^{-}$ ADU$^{-1}$	\\
Read Out Noise:	&	16 $e^{-}$ rms	\\
Binning used:	&	$1\times1$	\\
Typical seeing:	&	1.5 arcsec	\\
\enddata
\end{deluxetable}

\begin{deluxetable}{ccccr}
\tablecaption{Observation log of photometric
observations\label{tbl-3}} \tablewidth{0pt}
\tablehead{\colhead{Date(UT)} & \colhead{Band}& \colhead{Number of observations} &
\colhead{Time spans(h)} & \colhead{Time resolutions(min)}}
\startdata
2012 Jan 27	&	I	&	126	&	6.9 	&	3.2 	\\
2012 Jan 27	&	R	&	126	&	6.9 	&	3.2 	\\
2012 Jan 27	&	V	&	126	&	6.9 	&	3.2 	\\
2012 Jan 27	&	B	&	126	&	6.9 	&	3.2 	\\
2012 Jan 28	&	I	&	74	&	7.2 	&	5.9 	\\
2012 Jan 28	&	R	&	74	&	7.2 	&	5.9 	\\
2012 Jan 28	&	V	&	74	&	7.2 	&	5.6 	\\
2012 Jan 28	&	B	&	73	&	7.1 	&	5.9 	\\
2012 Jan 29	&	I	&	72	&	7.2 	&	5.9 	\\
2012 Jan 29	&	R	&	71	&	7.2 	&	5.9 	\\
2012 Jan 29	&	V	&	73	&	7.2 	&	5.9 	\\
2012 Jan 29	&	B	&	73	&	7.2 	&	5.9 	\\
2012 Jan 30	&	I	&	71	&	6.9 	&	5.9 	\\
2012 Jan 30	&	R	&	71	&	6.9 	&	5.9 	\\
2012 Jan 30	&	V	&	71	&	6.9 	&	5.9 	\\
2012 Jan 30	&	B	&	71	&	6.9 	&	5.9 	\\
2012 Jan 31	&	I	&	66	&	6.8 	&	5.9 	\\
2012 Jan 31	&	R	&	66	&	6.8 	&	5.9 	\\
2012 Jan 31	&	V	&	66	&	6.8 	&	5.9 	\\
2012 Jan 31	&	B	&	67	&	6.8 	&	5.9 	\\
2012 Feb 01	&	I	&	58	&	5.6 	&	5.9 	\\
2012 Feb 01	&	R	&	59	&	5.7 	&	5.9 	\\
2012 Feb 01	&	V	&	59	&	5.7 	&	5.9 	\\
2012 Feb 01	&	B	&	59	&	5.7 	&	5.9 	\\
2012 Feb 02	&	I	&	58	&	6.8 	&	5.9 	\\
2012 Feb 02	&	R	&	58	&	6.8 	&	5.9 	\\
2012 Feb 02	&	V	&	57	&	6.6 	&	5.9 	\\
2012 Feb 02	&	B	&	59	&	6.8 	&	5.9 	\\
2012 Feb 03	&	I	&	30	&	5.1 	&	3.7 	\\
2012 Feb 03	&	R	&	30	&	5.1 	&	3.7 	\\
2012 Feb 03	&	V	&	30	&	5.1 	&	3.7 	\\
2012 Feb 03	&	B	&	28	&	5.0 	&	3.7 	\\
2012 Feb 04	&	I	&	78	&	6.2 	&	4.7 	\\
2012 Feb 04	&	R	&	78	&	6.2 	&	4.7 	\\
2012 Feb 04	&	V	&	78	&	6.2 	&	4.7 	\\
2012 Feb 04	&	B	&	78	&	6.2 	&	4.7 	\\
2012 Feb 05	&	I	&	51	&	7.2 	&	4.7 	\\
2012 Feb 05	&	R	&	51	&	7.2 	&	4.7 	\\
2012 Feb 05	&	V	&	51	&	7.1 	&	4.7 	\\
2012 Feb 05	&	B	&	53	&	7.0 	&	4.7 	\\
2012 Feb 06	&	I	&	35	&	6.4 	&	11.4 	\\
2012 Feb 06	&	R	&	35	&	6.4 	&	11.4 	\\
2012 Feb 06	&	V	&	35	&	6.4 	&	11.4 	\\
2012 Feb 07	&	I	&	66	&	6.9 	&	4.5 	\\
2012 Feb 07	&	R	&	67	&	6.9 	&	4.5 	\\
2012 Feb 07	&	V	&	65	&	6.7 	&	4.5 	\\
2012 Feb 08	&	I	&	94	&	7.0 	&	4.5 	\\
2012 Feb 08	&	R	&	94	&	7.0 	&	4.5 	\\
2012 Feb 08	&	V	&	93	&	7.0 	&	4.5 	\\
\enddata
\end{deluxetable}

\begin{deluxetable}{cccr}
\tablecaption{Data of $B$ Band\label{tbl-2}} \tablewidth{0pt}
\tablehead{\colhead{Date(UT)} & \colhead{MJD} & \colhead{Magnitude}
& \colhead{$\sigma$}} \startdata
2012 Jan 27	&	55953.6895 	&	14.7430 	&	0.0225 	\\
2012 Jan 27	&	55953.6920 	&	14.7385 	&	0.0225 	\\
2012 Jan 27	&	55953.6943 	&	14.6920 	&	0.0225 	\\
2012 Jan 27	&	55953.6966 	&	14.7290 	&	0.0225 	\\
2012 Jan 27	&	55953.6988 	&	14.7315 	&	0.0225 	\\

\enddata
\tablecomments{Column (1) is the universal time (UT) of observation,
column (2) the corresponding modified Julian day (MJD), column (3)
the magnitude, column (4) the rms error. Table 3 is available in its entirety in the
electronic edition of the {\sl The Astronomical Journal}. A portion is shown here for guidance regarding its form
and content.}
\end{deluxetable}

\begin{deluxetable}{cccr}
\tablecaption{Data of $V$ Band\label{tbl-2}} \tablewidth{0pt}
\tablehead{\colhead{Date(UT)} & \colhead{MJD} & \colhead{Magnitude}
& \colhead{$\sigma$}} \startdata
2012 Jan 27	&	55953.6901 	&	14.2280 	&	0.0133 	\\
2012 Jan 27	&	55953.6924 	&	14.2300 	&	0.0133 	\\
2012 Jan 27	&	55953.6946 	&	14.2095 	&	0.0133 	\\
2012 Jan 27	&	55953.6969 	&	14.2405 	&	0.0133 	\\
2012 Jan 27	&	55953.6992 	&	14.2090 	&	0.0133 	\\

\enddata
\tablecomments{The meaning of each column is the same as that in Table 3. Table 4 is available in its entirety in the
electronic edition of the {\sl The Astronomical Journal}. A portion is shown here for guidance regarding its form
and content.}
\end{deluxetable}

\begin{deluxetable}{cccr}
\tablecaption{Data of $R$ Band\label{tbl-3}} \tablewidth{0pt}
\tablehead{\colhead{Date(UT)} & \colhead{MJD} & \colhead{Magnitude}
& \colhead{$\sigma$}} \startdata
2012 Jan 27	&	55953.6904 	&	13.8020 	&	0.0112 	\\
2012 Jan 27	&	55953.6927 	&	13.8100 	&	0.0112 	\\
2012 Jan 27	&	55953.6950 	&	13.8215 	&	0.0112 	\\
2012 Jan 27	&	55953.6973 	&	13.8045 	&	0.0112 	\\
2012 Jan 27	&	55953.6996 	&	13.8015 	&	0.0112 	\\

\enddata
\tablecomments{The meaning of each column is the same as that in Table 3. Table
5 is available in its entirety in the electronic edition of the {\sl
The Astronomical Journal}. A portion is shown here for
guidance regarding its form and content.}
\end{deluxetable}

\begin{deluxetable}{cccr}
\tablecaption{Data of $I$ Band\label{tbl-4}} \tablewidth{0pt}
\tablehead{\colhead{Date(UT)} & \colhead{MJD} & \colhead{Magnitude}
& \colhead{$\sigma$}} \startdata
2012 Jan 27	&	55953.6908 	&	13.2665 	&	0.0117 	\\
2012 Jan 27	&	55953.6931 	&	13.2760 	&	0.0117 	\\
2012 Jan 27	&	55953.6954 	&	13.2750 	&	0.0117 	\\
2012 Jan 27	&	55953.6976 	&	13.2590 	&	0.0117 	\\
2012 Jan 27	&	55953.6999 	&	13.2750 	&	0.0117 	\\
\enddata
\tablecomments{The meaning of each column is the same as that in Table 3. Table
6 is available in its entirety in the electronic edition of the {\sl
The Astronomical Journal}. A portion is shown here for
guidance regarding its form and content.}
\end{deluxetable}

\begin{deluxetable}{cccccccccr}
\small
\tablecaption{Results of IDV Observations of S5 0716+714\label{tbl-5}}
\tablewidth{0pt} \tablehead{\colhead{Date} & \colhead{Band}  & \colhead{$C$} & \colhead{$F$} &\colhead{$F_C(99)$}
&\colhead{$F_A$} &\colhead{$F_A(99)$} &\colhead{V/N} &\colhead{A(\%)} &\colhead{Ave(mag)}\\
\colhead{(1)} & \colhead{(2)}  & \colhead{(3)} &
\colhead{(4)} &\colhead{(5)} &\colhead{(6)} &\colhead{(7)}
&\colhead{(8)} &\colhead{(9)} &\colhead{(10)}} \startdata
2012 Jan 27	&	I	&	3.69 	&	13.60 	&	1.53 	&	52.89 	&	1.98 	&	V	&	17.12 	&	13.18 	\\
2012 Jan 27	&	R	&	4.25 	&	18.08 	&	1.53 	&	58.32 	&	1.98 	&	V	&	18.23 	&	13.72 	\\
2012 Jan 27	&	V	&	3.66 	&	13.38 	&	1.53 	&	36.13 	&	1.98 	&	V	&	18.10 	&	14.13 	\\
2012 Jan 27	&	B	&	2.78 	&	7.75 	&	1.53 	&	8.02 	&	1.98 	&	V	&	27.92 	&	14.63 	\\
2012 Jan 28	&	I	&	4.00 	&	15.99 	&	1.73 	&	27.60 	&	2.44 	&	V	&	15.61 	&	13.31 	\\
2012 Jan 28	&	R	&	3.93 	&	15.46 	&	1.73 	&	48.93 	&	2.44 	&	V	&	12.85 	&	13.94 	\\
2012 Jan 28	&	V	&	2.93 	&	8.60 	&	1.73 	&	31.00 	&	2.44 	&	V	&	25.09 	&	14.30 	\\
2012 Jan 28	&	B	&	2.73 	&	7.44 	&	1.74 	&	28.37 	&	2.45 	&	V	&	17.13 	&	14.80 	\\
2012 Jan 29	&	I	&	1.81 	&	3.33 	&	1.75 	&	5.09 	&	2.45 	&	PV	&	 	&	12.95 	\\
2012 Jan 29	&	R	&	2.60 	&	6.76 	&	1.75 	&	24.43 	&	2.46 	&	V	&	13.96 	&	13.43 	\\
2012 Jan 29	&	V	&	3.06 	&	9.37 	&	1.74 	&	26.68 	&	2.45 	&	V	&	12.81 	&	13.91 	\\
2012 Jan 29	&	B	&	2.26 	&	5.12 	&	1.74 	&	14.55 	&	2.45 	&	PV	&	 	&	14.41 	\\
2012 Jan 30	&	I	&	3.76 	&	14.15 	&	1.75 	&	57.76 	&	2.46 	&	V	&	13.60 	&	13.06 	\\
2012 Jan 30	&	R	&	4.43 	&	19.62 	&	1.75 	&	49.74 	&	2.46 	&	V	&	16.46 	&	13.60 	\\
2012 Jan 30	&	V	&	3.79 	&	14.33 	&	1.75 	&	42.76 	&	2.46 	&	V	&	17.16 	&	14.03 	\\
2012 Jan 30	&	B	&	2.75 	&	7.56 	&	1.75 	&	40.27 	&	2.46 	&	V	&	18.20 	&	14.52 	\\
2012 Jan 31	&	I	&	1.86 	&	3.46 	&	1.79 	&	6.01 	&	2.54 	&	PV	&		&	12.99 	\\
2012 Jan 31	&	R	&	1.69 	&	2.85 	&	1.79 	&	6.41 	&	2.54 	&	PV	&	 	&	13.51 	\\
2012 Jan 31	&	V	&	2.31 	&	5.33 	&	1.79 	&	5.78 	&	2.54 	&	PV	&	 	&	13.92 	\\
2012 Jan 31	&	B	&	1.82 	&	3.31 	&	1.78 	&	4.85 	&	2.53 	&	PV	&	 	&	14.39 	\\
2012 Feb 01	&	I	&	6.90 	&	47.66 	&	1.87 	&	164.48 	&	2.66 	&	V	&	17.07 	&	12.72 	\\
2012 Feb 01	&	R	&	8.26 	&	68.23 	&	1.86 	&	77.19 	&	2.65 	&	V	&	22.05 	&	13.22 	\\
2012 Feb 01	&	V	&	4.53 	&	20.54 	&	1.86 	&	123.93 	&	2.65 	&	V	&	18.06 	&	13.63 	\\
2012 Feb 01	&	B	&	1.96 	&	3.86 	&	1.86 	&	72.58 	&	2.65 	&	PV	&	 	&	14.11 	\\
2012 Feb 02	&	I	&	1.36 	&	1.86 	&	1.87 	&	24.90 	&	2.66 	&	PV	&	 	&	12.45 	\\
2012 Feb 02	&	R	&	1.37 	&	1.87 	&	1.87 	&	11.81 	&	2.66 	&	PV	&	 	&	12.92 	\\
2012 Feb 02	&	V	&	0.82 	&	0.68 	&	1.88 	&	12.49 	&	2.73 	&	PV	&	 	&	13.30 	\\
2012 Feb 02	&	B	&	0.77 	&	0.62 	&	1.86 	&	12.21 	&	2.65 	&	PV	&	 	&	13.74 	\\
2012 Feb 03	&	I	&	2.08 	&	4.33 	&	2.42 	&	22.41 	&	3.90 	&	PV	&	 	&	12.47 	\\
2012 Feb 03	&	R	&	1.95 	&	3.80 	&	2.42 	&	17.45 	&	3.90 	&	PV	&	 	&	12.99 	\\
2012 Feb 03	&	V	&	1.68 	&	2.84 	&	2.42 	&	8.02 	&	3.90 	&	PV	&	 	&	13.38 	\\
2012 Feb 03	&	B	&	1.81 	&	3.26 	&	2.51 	&	3.62 	&	3.99 	&	PV	&		&	13.86 	\\
2012 Feb 04	&	I	&	5.36 	&	28.77 	&	1.71 	&	92.85 	&	2.34 	&	V	&	19.55 	&	12.74 	\\
2012 Feb 04	&	R	&	3.64 	&	13.26 	&	1.71 	&	77.16 	&	2.34 	&	V	&	26.53 	&	13.29 	\\
2012 Feb 04	&	V	&	2.57 	&	6.62 	&	1.71 	&	30.70 	&	2.34 	&	V	&	28.24 	&	13.72 	\\
2012 Feb 04	&	B	&	1.36 	&	1.86 	&	1.71 	&	17.71 	&	2.34 	&	PV	&	 	&	14.20 	\\
2012 Feb 05	&	I	&	1.25 	&	1.58 	&	1.95 	&	15.88 	&	2.89 	&	PV	&	 	&	12.71 	\\
2012 Feb 05	&	R	&	1.21 	&	1.49 	&	1.95 	&	7.92 	&	2.89 	&	PV	&	 	&	13.21 	\\
2012 Feb 05	&	V	&	0.95 	&	0.94 	&	1.95 	&	4.20 	&	2.89 	&	PV	&	 	&	13.52 	\\
2012 Feb 05	&	B	&	1.43 	&	2.06 	&	1.92 	&	13.76 	&	2.85 	&	PV	&	 	&	14.13 	\\
2012 Feb 06	&	I	&	0.78 	&	0.61 	&	2.26 	&	1.00 	&	3.53 	&	N	&		&	12.40 	\\
2012 Feb 06	&	R	&	0.90 	&	0.80 	&	2.26 	&	5.29 	&	3.53 	&	PV	&	 	&	12.86 	\\
2012 Feb 06	&	V	&	0.79 	&	0.62 	&	2.26 	&	4.51 	&	3.53 	&	PV	&		&	13.21 	\\
2012 Feb 07	&	I	&	1.22 	&	1.54 	&	1.79 	&	41.82 	&	2.53 	&	PV	&		&	12.52 	\\
2012 Feb 07	&	R	&	1.44 	&	2.07 	&	1.78 	&	15.88 	&	2.53 	&	PV	&		&	13.06 	\\
2012 Feb 07	&	V	&	1.21 	&	1.45 	&	1.76 	&	5.87 	&	2.55 	&	PV	&	 	&	13.44 	\\
2012 Feb 08	&	I	&	2.74 	&	7.52 	&	1.63 	&	92.30 	&	2.21 	&	V	&	29.27 	&	12.31 	\\
2012 Feb 08	&	R	&	2.97 	&	8.84 	&	1.63 	&	110.20 	&	2.21 	&	V	&	33.22 	&	12.81 	\\
2012 Feb 08	&	V	&	1.89 	&	3.58 	&	1.63 	&	27.82 	&	2.22 	&	PV	&	 	&	13.21 	\\
\enddata
\tablecomments{Column 1 is the date of observation, column 2 the
observed band, column 3 the
value of $C$ test, column 4 the average $F$ value, column 5 the
critical $F$ value with 99 per cent confidence level, column 6 the
$F$ value of ANOVA, column 7 the critical $F$ value of ANOVA with 99
per cent confidence level, column 8 the variability status (V:
variable, PV: probable variable, N: non-variable), column 9-10 the
variability amplitude and daily average magnitudes respectively.}
\end{deluxetable}

\begin{deluxetable}{ccccr}
\small
\tablecaption{Results of change rate. \label{tbl-4}} \tablewidth{0pt}
\tablehead{\colhead{Date(UT)} & \colhead{Band} & \colhead{MJD$^\ast$} & \colhead{Magnitude$^\ast$} &\colhead{CR(Mag/h)}} \startdata
2012 Jan 27	&	B	&	55953.6895 	&	14.743 	&	0.037	\\
	&		&	55953.8845 	&	14.517 	&	-0.075	\\
	&		&	55953.9619 	&	14.782 	&		\\
	&	V	&	55953.7152 	&	14.248 	&	0.039	\\
	&		&	55953.8573 	&	14.066 	&	-0.026	\\
	&		&	55953.9668 	&	14.188 	&		\\
	&	R	&	55953.7064 	&	13.825 	&	0.038	\\
	&		&	55953.8714 	&	13.642 	&	-0.04	\\
	&		&	55953.9671 	&	13.763 	&		\\
	&	I	&	55953.7091 	&	13.291 	&	0.03	\\
	&		&	55953.8879 	&	13.119 	&	-0.031	\\
	&		&	55953.9744 	&	13.225 	&		\\
2012 Jan 28	&	B	&	55954.6851 	&	14.889 	&	0.034	\\
	&		&	55954.8249 	&	14.749 	&	-0.032	\\
	&		&	55954.8988 	&	14.828 	&		\\
	&	V	&	55954.6859 	&	14.380 	&	0.037	\\
	&		&	55954.8052 	&	14.255 	&	-0.032	\\
	&		&	55954.8586 	&	14.322 	&		\\
	&	R	&	55954.6865 	&	13.942 	&	0.031	\\
	&		&	55954.8346 	&	13.821 	&	-0.054	\\
	&		&	55954.8674 	&	13.882 	&		\\
	&	I	&	55954.6871 	&	13.380 	&	0.029	\\
	&		&	55954.8352 	&	13.281 	&		\\
	&		&	55954.9049 	&	13.328 	&		\\
2012 Jan 29	&	V	&	55955.6805 	&	13.831 	&	-0.026	\\
	&		&	55955.8317 	&	13.960 	&	0.027	\\
	&		&	55955.9753 	&	13.842 	&		\\
	&	R	&	55955.6811 	&	13.429 	&	-0.02	\\
	&		&	55955.8405 	&	13.534 	&	0.03	\\
	&		&	55955.9718 	&	13.434 	&		\\
2012 Jan 30	&	B	&	55956.6878 	&	14.401 	&	-0.044	\\
	&		&	55956.8243 	&	14.584 	&	0.021	\\
	&		&	55956.9602 	&	14.462 	&		\\
	&	V	&	55956.6886 	&	13.924 	&	-0.042	\\
	&		&	55956.8333 	&	14.096 	&	0.032	\\
	&		&	55956.9487 	&	13.976 	&		\\
	&	R	&	55956.6892 	&	13.506 	&	-0.038	\\
	&		&	55956.8298 	&	13.671 	&	0.025	\\
	&		&	55956.9657 	&	13.569 	&		\\
	&	I	&	55956.6898 	&	12.975 	&	-0.032	\\
	&		&	55956.8345 	&	13.111 	&	0.024	\\
	&		&	55956.9745 	&	13.013 	&		\\
2012 Feb 01	&	V	&	55958.8348 	&	13.694 	&	0.054	\\
	&		&	55958.9539 	&	13.522 	&		\\
	&	R	&	55958.8313 	&	13.285 	&	0.052	\\
	&		&	55958.9586 	&	13.136 	&		\\
	&	I	&	55958.8319 	&	12.773 	&	0.043	\\
	&		&	55958.9798 	&	12.607 	&		\\
2012 Feb 04	&	V	&	55961.7198 	&	13.616 	&	-0.019	\\
	&		&	55961.9772 	&	13.899 	&		\\
	&	R	&	55961.7205 	&	13.185 	&	-0.024	\\
	&		&	55961.9778 	&	13.451 	&		\\
	&	I	&	55961.7211 	&	12.636 	&	-0.027	\\
	&		&	55961.9718 	&	12.868 	&		\\
2012 Feb 08	&	R	&	55965.6871 	&	12.979 	&	0.039	\\
	&		&	55965.9327 	&	12.654 	&		\\
	&	I	&	55965.6911 	&	12.457 	&	0.034	\\
	&		&	55965.9430 	&	12.162 	&		\\
\enddata
\tablecomments{The `$\ast$' stands for increasing/decreasing time points and corresponding magnitudes; When calculating the change rate, we consider the coefficients of correlation $>0.7$; for the CR, the positive and negative signs are increasing brightness and decreasing brightness respectively.}
\end{deluxetable}

\begin{deluxetable}{cccr}
\small
\tablecaption{Results of Spearman rank analysis. \label{tbl-4}} \tablewidth{0pt}
\tablehead{\colhead{Date(UT)} & \colhead{$N$} & \colhead{$r$} & \colhead{$P$}} \startdata
2012 Jan 27 & 126 & 0.036 & 0.69 \\
2012 Jan 28 & 70 & 0.38 & 0.001 \\
2012 Jan 29 & 69 & 0.38 & 0.001 \\
2012 Jan 30 & 71 & 0.43 & 0.0002 \\
2012 Jan 31 & 66 & 0.25 & 0.04 \\
2012 Feb 01 & 57 & 0.14 & 0.32 \\
2012 Feb 02 & 56 & -0.05 & 0.73 \\
2012 Feb 03 & 30 & 0.53 & 0.002 \\
2012 Feb 04 & 77 & -0.27 & 0.02 \\
2012 Feb 05 & 51 & 0.25 & 0.07 \\
2012 Feb 06 & 35 & 0.41 & 0.01 \\
2012 Feb 07 & 65 & 0.36 & 0.003 \\
2012 Feb 08 & 92 & 0.29 & 0.006 \\
2012 Jan 28-Feb 08 & 866 & 0.56 & $1\times10^{-6}$ \\
\enddata
\tablecomments{$N$ is number of data; $r$ is the
coefficient of correlation; $P$ is the significance level.}
\end{deluxetable}

\end{document}